\documentclass{aa}
\usepackage[varg]{txfonts}
\usepackage{graphicx}
\usepackage{ulem}
\usepackage{color}

\usepackage{natbib}
\bibpunct{(}{)}{;}{a}{}{,} % to follow the A&A style

\begin{document}

\title{Formation of close-in super-Earths in evolving protoplanetary disks due to disk winds}
\titlerunning{Formation of close-in super-Earths}
\author{Masahiro Ogihara\inst{\ref{inst1}}
\and Eiichiro Kokubo\inst{\ref{inst1}}
\and Takeru K. Suzuki\inst{\ref{inst2}}
\and Alessandro Morbidelli\inst{\ref{inst3}}
}
\institute{Division of Theoretical Astronomy, National Astronomical Observatory of Japan, 2-21-1, Osawa, Mitaka, 181-8588 Tokyo, Japan \email{masahiro.ogihara@nao.ac.jp}\label{inst1}
\and School of Arts \& Sciences, University of Tokyo, 3-8-1, Komaba, Meguro, 153-8902 Tokyo, Japan\label{inst2}
\and Laboratoire Lagrange, Universit\'e C\^ote d'Azur, Observatoire de la C\^ote d'Azur, CNRS,
Blvd de l'Observatoire, CS 34229, 06304 Nice Cedex 4, France\label{inst3}
}
\date{Received 29 January 2018 / Accepted 19 March 2018}

%limited to 300 words
\abstract 
%Context
{
Planets with masses larger than about $0.1~M_\oplus$ undergo rapid inward migration (type I migration) in a standard protoplanetary disk. Recent magnetohydrodynamical simulations revealed the presence of magnetically driven disk winds, which would alter the disk profile and the type I migration in the close-in region. 
} 
%Aims
{
We investigate orbital evolution of planetary embryos in disks that viscously evolve under the effects of disk winds. The aim is to discuss effects of altered disk profiles on type I migration. In addition, we aim to examine whether observed distributions of close-in super-Earths can be reproduced by simulations that include effects of disk winds.
} 
%Methods
{
We perform \textit{N}-body simulations of super-Earth formation from planetary embryos, in which a recent model for disk evolution is used. We explore a wide range of parameters and draw general trends. We also carry out \textit{N}-body simulations of close-in super-Earth formation from embryos in such disks under various conditions.
} 
%Results
{
We find that the type I migration is significantly suppressed in many cases. Even in cases in which inward migration occurs, the migration timescale is lengthened to 1~Myr, which mitigates the type I migration problem. This is because the gas surface density is decreased and has a flatter profile in the close-in region due to disk winds. 
We find that when the type I migration is significantly suppressed, planets undergo late orbital instability during the gas depletion, leading to a non-resonant configuration. We also find that observed distributions of close-in super-Earths (e.g., period ratio, mass ratio) can be reproduced. In addition, we show that in some results of simulations,  systems with a chain of resonant planets, like the TRAPPIST-1 system, form. 
} 
%Conclusions
{}
\keywords{Planets and satellites: formation -- Protoplanetary disks -- Planet-disk interactions -- Methods: numerical}
\maketitle

\section{Introduction}

Recent magnetohydrodynamic (MHD) simulations for protoplanetary disks revealed the existence of magnetically driven disk winds, which includes magnetorotational instability (MRI)-driven disk winds and magnetocentrifugal disk winds (e.g., \citealt{suzuki_inutsuka09}; \citealt{fromang_etal13}; \citealt{bethune_etal17}). \citet{suzuki_etal10} investigated the global disk evolution with mass loss due to MRI-driven disk winds, and found that  disk profiles in the close-in region can be altered from a power-law distribution. Inside $\simeq 1 {~\rm au}$ from the central star, the slope of gas surface density of the disk can be flat or even positive. In such a case, several processes in planet formation would be altered from previously investigated ones. As an example, this would affect formation of terrestrial planets and their orbital evolution. In fact, we showed, using \textit{N}-body simulations, that orbital migration of Earth-sized planets (type I migration) is modified in disks evolving via disk winds \citep{ogihara_etal15b,ogihara_etal15c}.

The global disk evolution model by \citet{suzuki_etal10} was further developed by \citet{suzuki_etal16} to incorporate the effects of disk wind torque (magnetic braking), which triggers mass accretion \citep{bai_stone13,gressel_etal15,simon_etal15}, as well as mass loss and energetics. They found that the mass accretion induced by the wind torque (wind-driven accretion) is important; in MRI-inactive disks, the wind-driven accretion can dominate over the turbulence-driven accretion.

In this paper, we investigate orbital evolution of planetary embryos in the latest disk model with magnetically driven disk winds developed by \citet{suzuki_etal16} by performing \textit{N}-body simulations. There exists uncertainties in the disk evolution model. Therefore, in \citet{suzuki_etal16}, they used several parameters and showed possible disk evolution. In this work, we also employ a range of parameters and perform \textit{N}-body simulations in various disks to look for general trends of terrestrial planet formation in disks with disk winds. In particular, we aim to reveal how disk winds affect type I migration. We note that in a separate paper (\citealt{ogihara_etal18}; hereafter Paper~I), we demonstrated that the localized orbital configuration of  terrestrial planets in the solar system is reproduced by simulation with specific model parameters.

Subsequently, we apply our model to in-situ formation of close-in super-Earth systems. Observations of extrasolar planets have revealed a number of close-in super-Earths. As of August 2017, the number of confirmed multiple close-in super-Earth systems is 432, which harbor 1087 planets with masses $M < 100~M_\oplus$  (or with radii $R < 10~R_\oplus$) and semimajor axes $a < 1 {\rm ~au}$ (or with orbital periods $P < 200 {\rm ~day}$). Given the large population, we can obtain statistical information of close-in super-Earths (e.g., period ratio distribution, mass distribution).

Formation models of close-in super-Earths are roughly divided into two groups\footnote{There are several ways to define the formation model. In this paper, we define the model according to the region where planetary embryos that eventually grow to close-in super-Earths form. If planetary embryos are located in the close-in region, we call the model an ``in-situ formation'' model.}; namely, in-situ formation models (e.g., \citealt{hansen_murray12,hansen_murray13}; \citealt{ogihara_etal15a}) and migration models (e.g., \citealt{cossou_etal14}; \citealt{izidoro_etal17}). Here we focus on the in-situ formation model. \citet{hansen_murray12,hansen_murray13} performed \textit{N}-body simulations of in-situ formation of close-in super-Earths from massive planetary embryos in a gas-free environment, and showed that several observed properties (e.g., distributions of orbital periods) can be reproduced. They ignored the existence of gas; however, a gas disk should exist during the growth of planetary embryos because the growth timescale of protoplanets is much shorter than the disk lifetime.

\citet{ogihara_etal15a} investigated the in-situ accretion of close-in super-Earths using \textit{N}-body simulations that include effects of a gas disk assuming a simple power-law disk based on the minimum-mass solar nebular (MMSN) model (\citealt{weidenschilling77}; \citealt{hayashi81}). They found that planets grow and migrate very rapidly, and observed that distributions of period ratio and mass ratio cannot be reproduced by their simulations. This is because super-Earths obtain a relatively compact configuration immediately outside the disk inner edge due to rapid inward migration. Population synthesis calculations also show that planets tend to pile up near the disk inner edge assuming a rapid inward type I migration (e.g., \citealt{ida_lin08}), while the observed distribution of close-in super-Earths does not show such a pileup.

\citet{ogihara_etal15c} investigated the in-situ formation of close-in super-Earths using the disk model of \citet{suzuki_etal10} which considered MRI-driven disk winds. They pointed out that type I migration can be suppressed in some cases (e.g., flat gas surface density profile), which would help in reproducing the observed properties of close-in super-Earths.

In this work, formation of close-in super-Earths is investigated using the recent disk model of magnetically driven disk winds \citep{suzuki_etal16} in combination with \textit{N}-body simulations. One of the aims of this work is to discuss whether observed distributions of close-in super-Earths can be reproduced by simulations.
Considering the effects of magnetically driven disk winds, \citet{ogihara_etal17} investigated the effects of global gas flows on type I migration, focusing on the desaturation of corotation torque. In this paper, we do not consider the effects of global gas flows on type I migration, but we explore a wide range of simulation parameters and present a detailed analysis of final configurations.

We note that in this paper the disk profile is altered due to magnetically driven disk winds. However, one can apply our results to more general cases with altered disk profiles. For example, disks obtain flat or positive profiles if the gas density is set at zero at the inner boundary \citep{lynden-bell_pringle74}.

The paper is organized as follows. In Section~\ref{sec:model}, we present the numerical models of disk evolution, type I migration, and \textit{N}-body simulation. In Section~\ref{sec:various}, we show how the type I migration is altered in an evolving disk due to disk winds by \textit{N}-body simulations from the isolation-mass protoplanets.
In Section~\ref{sec:SE}, we apply our model to the formation of close-in super-Earths and investigate whether observed orbital distributions are reproduced. In Section~\ref{sec:discussion}, we discuss results of simulations of super-Earth formation. In Section~\ref{sec:conclusions}, we make concluding remarks.

\section{Model}
\label{sec:model}
\subsection{Disk evolution by \citet{suzuki_etal16}}
\label{sec:disk_model}
According to \citet{suzuki_etal16}, evolution of gas surface density is calculated by solving the diffusion equation:
\begin{eqnarray}
\label{eq:diffusion}
\frac{\partial \Sigma_{\rm g}}{\partial t} = \frac{1}{r} \frac{\partial}{\partial r} \left[\frac{2}{r\Omega} \left\{ \frac{\partial}{\partial r} (r^2 \Sigma_{\rm g} \overline{\alpha_{r,\phi}} c_{\rm s}^2) + r^2 \overline{\alpha_{\phi,z}} \frac{\Sigma_{\rm g} H \Omega^2}{2 \sqrt{\pi}} \right\} \right] \nonumber \\
 - C_{\rm w} \frac{\Sigma_{\rm g} \Omega}{\sqrt{2 \pi}},
\end{eqnarray}
where $\Omega$ is the Keplerian frequency. Here there are three parameters; $\overline{\alpha_{r,\phi}}$ indicates an effective turbulent viscosity based on a \citet{shakura_sunyaev73} $\alpha$ viscosity parameterization. Parameters $\overline{\alpha_{\phi,z}}$ and $C_{\rm w}$ are used as measures of the angular momentum loss due to the wind torque and mass loss due to disk winds, respectively. For the parameters, we employ the same values as in \citet{suzuki_etal16}, which are based on MHD simulations (\citealt{suzuki_etal10}; \citealt{bai13}). We consider two cases for $\overline{\alpha_{r,\phi}}$; $\overline{\alpha_{r,\phi}} = 8\times 10^{-5}$ represents MRI-inactive disks, and $\overline{\alpha_{r,\phi}} = 8\times 10^{-3}$ represents MRI-active disks. MRI-inactive disks correspond to cases that take into account a dead zone by non-ideal MHD effects. Here radially constant values are assumed. For the description of $\overline{\alpha_{\phi,z}}$, we usually use
 \begin{eqnarray}
\label{eq:bai}
\overline{\alpha_{\phi,z}} = 10^{-5} \left(\frac{\Sigma_{\rm g}}{\Sigma_{\rm g,ini}}\right)^{-0.66}.
\end{eqnarray}
In some simulations, a constant $\overline{\alpha_{\phi,z}}$ is used ($\overline{\alpha_{\phi,z}} = 10^{-4}$). We set upper limits on $C_{\rm w}$ as $C_{\rm w} = 1 \times 10^{-5}$ and $2 \times 10^{-5}$ for inactive and active disks, respectively. In simulations, values of $C_{\rm w}$ are generally smaller than the upper limits because the parameter $C_{\rm w}$ is constrained by energetics (see Figs.~2 and 6 in \citealt{suzuki_etal16}). We usually use the ``strong DW'' regime (Eqs.~18 and 19 in \citealt{suzuki_etal16}), while several simulations are performed using the ``weak DW'' regime (Eqs.~20 and 21 in \citealt{suzuki_etal16}).

The initial gas surface density profile is given by 
\begin{eqnarray}
\Sigma_{\rm g,ini} = 1.7 \times 10^4 \left(\frac{r}{1 {\rm ~au}}\right)^{-3/2} \exp\left(\frac{-r}{30 {\rm ~au}}\right) ~{\rm g~cm^{-2}}.
\end{eqnarray}
The gas surface density at $r = 1$~au is ten times more massive than the MMSN model in order to consider early disk evolution.
We note that in this work we consider late stage of planet formation and start \textit{N}-body simulations with planetary embryos. Taking into account the growth time of the embryos, we  refer to $t = t_0$ in the disk evolution as ``initial ($t=0$)'' in \textit{N}-body simulations. We use $t_0 = 0.1~{\rm Myr}$ for our standard model in this paper. For temperature evolution, we refer readers to Section~2.4 of \citet{suzuki_etal16}.

\begin{figure}
\begin{center}
\resizebox{0.9 \hsize}{!}{\includegraphics{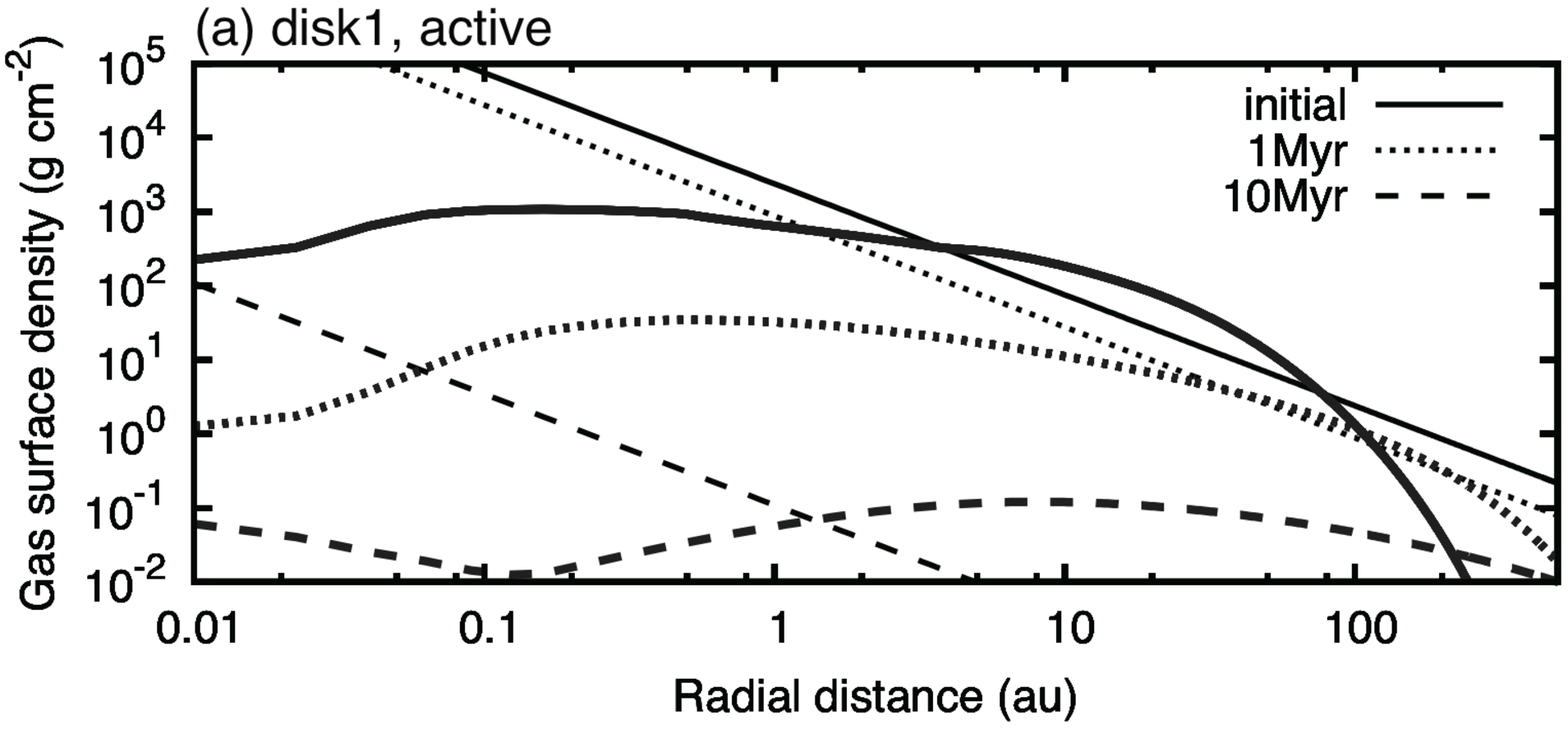}}
\resizebox{0.9 \hsize}{!}{\includegraphics{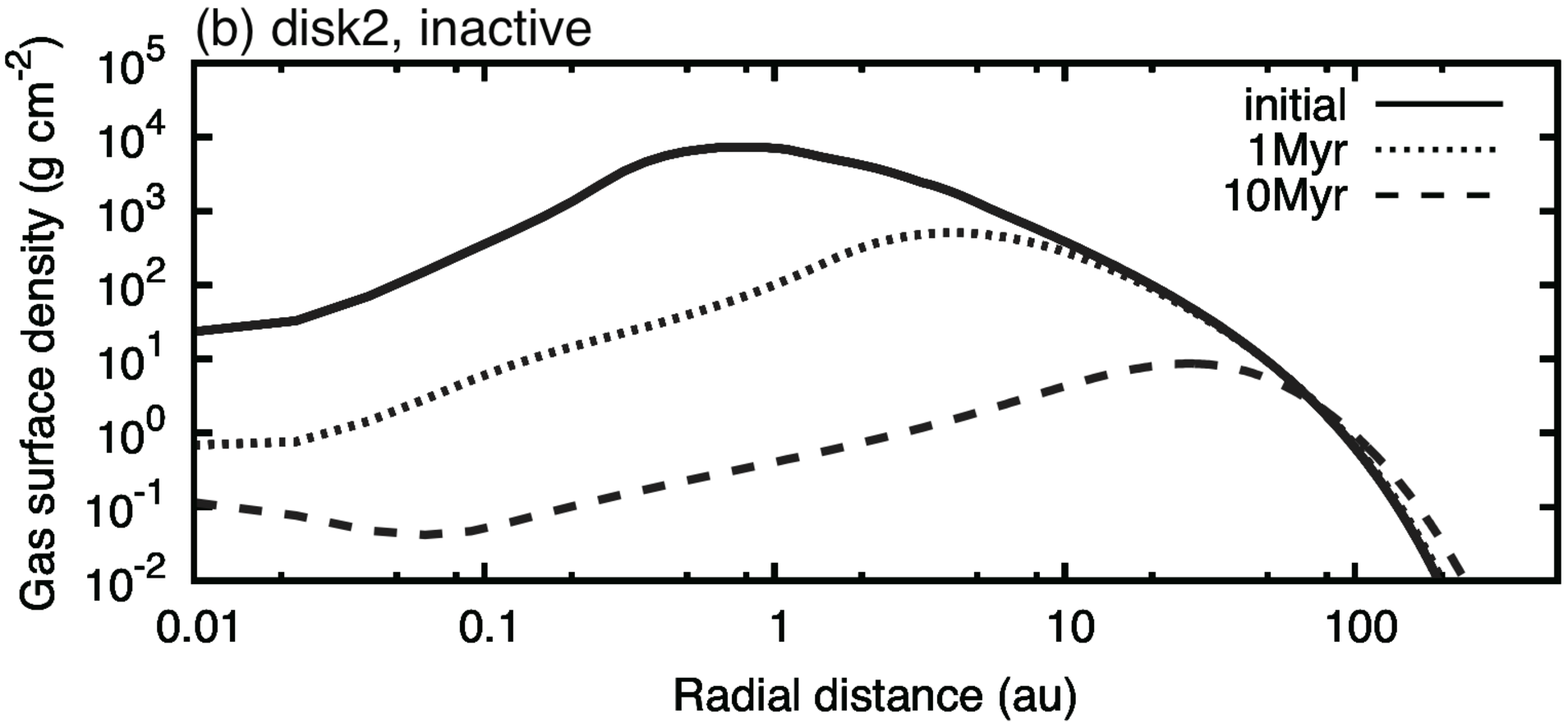}}
\end{center}
\caption{Time evolution of gas surface density of MRI-active ($\overline{\alpha_{r,\phi}}=8 \times 10^{-3}$, panel~(a)) and MRI-inactive disks ($\overline{\alpha_{r,\phi}}=8 \times 10^{-5}$, panel~(b)) based on \citet{suzuki_etal16}. The gas surface density evolution for disk0 which is a power-law distribution based on the MMSN model is also shown in panel~(a). The initial state corresponds to the time of 0.1~Myr after the disk evolution starts in \citet{suzuki_etal16} ($t_0 = 0.1~{\rm Myr}$).
}
\label{fig:r_sigma}
\end{figure}

Table~\ref{tbl:list1} summarizes all disk models used in this paper. Figure~\ref{fig:r_sigma} shows the evolution of gas surface density. The thin lines in Fig.~\ref{fig:r_sigma}(a) represent disk0 model, which is a simple power-law disk based on the MMSN model. Under the assumption that the disk dissipates after a few Myr, an exponential decay with a timescale of 1~Myr is considered. Models of disk1 and disk2 are based on the disk evolution in \citet{suzuki_etal16} (see their Figs.~1 and 5).
Figure~\ref{fig:r_sigma}(a) and (b) shows the case of an MRI-active disk (disk1; $\overline{\alpha_{r,\phi}}= 8 \times 10^{-3}$) and an MRI-inactive disk (disk2; $\overline{\alpha_{r,\phi}}= 8 \times 10^{-5}$). In Fig.~\ref{fig:r_sigma}(b), the surface density slope is positive inside a certain $r$, which moves outward with time. In MRI-active disks in Fig.~\ref{fig:r_sigma}(a), a turbulence-driven accretion plays a certain role in disk evolution, and the region of positive slope ($p>0$) is limited to $r \lesssim 0.2 {\rm ~au}$. Compared to cases for MRI-inactive disks, disks give more flat distributions at $r \simeq 0.1-1 {\rm au}$.
In this paper, we mainly use disk1 and disk2 models and investigate the orbital evolution of protoplanets (Section~\ref{sec:various}) and formation of close-in super-Earths (Section~\ref{sec:SE}). In Section~\ref{sec:various}, other disk models (disk0, disk3 - disk10) are also used. However, typical orbital evolution can be understood by disk1 and disk2 models.

\begin{table}
\caption{Summary of disk models. The first column gives the model name, followed by the parameter for effective turbulent viscosity $\overline{\alpha_{r,\phi}}$ and the angular momentum loss due to the wind torque $\overline{\alpha_{\phi,z}}$. The fourth column indicates the regime which is introduced by \citet{suzuki_etal16} and constrains the parameter $C_{\rm w}$. The evolution of gas surface density is shown in Fig.~\ref{fig:r_sigma} for disk0 - disk2 models and in Fig.~\ref{fig:r_sigma_app} for disk3 - disk10 models.}
\label{tbl:list1}
\centering
\begin{tabular}{l l l l l}
\hline\hline
Model& $\overline{\alpha_{r,\phi}}$ & $\overline{\alpha_{\phi,z}}$ &  Energetics&Comments\\
\hline 
disk0&	-&	- &	-&	power-law\\
disk1&	active&	Eq.~(\ref{eq:bai})&	strong&	standard\\
disk2&	inactive&	Eq.~(\ref{eq:bai})&	strong&	standard\\
\hline
disk3&	active&	Eq.~(\ref{eq:bai})&	weak&	-\\
disk4&	active&	$10^{-4}$&	strong&	-\\
disk5&	inactive&	Eq.~(\ref{eq:bai})&	weak&	-\\
disk6&	inactive&	$10^{-4}$&	strong&	-\\
disk7&	active&	Eq.~(\ref{eq:bai})&	strong&	$t_0 = 0~{\rm yr}$\\
disk8&	active&	Eq.~(\ref{eq:bai})&	strong&	$t_0 = 1~{\rm Myr}$\\
disk9&	inactive&	Eq.~(\ref{eq:bai})&	strong&	$t_0 = 0~{\rm yr}$\\
disk10&	inactive&	Eq.~(\ref{eq:bai})&	strong&	$t_0 = 1~{\rm Myr}$\\
\hline
\end{tabular}
\end{table}

%%Fig.1

\subsection{Type I migration}
\label{sec:typeI}

Planets with masses larger than approximately $0.1 ~M_\oplus$ exhibit orbital evolution mainly due to tidal interaction with the disk (type I migration; e.g., \citealt{goldreich_tremain80}; \citealt{ward86}). To incorporate the effect of type I migration, we directly implement torques in the equation of motion for a particle. The type I migration torque is given by
\begin{eqnarray}
\Gamma = \frac{\beta}{2} \Gamma_0,
\end{eqnarray}
where $\beta$ is a coefficient that determines the direction and rate of migration. The torque is normalized by a commonly used value of
\begin{eqnarray}
\Gamma_0 = \left( \frac{M}{M_*} \right) \left( \frac{\Sigma_{\rm g} r^2}{M_*} \right) \left( \frac{c_{\rm s}}{v_{\rm K}} \right)^{-2} M v_{\rm K}^2,
\end{eqnarray}
where $M$, $M_*$, and $v_{\rm K}$ are the mass of the planet, the mass of the host star, and the Keplerian velocity, respectively.

A detailed description of the coefficient $\beta$ is given by Eqs.~(11)-(13) in \citet{ogihara_etal15b}, which is based on \citet{paardekooper_etal11}. We here briefly review the dependence of migration on the surface density slope. The type I migration torque is expressed by the sum of the Lindblad torque and the corotation torque; $\Gamma = \Gamma_{\rm L} + \Gamma_{\rm c}$. Each torque is given by (\citealt{paardekooper_etal10})
\begin{eqnarray}
\Gamma_{\rm L} = (-1.8 - 0.07p + 1.2q) \Gamma_0,\\
\Gamma_{\rm c} = (1.2 + 2.4p - 4.0q) \Gamma_0,
\end{eqnarray}
where $p = \mathrm{d}\ln \Sigma_{\rm g}/ \mathrm{d}\ln r$ is the local surface density slope and $q= \mathrm{d}\ln T/ \mathrm{d}\ln r$ is the local temperature slope. In optically thin disks with radiative equilibrium governed by the radiation from the central star, the temperature slope is given by $q = -1/2$. Here we assume that the adiabatic index is 1.4. Then the torque is obtained simply by summing the two torques as $\Gamma = (-0.6 +2.3p -2.8q) \Gamma_0$. Planets tend to migrate inward for negative surface density slope (e.g., MMSN), while the migration can be outward in disks with positive slope. When the slope is flat, the migration can be suppressed.

The estimate described here is based on the assumption that the corotation torque is fully unsaturated. However, the corotation torque is prone to saturation, and the direction and rate of migration depend on saturation level (e.g., \citealt{paardekooper_etal11}). Some degree of diffusion is needed to avoid the saturation; the diffusion timescale should be comparable to the horseshoe libration time to keep a vortensity gradient in the corotation region.
The actual torque is given by Eqs.~(50)-(53) in \citet{paardekooper_etal11}, in which the saturation of corotation torque is calculated by using parameters of $p_\nu$ and $p_\chi$ with the help of Eqs.~(30) and (31) in \citet{paardekooper_etal11}. Here $\nu$ and $\chi$ represent viscous and thermal coefficients, respectively.
For the prescription of the viscous diffusivity, we consider the turbulent viscosity, in which $\nu = \overline{\alpha_{r, \phi}} (H/r)^2 r^2 \Omega / 3$ is used. The thermal diffusivity $\chi$ is calculated by Eq.~(34) in \citet{paardekooper_etal11}.
\citet{ogihara_etal17} also took into account the desaturation by the gas flow that includes $\overline{\alpha_{\phi,z}}$, and found that the desaturation due to the gas flow could change orbital evolution for MRI-inactive disks. Similar effects are investigated by \citet{mcnally_etal17}. However, it is still uncertain how the gas flow affects the desaturation. Thus we do not consider the effect of gas flows on desaturation in this paper.
The diffusion timescale depends on disk evolution and the libration timescale depends on planetary mass, which make the evaluation of the total toque complicated. ``Migration maps'' are useful to understand the orbital evolution of planets in results of \textit{N}-body simulations (e.g., \citealt{cossou_etal14}; \citealt{ogihara_etal15b}; \citealt{izidoro_etal17}). We discuss results of \textit{N}-body simulations showing migration maps in Section~\ref{sec:various}.

In addition to the $a$-damping (type I migration) due to the tidal interaction between planets and the gas disk, we consider the tidal damping of eccentricity and inclination. Regarding the prescription of damping forces, we use Eqs.~(38)-(40) in \citet{tanaka_ward04} (a typo is corrected by \citet{ogihara_etal07}) with a correction factor according to Eqs.~(11) and (12) in \citet{cresswell_nelson08} for planets with high eccentricities and inclinations.

\subsection{\textit{N}-body method and initial solid distribution}

The method of \textit{N}-body simulations are basically the same as in previous studies (e.g., \citealt{ogihara_etal17}). Orbital evolution is calculated by a fourth-order Hermite scheme with a hierarchical individual time step. When planets collide, perfect accretion is assumed. We assume an internal density of $\rho = 3 {\rm ~g~cm^{-3}}$ in determining the physical radius of a planet.

We start \textit{N}-body simulations with planetary embryos that are distributed in a ring-like region between $r_{\rm in}$ and $r_{\rm out}$ from the central star. We use $r_{\rm in} = 0.1 {\rm ~au}$ and $r_{\rm out} = 2 {\rm ~au}$ for our standard model. The solid distribution is based on a power-law distribution:
\begin{eqnarray}
\label{eq:solid}
\Sigma_{\rm d} = \Sigma_{{\rm d},0} \left( \frac{r}{1 {\rm ~au}}\right)^s  {\rm ~g~cm^{-2}},
\end{eqnarray}
where the scaling factor for solid amount ($\Sigma_{{\rm d},0}$) and the solid slope $(s)$ are parameters. 
In Section~\ref{sec:various}, simulations start with protoplanets with the isolation mass $(M_{\rm iso} = 2\pi a \Delta a \Sigma_{\rm d})$ except for a few runs. The number of initial particles is typically $N = 20-50$. When calculating the isolation mass, the orbital separation is assumed to be ten times the mutual Hill radius ($\Delta a = 10~r_{\rm H}$; \citealt{kokubo_ida98}).
On the other hand, in Section~\ref{sec:SE}, planetary embryos with $M = 0.2 ~M_\oplus$ are randomly distributed according to the initial power-law distribution of Eq.~(\ref{eq:solid}). The number of particles is $N = 100-400$.

\section{Orbital evolution of planetary embryos}
\label{sec:various}

We see orbital evolution of planetary embryos in a disk with an altered density profile due to disk winds. We are especially interested in how type I migration is altered, therefore we start \textit{N}-body simulations with isolation-mass embryos ($M\sim 0.1 - 1 ~M_\oplus$). We focus on the orbital migration in this section; therefore the orbital stability after the gas disk depletion and final properties of the system (e.g., planetary mass, period ratio distribution) are investigated in Section~\ref{sec:SE}.
In this section, we employ three different disk models (namely disk0, disk1 and disk2) and perform \textit{N}-body simulations. In Appendix~\ref{sec:app1}, we show results of \textit{N}-body simulations under various disk models (disk3-disk10). 
For each model, three simulations with different amounts of solid material are performed $(\Sigma_{{\rm d},0} = 15, 30, 75)$. The total solid mass is $M_{\rm tot} \simeq 5, 10,$ and 25 $M_\oplus$ for $\Sigma_{{\rm d},0} = 15, 30,$ and 75, respectively. The initial solid-to-gas surface density ratio at 1~au is 0.02, 0.04, and 0.1 for disk1, while that for disk2 is 0.002, 0.004, and 0.01, respectively.
The solid density slope in Eq.~(\ref{eq:solid}) is fixed at -3/2. In addition, we also perform simulations with different solid density slopes ($s = -1$ and -2). In some simulations, we do not use the isolation mass but place equal-mass embryos with $M = 0.1 ~M_\oplus$. However, we find that the type I migration does not depend significantly on the initial solid distribution. A brief summary of results is given in Appendix~\ref{sec:app1}.

%%Fig.2
\begin{figure}
\resizebox{0.8 \hsize}{!}{\includegraphics{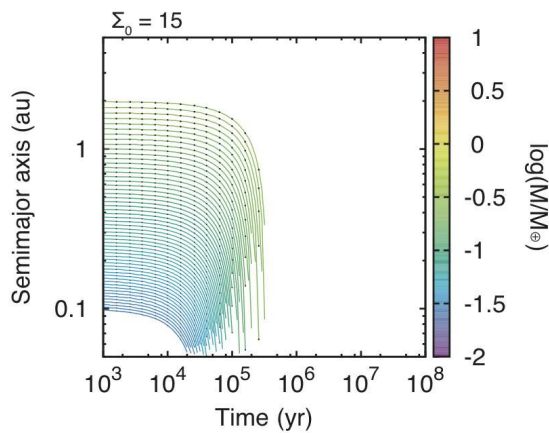}}
\caption{Time evolution of semimajor axis for disk0 model assuming $\Sigma_0 = 15$. The filled circles connected with solid lines represent the size of particles. The color of the lines indicates the mass (color bar).
}
\label{fig:t_a1}
\end{figure}

First, we show an example of orbital evolution in a power-law disk of disk0 in Fig.~\ref{fig:t_a1}. For initial distribution of protoplanets, we use our standard model (initial1) and $\Sigma_{{\rm d},0}=15$ is used. The masses of the smallest planet at 0.1~au and the largest planet at 2~au are $0.029~M_\oplus$ and $0.27~M_\oplus$, respectively. The color bar indicates the planetary mass. All planetary embryos undergo rapid inward migration before gas dispersal, and no planets remain after $\simeq$ 0.3 Myr. When planets get close to the star ($a<0.05 {\rm ~au}$), they are removed from simulations. Therefore the type I migration is problematic, if Mars-sized protoplanets form before the gas dispersal phase.  In cases of larger $\Sigma_{{\rm d},0}$, the isolation mass is larger and inward migration of such bodies is faster, leading to earlier loss of planets onto the star. We note that Mars would have grown to half of its final mass in $\sim 1.8$ Myr \citep{dauphas_pourmand11}. So Mars would have avoided inward migration even for disk0 model.

%%Fig.3
\begin{figure}
\resizebox{0.8 \hsize}{!}{\includegraphics{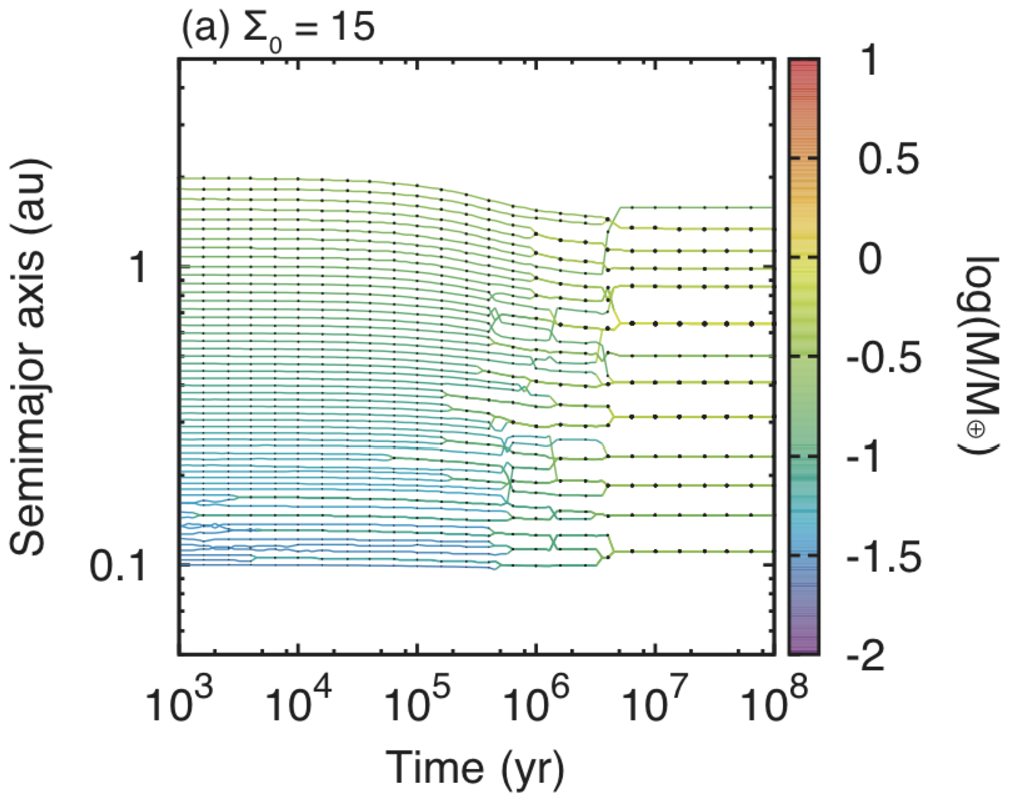}}
\resizebox{0.8 \hsize}{!}{\includegraphics{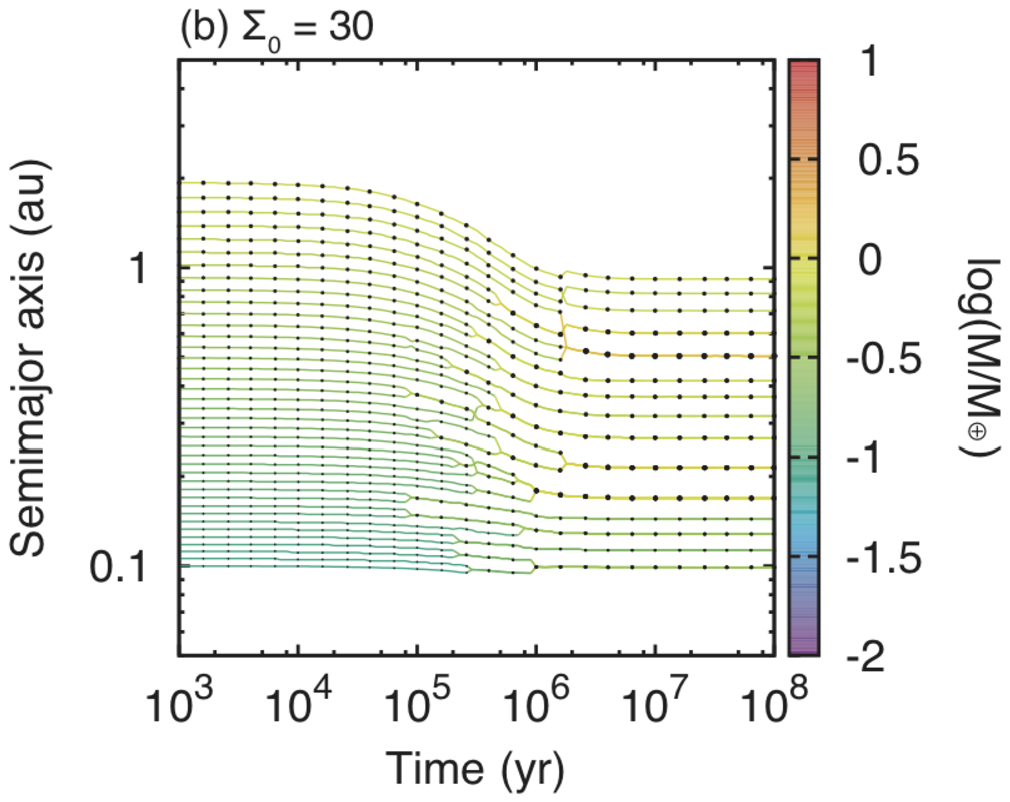}}
\resizebox{0.8 \hsize}{!}{\includegraphics{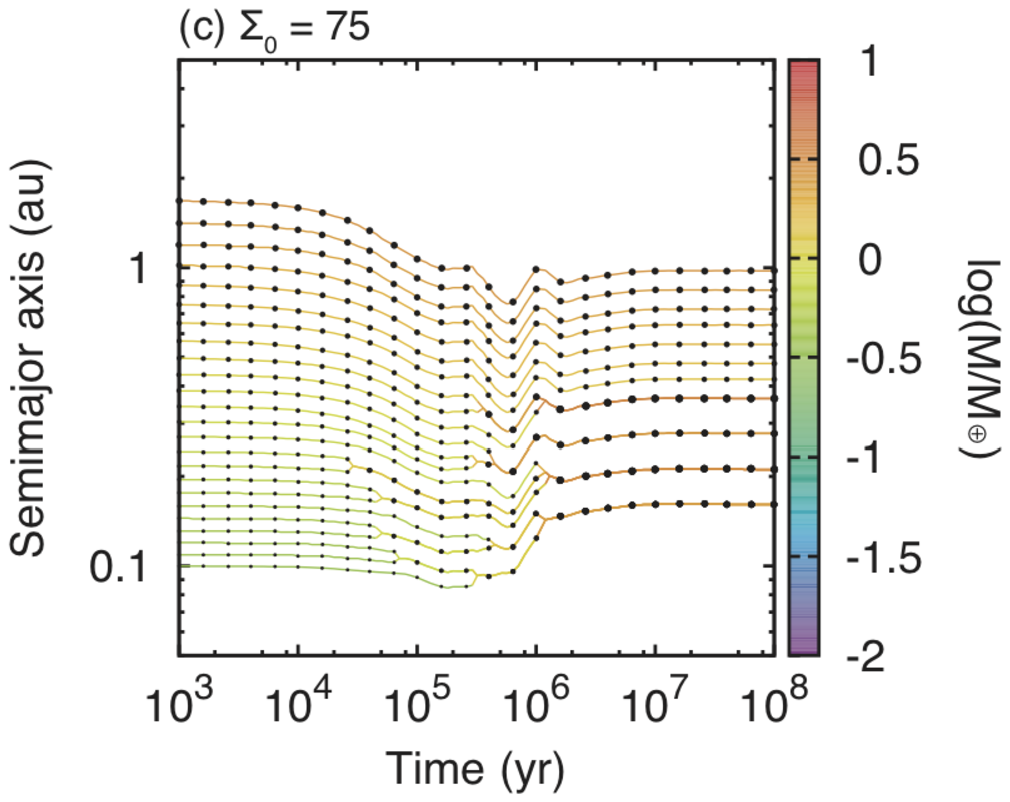}}
\caption{As in Fig.~\ref{fig:t_a1} bur for disk1 model (MRI active) with flat disk profile.
}
\label{fig:t_a3}
\end{figure}

Figure~\ref{fig:t_a3} shows results of simulations for our standard model of the MRI-active case (disk1). The evolution of gas surface density for disk1 is shown in Fig.~\ref{fig:r_sigma}(a). The surface density distribution is almost flat at $r \simeq 0.1-1 {\rm ~au}$. In Fig.~\ref{fig:t_a3}, it is observed that no planets undergo significant migration; even massive embryos in panels~(b) and (c) do not undergo migration. Compared to results for disk0 model, we find that the type I migration is  markedly different. This is because the positive corotation dominates over the negative Lindblad torque in a disk with flat disk profile. In addition, in MRI-active disk ($\overline{\alpha_{r,\phi}}=8 \times 10^{-3}$), the desaturation of corotation torque is effective for larger planets with $M \simeq 1-10 ~M_\oplus$, meaning that planets do not undergo inward migration even for high-mass cases ($\Sigma_{{\rm d},0}=30$ and 75).

%%Fig.4
\begin{figure}
\resizebox{0.8 \hsize}{!}{\includegraphics{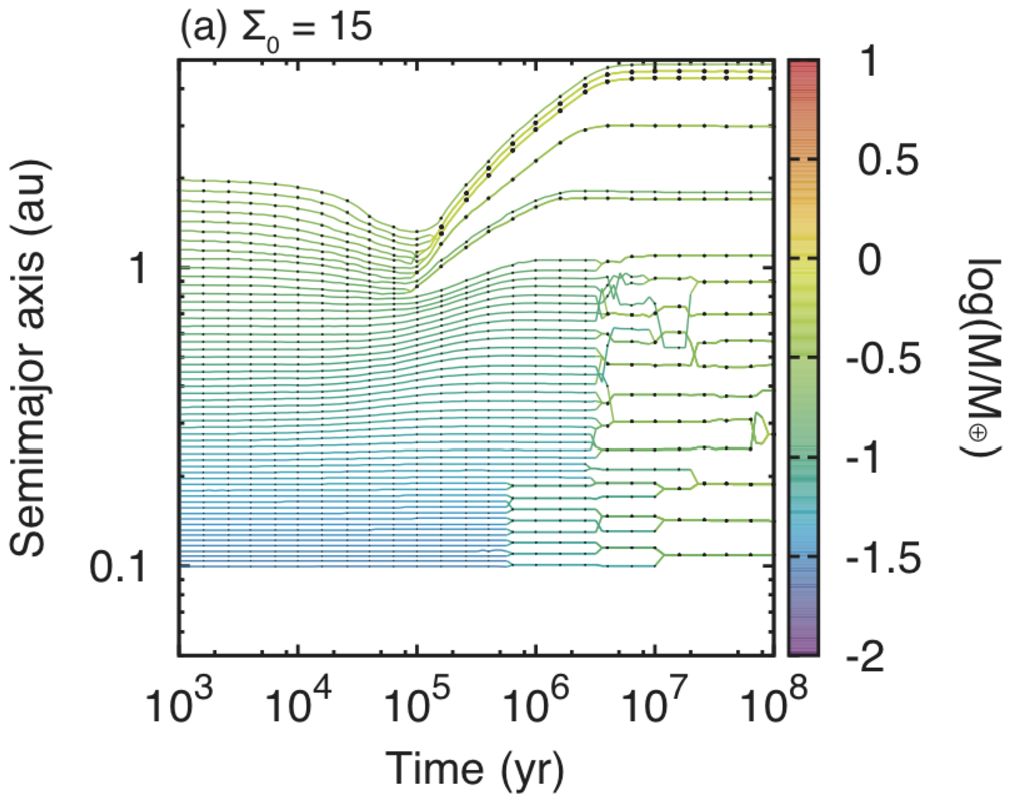}}
\resizebox{0.8 \hsize}{!}{\includegraphics{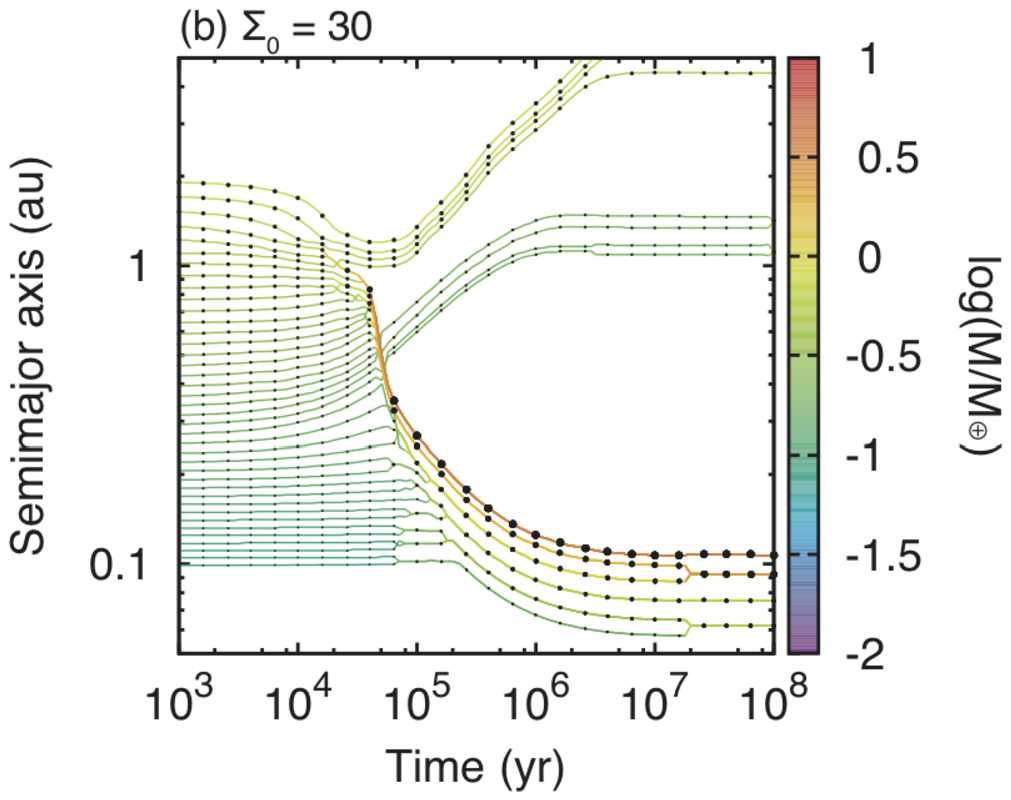}}
\resizebox{0.8 \hsize}{!}{\includegraphics{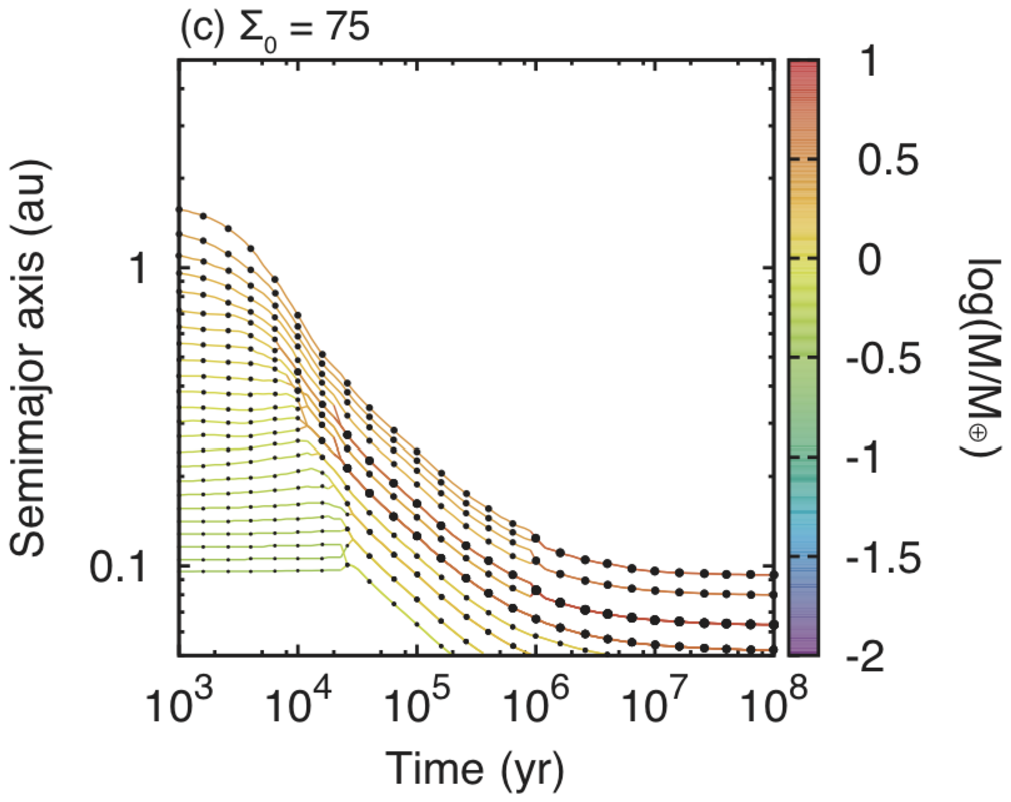}}
\caption{As in Fig.~\ref{fig:t_a1} bur for disk2 model (MRI inactive) with positive disk slope.
}
\label{fig:t_a2}
\end{figure}

Figure~\ref{fig:t_a2} shows results of our standard model for MRI-inactive disks (disk2). The evolution of gas surface density is shown in Fig.~\ref{fig:r_sigma}(b). In panel~(a) of Fig.~\ref{fig:t_a2}, protoplanets do not undergo significant migration in contrast to the results for disk0 model (Fig.~\ref{fig:t_a1}). This is because, inside $r \simeq 1 {\rm ~au}$ the surface density slope is positive and the gas density decreases owing to disk winds (Fig.~\ref{fig:r_sigma}(b)).
In panels (b) and (c) of Fig.~\ref{fig:t_a2}, in which simulations are started with larger embryos, some planets migrate inward. In both panels, embryos grow to relatively large planets ($M \gtrsim 2~M_\oplus$) and undergo migration, which shepherd interior smaller objects toward the star\footnote{At $r = 0.5~{\rm au}$ and $t \simeq 4 \times 10^4 {\rm ~yr}$ in Fig.~\ref{fig:t_a2}(b), the migration speed for planets with $M\simeq 1-2 ~M_\oplus$ is high. Thus, larger inwardly migrating planets pass through several smaller embryos with  $M\simeq 0.2 ~M_\oplus$.}. Similar evolution is seen in Fig.~8(a) of \citet{ogihara_etal17}. 
We note that as shown in \citet{ogihara_etal17}, when there exists an accretion flow at the midplane (e.g., wind-driven accretion; case~B in \citealt{ogihara_etal17}), desaturation of corotation torque is expected, which prevents inward migration of most planets including larger ones ($M > M_\oplus$). Thus, in this case, inward migration seen in Fig.~\ref{fig:t_a2}(b) and (c) would not occur (see Figs.~8(b) in \citealt{ogihara_etal17}). However, \citet{ogihara_etal17} did not perform hydrodynamical simulations and actual effects of radial gas flow on type I migration are not yet clearly understood; therefore we do not consider the effects of gas flow in this paper. Nevertheless, it is noteworthy that the inward migration is much slower than that in the disk0 model (see Fig.~\ref{fig:t_a1}), and even massive planets with $M \gtrsim 2~M_\oplus$ in Fig.~\ref{fig:t_a2}(c) do not fall onto the star.

\begin{figure}
\resizebox{1.0 \hsize}{!}{\includegraphics{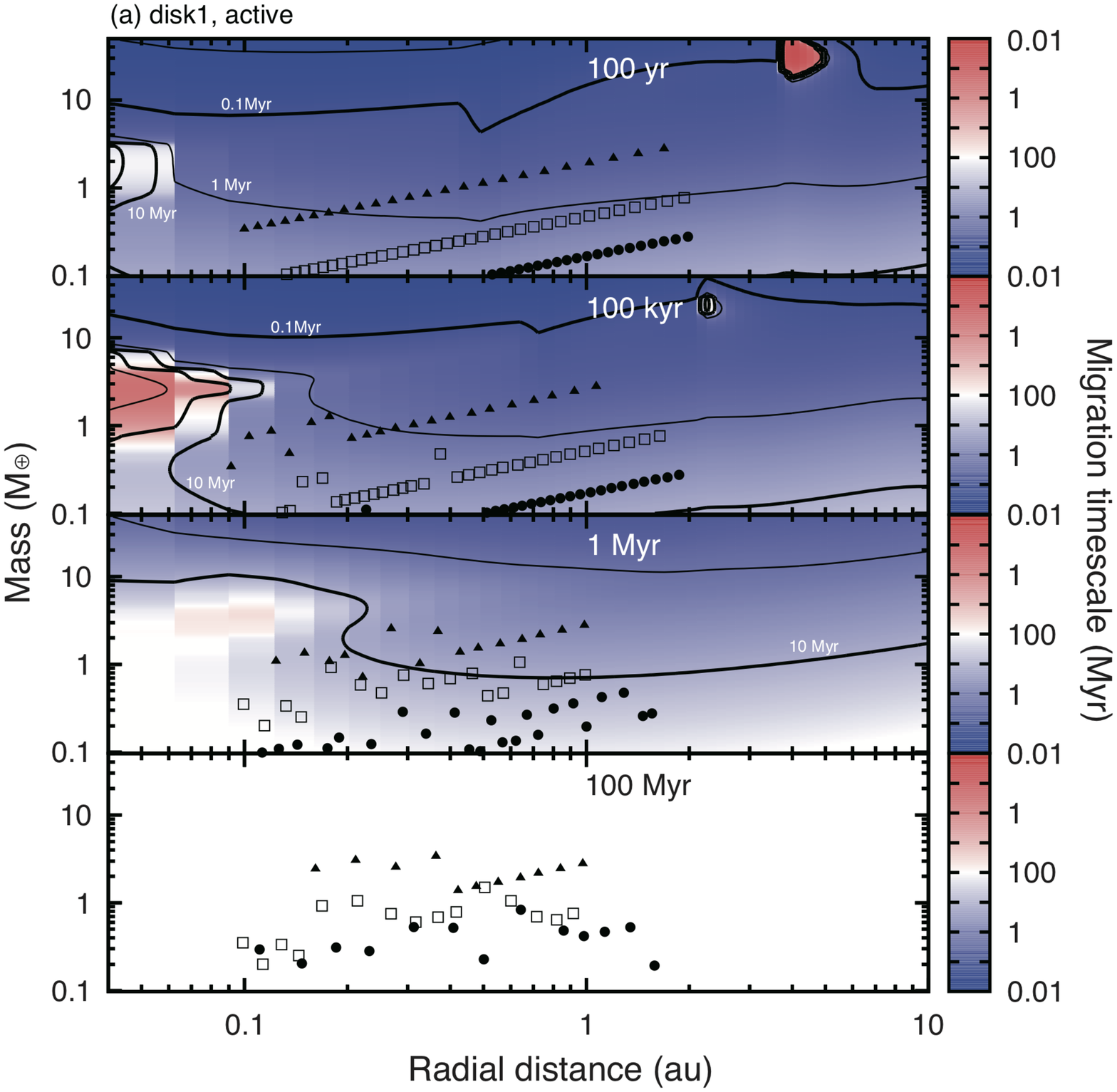}}\\
\resizebox{1.0 \hsize}{!}{\includegraphics{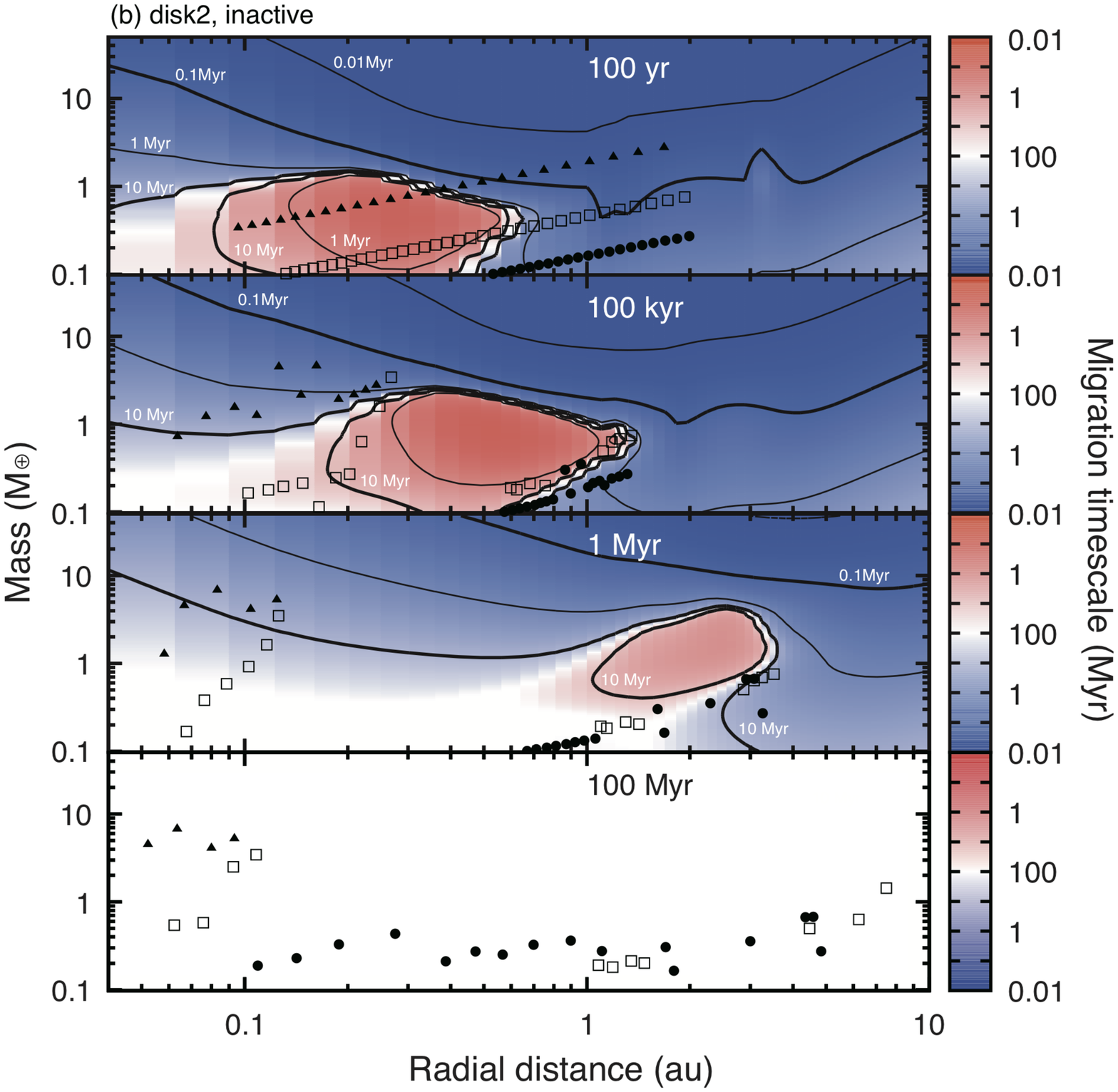}}
\caption{Migration maps for disk1 model (a) and disk2 model (b). The color indicates the migration timescale which is calculated using \citet{paardekooper_etal11}. In the blue region, particles migrate inward, while they undergo outward migration in the red region. Snapshots of simulations are overplotted by filled circles ($\Sigma_0 = 15$), open squares ($\Sigma_0 = 30$), and filled triangles ($\Sigma_0 = 75$). When the disk has a flat slope (panel~a), the type I migration can be significantly suppressed.
}
\label{fig:map1}
\end{figure}

As seen above, the direction and rate of migration depend on the planetary mass and disk properties, which makes the prediction of orbital evolution difficult. Migration maps help us to understand results. Figure~\ref{fig:map1}(a) shows the migration map for disk1, in which the color indicates the migration timescale; blue and red regions correspond to inward migration and outward migration regions, respectively. Snapshots of systems of \textit{N}-body simulations are overplotted, in which filled circles, open squares, and filled triangles correspond to Fig.~\ref{fig:t_a3}(a), (b), and (c), respectively. If the migration timescale is longer than 1 Myr ($\sim$ disk lifetime), significant migration does not occur. 
Only massive super-Earths ($M \gtrsim 10~M_\oplus$) can undergo inward migration faster than 0.1~Myr for disk1 model. The mass of the most massive planet is less than 10 $M_\oplus$. Therefore no significant migrations are observed in Fig.~\ref{fig:t_a3}.
As described in Section~\ref{sec:typeI}, the formula of fully unsaturated torque is reduced to a simpler form of $\Gamma = (-0.6 +2.3p -2.8q) \Gamma_0$. Assuming $q = -0.5$, which gives a good approximation for disks with disk winds (see Figs.~1 and 5 of \citealt{suzuki_etal16}), $|\Gamma / \Gamma_0|$ has a minimum for $p \simeq -0.3$. This means that if the density distribution has a flat gradient, the type I migration can be suppressed. 

%%Fig.5

Figure~\ref{fig:map1}(b) shows the migration map for the MRI-inactive disk (disk2). Snapshots of the system are shown by filled circles (Fig.~\ref{fig:t_a2}(a)), open squares (Fig.~\ref{fig:t_a2}(b)), and filled triangles (Fig.~\ref{fig:t_a2}(c)). At $t = 0.1 {\rm ~Myr}$, migration region, in which the timescale is shorter than 0.1 Myr, is limited in the high-mass ($M>1~M_\oplus$) region. This explains why low-mass planets do not undergo migration in Fig.~\ref{fig:t_a2}(a). Comparing the contour of 0.1~Myr in Fig.~\ref{fig:map1}(a) and (b), relatively large planets ($M \gtrsim 2 ~M_\oplus$) migrate inward passing over the outward migration region (red) only in Fig.~\ref{fig:map1}(b). This inward migration is seen in Fig.~\ref{fig:t_a2}(b) and (c).
The type I migration torque can be divided into the Lindbald torque and the corotation torque. When the corotation torque dominates over the Lindblad torque, the type I migration can be suppressed. In a low viscosity disk of disk2, the corotation torque is saturated for $M \gtrsim 2~M_\oplus$; thus such planets undergo inward migration.
The difference in type I migration between low-mass embryo (Fig.~\ref{fig:t_a2}(a)) and high-mass embryo cases (Fig.~\ref{fig:t_a2}(c)) can also be interpreted as the low viscosity assumed in MRI-inactive disks being enough to desaturate the corotation torque of small-mass embryos in Fig.~\ref{fig:t_a2}(a), but not the large corotation torque of massive embryos.

To summarize the results of this section, we confirm that disk winds play a crucial role in slowing down the type I migration using the disk evolution model of \citet{suzuki_etal16} compared to the power-law disk model based on the MMSN. 
This is because the disk profile is altered due to the effects of disk winds and the corotation torque can dominate over the Lindblad torque. We note that in MRI-inactive (low-viscosity) disks, the corotation torque can saturate for $M \gtrsim 1~M_\oplus$ and massive planets can migrate inward. Even in this case, the migration rate is smaller than that in the disk0 model.
Although \citet{ogihara_etal15b,ogihara_etal17} already found that type I migration can be suppressed by effects of disk winds, we can draw more general conclusions that migration can be suppressed in most cases.
As stated earlier in this section, we perform additional simulations with different initial solid distributions. However,  results have no clear dependence on the initial distribution (see also Appendix~\ref{sec:app1}).
We also note that the type I migration is altered because the density slope is changed. Our results can be applied to more general cases of altered disk profiles without disk winds.

\section{Formation of close-in super-Earth systems}
\label{sec:SE}

Here we perform \textit{N}-body simulations of in situ formation of close-in super-Earths for disk1 and disk2 models. To compare with observed statistical measures, ten runs of simulations with different initial positions of particles are performed for each model.

For disk evolution, we use disk1 and disk2 models (see Fig.~\ref{fig:r_sigma} for $\Sigma_{\rm g}$ evolution). As an initial solid distribution of typical simulations, planetary embryos with $M = 0.2 ~M_\oplus$ are distributed between $r=0.1~{\rm au}$ and $2~{\rm au}$. In our standard model, $\Sigma_{{\rm d},0} = 159$ is used, which is about the same as in \citet{hansen_murray12} and \citet{ogihara_etal15a}. In this case, the total solid mass between 0.1 and 2~au is $80 ~M_\oplus$. In some simulations, the initial solid distribution is varied (e.g., total mass, width of solid distribution, solid density slope).

Table~\ref{tbl:list3} lists the parameters for each model. Model1 and model2 are our standard models in MRI-active disks with flat disk profile and MRI-inactive disks with positive disk slope, respectively. In model3 - model6, the total mass of protoplanets is decreased. In model7 - model9, simulations are started from a narrower ring. In model10 and model11, the slope of initial solid distribution ($s$ in Eq.~\ref{eq:solid}) is varied. Results of simulations for model7-model15 are basically shown only in Appendix~\ref{sec:app3}.

\begin{table}
\caption{Summary of models. The first column indicates the model name. The second column indicates the disk model. In disk1 and disk2 models, $\overline{\alpha_{r,\phi}} = 8 \times 10^{-3}$ and $8 \times 10^{-5}$ are used, respectively. The third column shows the total mass of initial planetary embryos.}
\label{tbl:list3}
\centering
\begin{tabular}{l l l l}
\hline\hline
Model& Disk model & $M_{\rm tot} (M_\oplus)$ &Comments\\
\hline 
model1&	disk1 & 80 &	standard model\\
model2&	disk2 & 80 &	 standard model\\
model3&	disk1 & 40 &	\\
model4&	disk1 & 20 &	\\
model5&	disk2 & 40 &	 \\
model6&	disk2 & 20 & \\
\hline
model7&	disk1 & 50 & $r_{\rm out} = 1 {\rm ~au}$	\\
model8&	disk1 & 30 & $r_{\rm out} = 0.5 {\rm ~au}$\\
model9&	disk1 & 17 & $r_{\rm out} = 0.3 {\rm ~au}$\\
model10&	disk1 & 80 & solid slope = -1	 \\
model11&	disk1 & 80 & solid slope = -2	 \\
\hline
\end{tabular}
\end{table}

\subsection{General results}

We first see results of \textit{N}-body simulations in our standard setting.
  
\subsubsection{Active disk with flat disk profile (model1)}
\label{sec:model1}

%%Fig.6
\begin{figure}
\resizebox{0.8 \hsize}{!}{\includegraphics{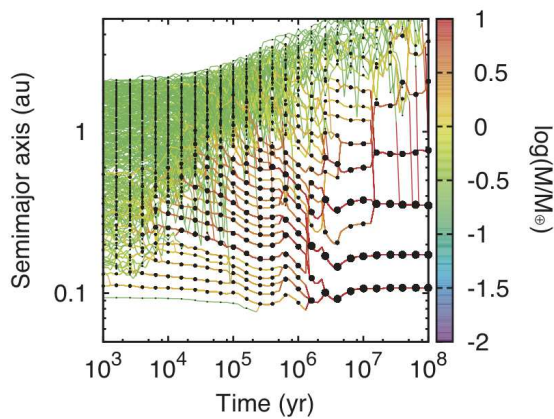}}
\caption{Time evolution of semimajor axis for simulation of close-in super-Earth formation that uses model1.
}
\label{fig:t_a11}
\end{figure}

Figure~\ref{fig:t_a11} shows the time evolution of semimajor axis for our standard model (model1) in an MRI-active disk (see Fig.~\ref{fig:r_sigma} for disk evolution). Growth of planets proceeds from the inner region, and the mass of grown planets is $\simeq 3 ~M_\oplus$ inside $r \simeq 1 {\rm ~au}$ at $t = 0.1 {\rm ~Myr}$. These planets should undergo rapid inward migration in the power-law disk model based on the MMSN; however the type I migration is significantly suppressed in a flat disk profile with effective desaturation of positive corotation torque. Most planets are in relatively close mean-motion resonances (e.g., 6:5, 5:4) before the gas depletion and the orbital separation between planets is $\simeq 6-10~r_{\rm H}$, which are disrupted by orbit crossings during the gas dissipation phase ($t \gtrsim 1 {\rm ~Myr}$). After the stage of giant impacts, several planets remain keeping their orbital separations as $\Delta \simeq 10-30 ~r_{\rm H}$. The system is stable over 100 Myr according to dynamical simulations on orbital stability (e.g., \citealt{chambers_etal96}; \citealt{yoshinaga_etal99}).

%%Fig.7
\begin{figure*}
\center{
\resizebox{0.45 \hsize}{!}{\includegraphics{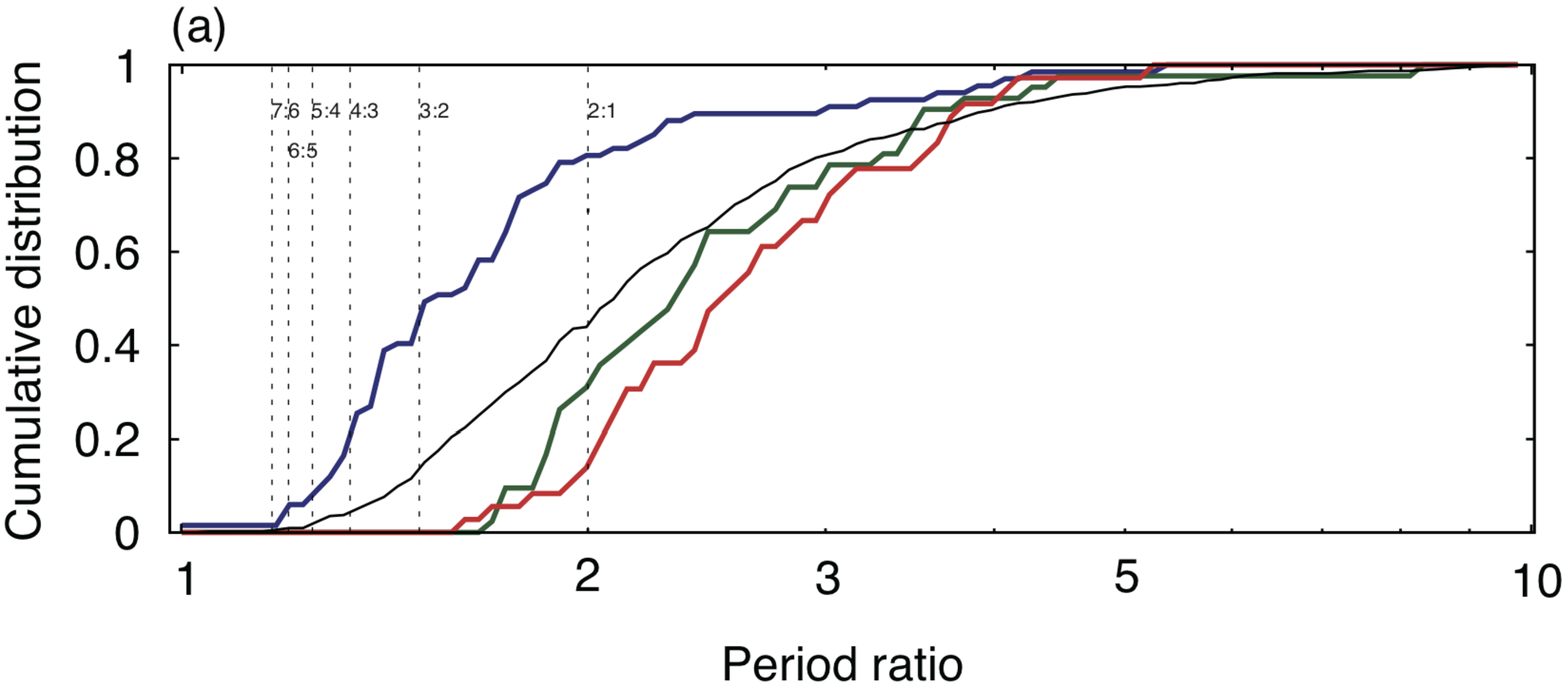}}
\resizebox{0.45 \hsize}{!}{\includegraphics{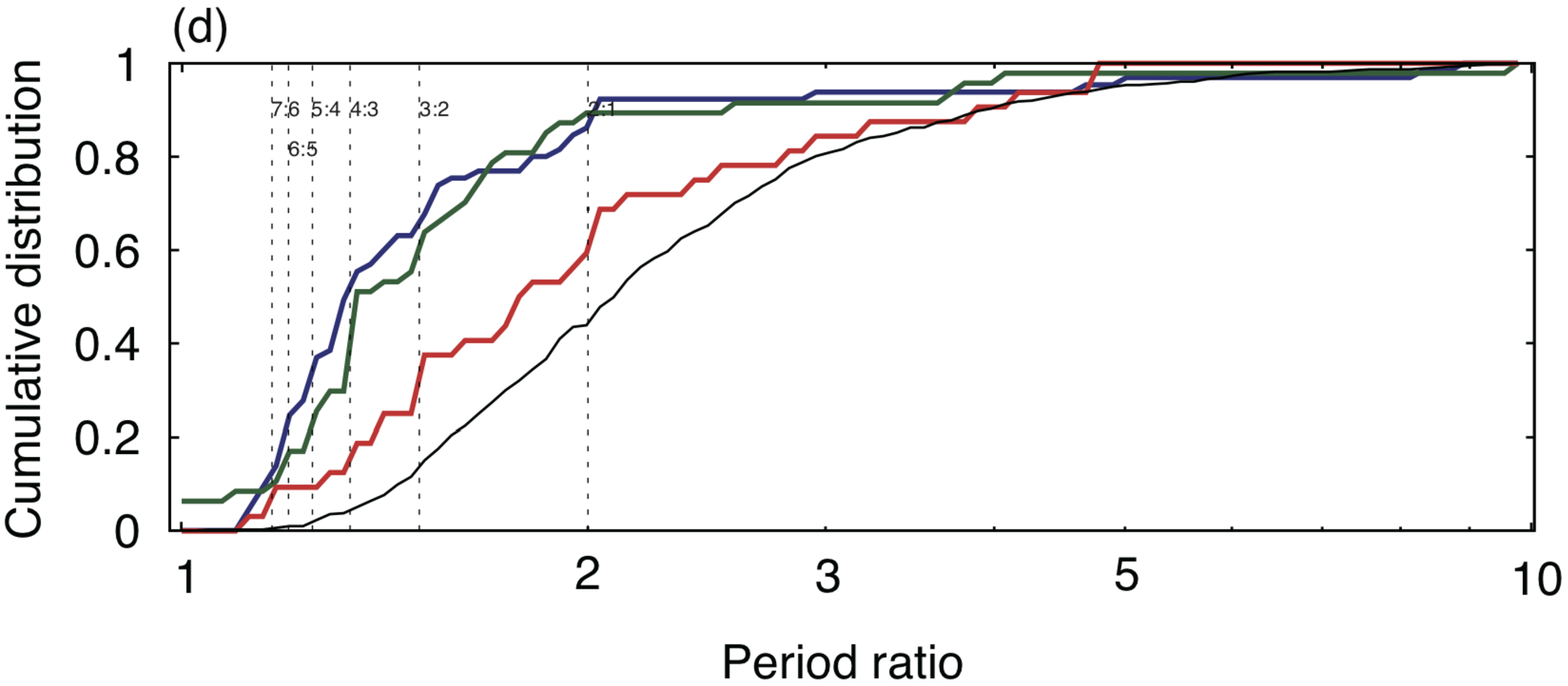}}
\resizebox{0.45 \hsize}{!}{\includegraphics{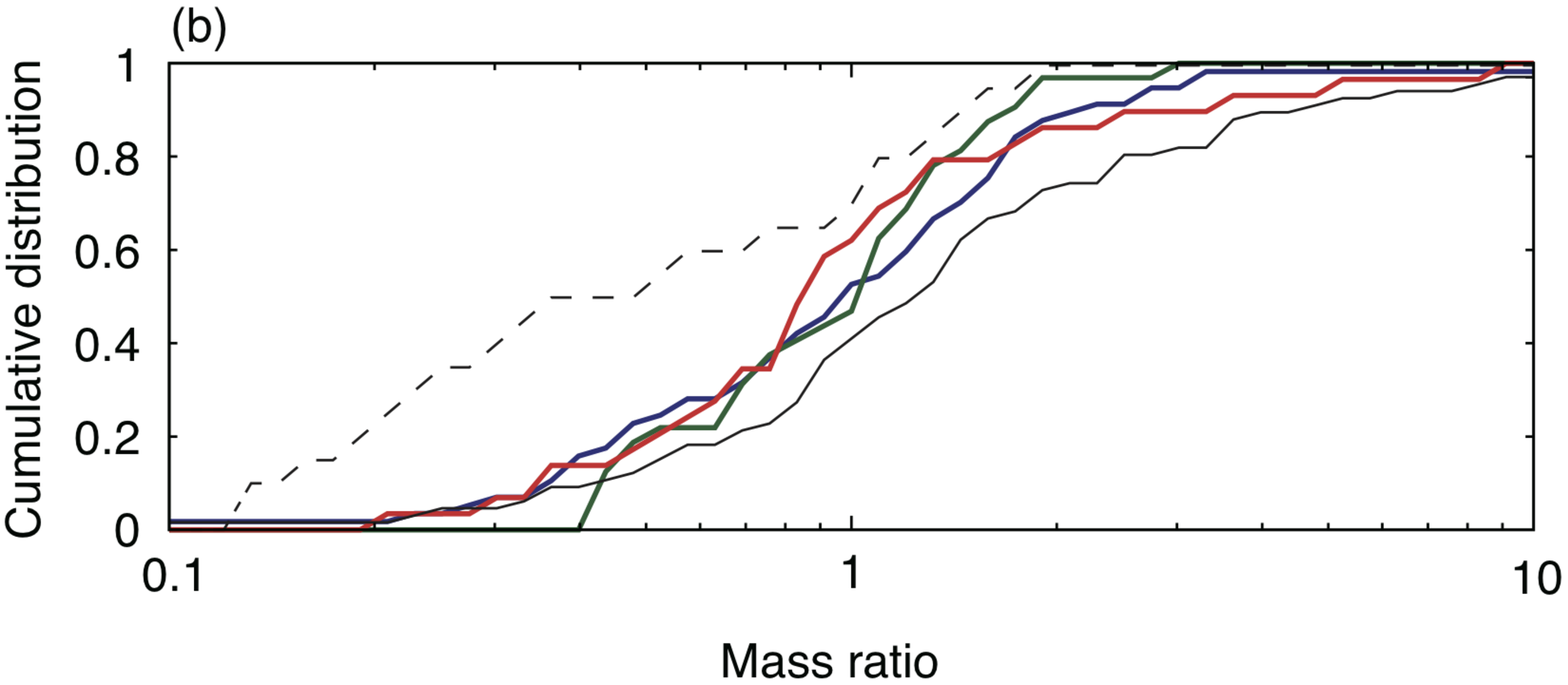}}
\resizebox{0.45 \hsize}{!}{\includegraphics{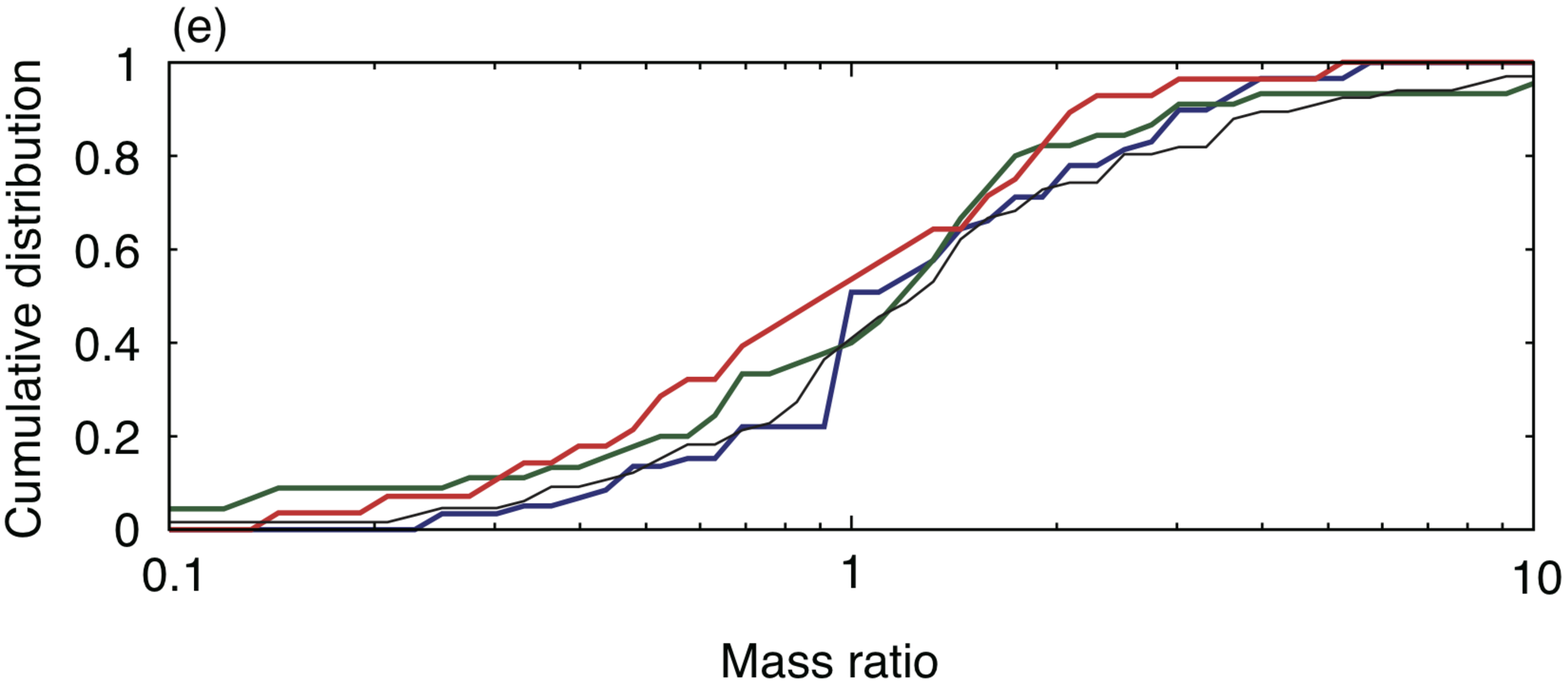}}
\resizebox{0.45 \hsize}{!}{\includegraphics{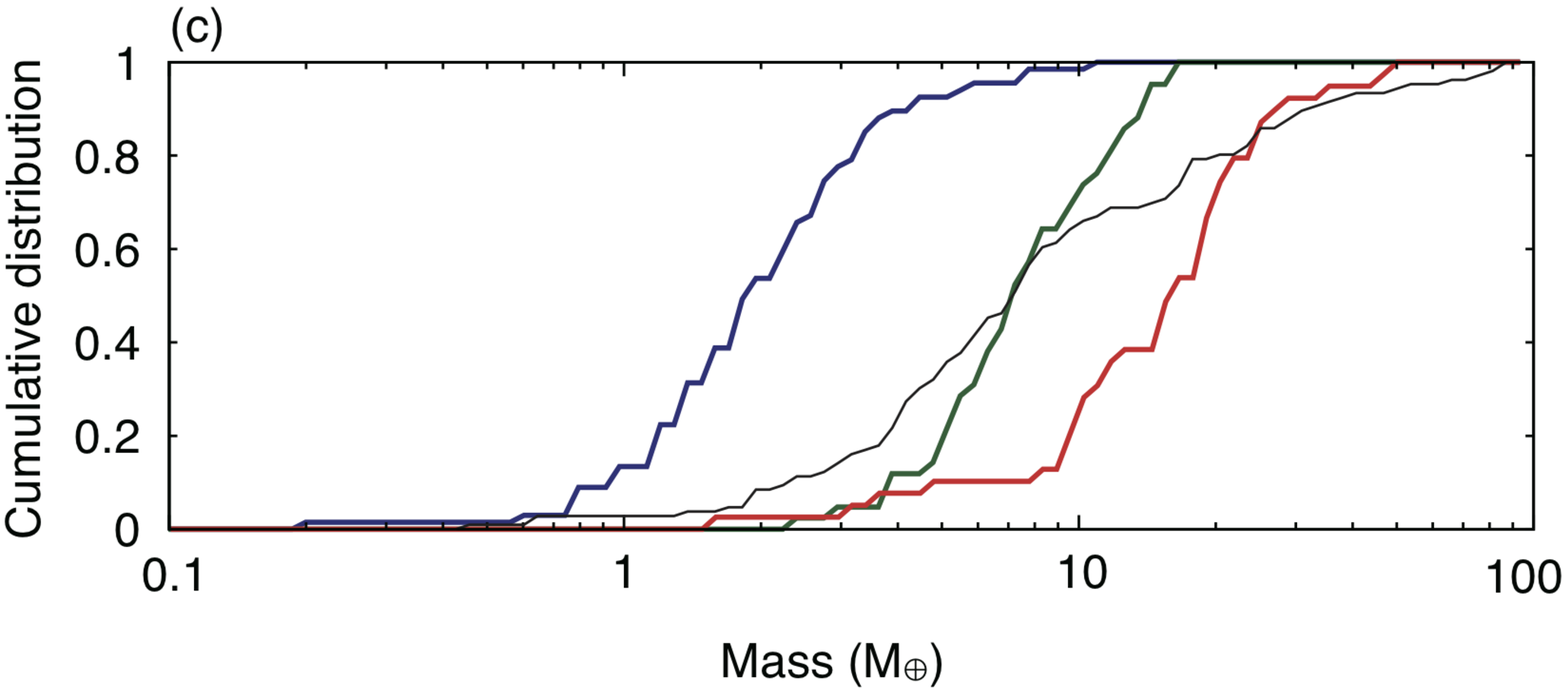}}
\resizebox{0.45 \hsize}{!}{\includegraphics{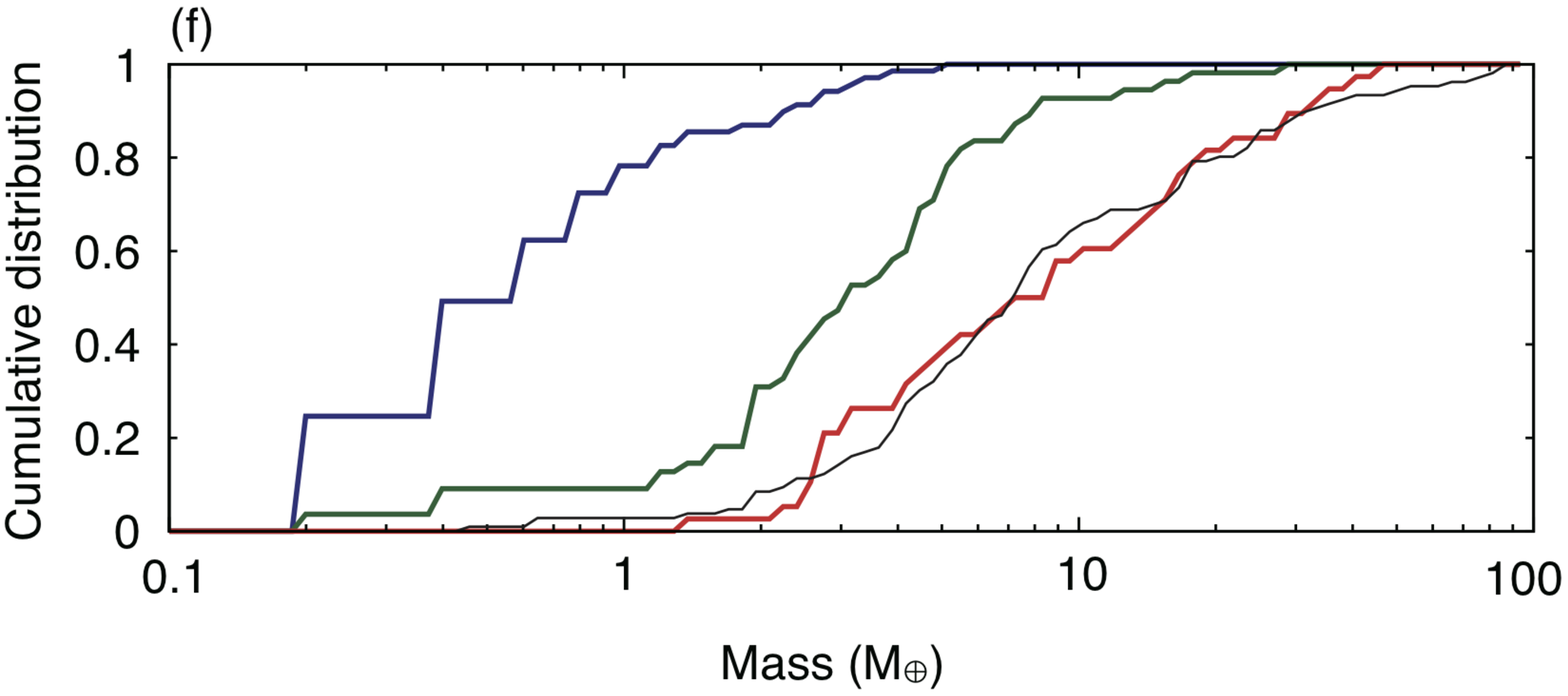}}
}
\caption{Cumulative distributions of final orbital configuration. Panel~(a) shows the period ratio of adjacent pair ($P_{\rm out} / P_{\rm in}$), in which vertical lines indicate locations of first-order mean-motion resonances (MMRs). Panel~(b) shows the mass ratio $M_{\rm out} / M_{\rm in}$. Panel~(c) shows the planetary mass distribution. Black solid lines in each panel represent observed distributions of close-in super-Earths. Red, green, and blue lines indicate results for model1, model3, and model4 in MRI-active disks, respectively. Black dashed line in panel~(b) shows the result of \citet{ogihara_etal15a} for a power-law disk based on the MMSN. Panels~(d)-(f) are same as panels~(a)-(c) but for MRI-inactive disks. Red, green, and blue lines in panels~(d)-(f) indicate results for model2, model5, and model6, respectively. 
}
\label{fig:sum1}
\end{figure*}

Figure~\ref{fig:sum1}(a)-(c) shows a summary of all simulations, in which red lines indicate results for model1. In panel~(a), cumulative distributions of period ratios of adjacent pairs are shown. The solid black line indicates the period ratio distribution of observed close-in super-Earths. The observational data are sourced from the NASA Exoplanet Archive, and 1087 confirmed planets in multiple close-in super-Earth systems are used.
Dashed vertical lines correspond to locations of first-order mean-motion resonances. We find that in results for model1 (the red line), final planets are more separated than in observed systems. A discussion of the orbital separation is given in Section~\ref{sec:total_mass}. In panel~(b) of Fig.~\ref{fig:sum1}, cumulative distributions of mass ratios ($M_{\rm out}/M_{\rm in}$) of adjacent pairs are shown. The black line and red line correspond to distributions of observed close-in super-Earths and results of model1, respectively. The black dashed line shows results of fiducial simulations of \citet{ogihara_etal15a}, in which disk profiles are based on the MMSN model and type I migration is relatively rapid. They found that inner planets are basically larger than outer ones and ``a mass ranking'' is observed. On the other hand, results for model1 show that there is no clear mass ranking. The mass-ratio distribution is almost comparable to the observed distribution. We discuss the reason for this in Section~\ref{sec:mass_ranking}. In panel~(c), cumulative distributions of the planetary mass are shown. We find here that the planet mass is larger than observed mass of close-in super-Earths. This is simply because the total mass of initial particles is large for model1 ($M_{\rm tot} = 80~M_\oplus$). As we present results of simulations with smaller $M_{\rm tot}$ in Sect.~\ref{sec:total_mass0}, the mass distribution shifts to smaller mass.

\subsubsection{Inactive disk with positive disk slope (model2)}
\label{sec:model2}

%%Fig.8
\begin{figure}
\resizebox{0.8 \hsize}{!}{\includegraphics{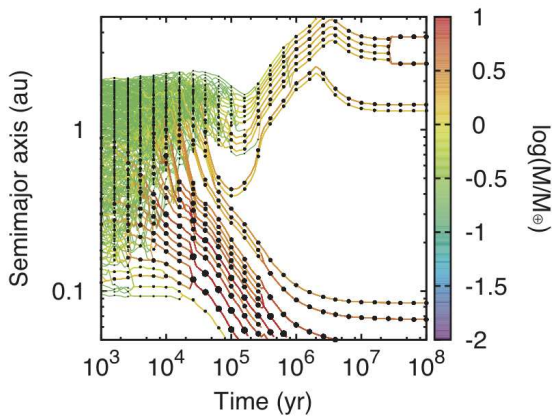}}
\caption{As in Fig.~\ref{fig:t_a11} but for model2 in an MRI-inactive disk.
}
\label{fig:t_a33}
\end{figure}

Figure~\ref{fig:t_a33} shows the result of a simulation in an MRI-inactive disk (see Fig.~\ref{fig:r_sigma} for disk evolution). As seen in Section~\ref{sec:various}, planets are more affected by migration than in active disks. Inside $r \simeq 1 {\rm ~au}$, planets grow to $M \simeq 5 ~M_\oplus$ and can undergo inward migration, while planets formed at $r \gtrsim 1 {\rm ~au}$ have masses of $\simeq 1 ~M_\oplus$ and migrate outward. In the result of Fig.~\ref{fig:t_a33}, many planets are lost to the central star. 
At the end of the simulation, planets are divided into two groups: interior to 0.1~au and exterior to 1~au. We do not observe such a bimodal distribution in observed super-Earth systems.
We see the chaotic nature of \textit{N}-body simulations, which causes differences in final orbital configurations between different runs using the same disk model. In some of the runs, planets that remain inside $r \simeq 1 {\rm ~au}$ at the end of the simulation do not undergo significant inward migration. In these cases, the evolution is more or less similar to that shown in Fig.~\ref{fig:t_a11}.

Figure~\ref{fig:sum1}(d)-(f) shows cumulative distributions of final orbital configurations in inactive disks. Red lines indicate distributions of ten runs of simulations for model2. The period ratio distribution and the mass distribution are different from those for the active case (model1). The period ratio distribution for model2 shows that several planets are in mean-motion resonances in the final state, and that final configurations are more compact than observed super-Earths. This is because in some simulations, planets do not undergo orbital instability after the gas disk dispersal, and the orbital separation between planets in such systems is smaller at the end. In four runs of all ten simulations, many planets are lost to the star as shown in Fig.~\ref{fig:t_a33}. In this case, the averaged number of planets at the end is 2.75. While in another six runs, in which planets do not undergo significant inward migration, the average number of final planets is 4.3. The cumulative distribution shown in Fig.~\ref{fig:sum1} mostly represents the latter case (not shown in Fig.~\ref{fig:t_a33}). For the mass-ratio distribution, a mass ranking is not seen and there is no significant difference between model1 and model2. The observed mass distribution is well fitted by the distribution of model2.
Looking at distributions in red for model2 in Fig.~\ref{fig:sum1}(d)-(f), we may conclude that results for inactive disks (disk2) can reproduce observed distributions better than those for active disks (disk1). However, special care would have to be taken when drawing conclusions. First, as stated above, many planets are in mean-motion resonances in results for model2, which is inconsistent with observations. Second, the bimodal distribution seen in Fig.~\ref{fig:t_a33} is not consistent with observed $a$ distribution of super-Earths. Third, although there is a wide variety of planetary masses in the observed distribution, this variety should not be reproduced by only one set of settings. In fact, as is discussed in Section~\ref{sec:similar_mass}, planets observed in the same system tend to have similar masses. Therefore, at least when only considering MRI-inactive disks, we cannot reproduce several observed properties altogether. As is seen in Section~\ref{sec:total_mass0}, observed distributions are more likely to be reproduced by results in active disks without significant migration.

\subsection{Dependence on total mass}
\label{sec:total_mass0}

Here we present results of \textit{N}-body simulations, in which the total mass of initial embryos is reduced from our standard model, and discuss their dependence on final orbital configurations.

\subsubsection{Active disk with smaller total mass (model3, model4)}
\label{sec:active_smaller}

%%Fig.9
\begin{figure}
\resizebox{0.8 \hsize}{!}{\includegraphics{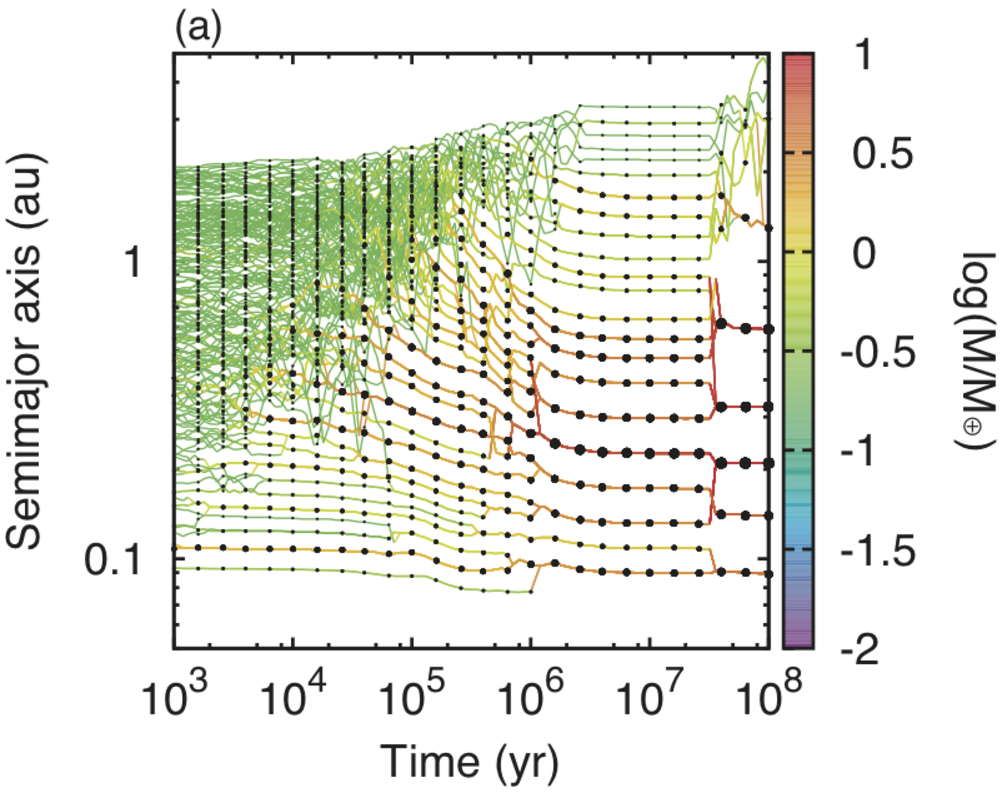}}
\resizebox{0.8 \hsize}{!}{\includegraphics{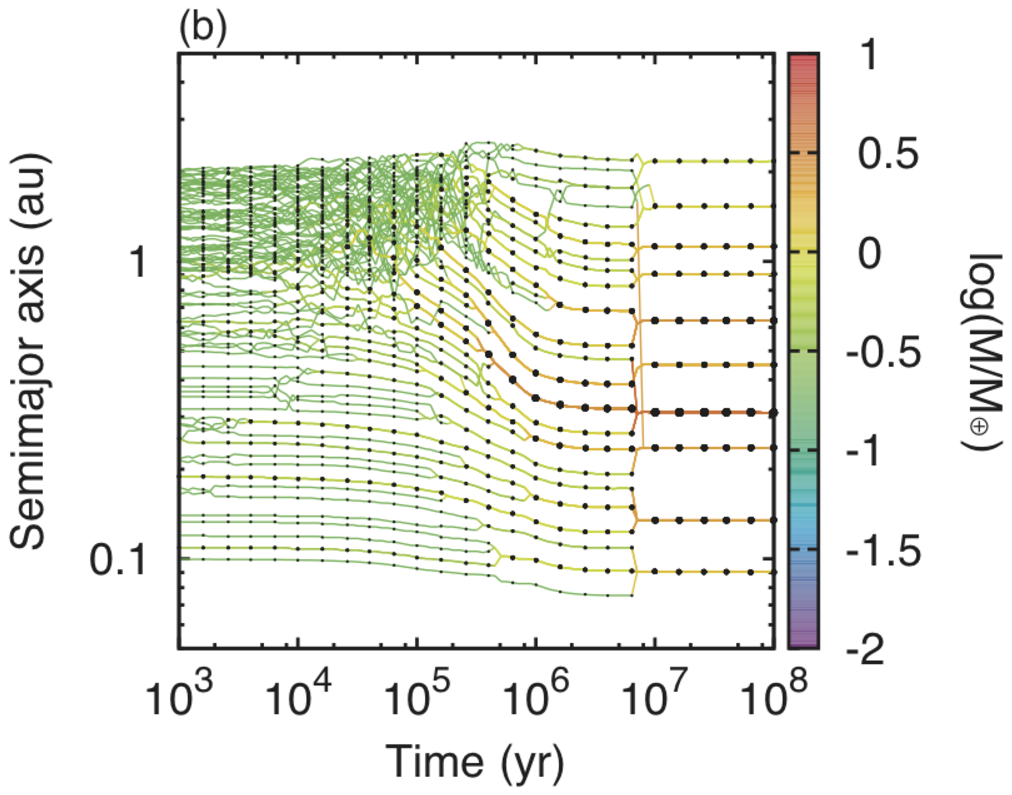}}
\caption{As in Fig.~\ref{fig:t_a11} but for model3 (a) and model4 (b), in which the initial total mass is decreased.
}
\label{fig:t_a44}
\end{figure}

We see in Fig.~\ref{fig:sum1}(c) that the planetary mass in the result of simulations for model1 is systematically larger than the mass of observed close-in super-Earths. We perform \textit{N}-body simulations with smaller total mass. Figure~\ref{fig:t_a44}(a) and (b) show results of typical simulations for model3 ($M_{\rm tot} = 40~M_\oplus$) and model4 ($M_{\rm tot} = 20~M_\oplus$), respectively. The general picture of planet formation is basically the same as that for model1. Through mutual collisions of planets during the gas dissipation phase ($t \gtrsim 1~{\rm Myr}$), final orbital configurations are determined. However, we observe distinct differences in cumulative orbital distributions in Fig.~\ref{fig:sum1}. Obviously, in panel~(c), the planetary mass is dependent on the initial total mass. The mass of planets is less massive in results for model4 (blue line). Interestingly, there is also a clear trend for the period ratio distribution in panel~(a). The period ratio yields smaller values for simulations with smaller total mass (model4). From Fig.~\ref{fig:sum1}, we find that the result for model3 provides the best match among our simulations. We also present a brief discussion on the dependence of the period ratio on the total mass in Sect.~\ref{sec:total_mass}. We note that planets are temporarily captured in mean-motion resonances while the gas is present; however, they lose commensurabilities via giant impacts during gas depletion.
We also note that observed distributions can be biased due to lower detection efficiency of low-mass planets. Therefore, both the period-ratio distribution and the mass distribution would shift towards the left when future observations detect smaller planets. In this case, results for a case with smaller total mass (model4) would better reproduce observed distributions.

\subsubsection{Inactive disk with smaller total mass (model5, model6)}

\begin{figure}
\resizebox{0.8 \hsize}{!}{\includegraphics{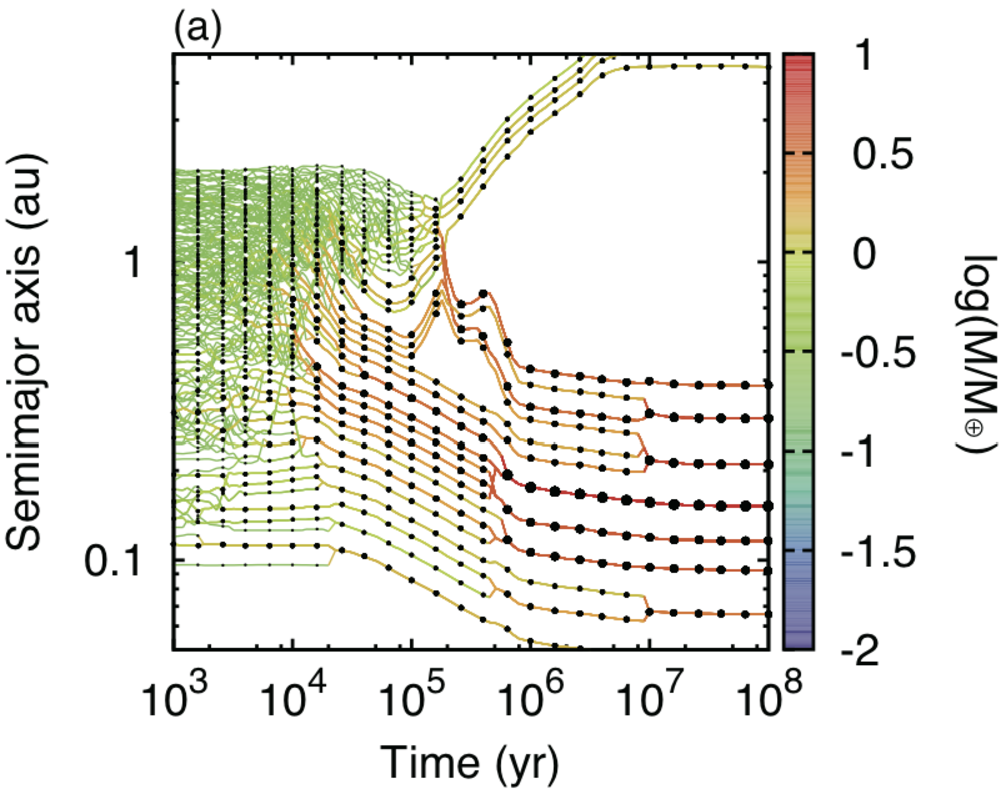}}
\resizebox{0.8 \hsize}{!}{\includegraphics{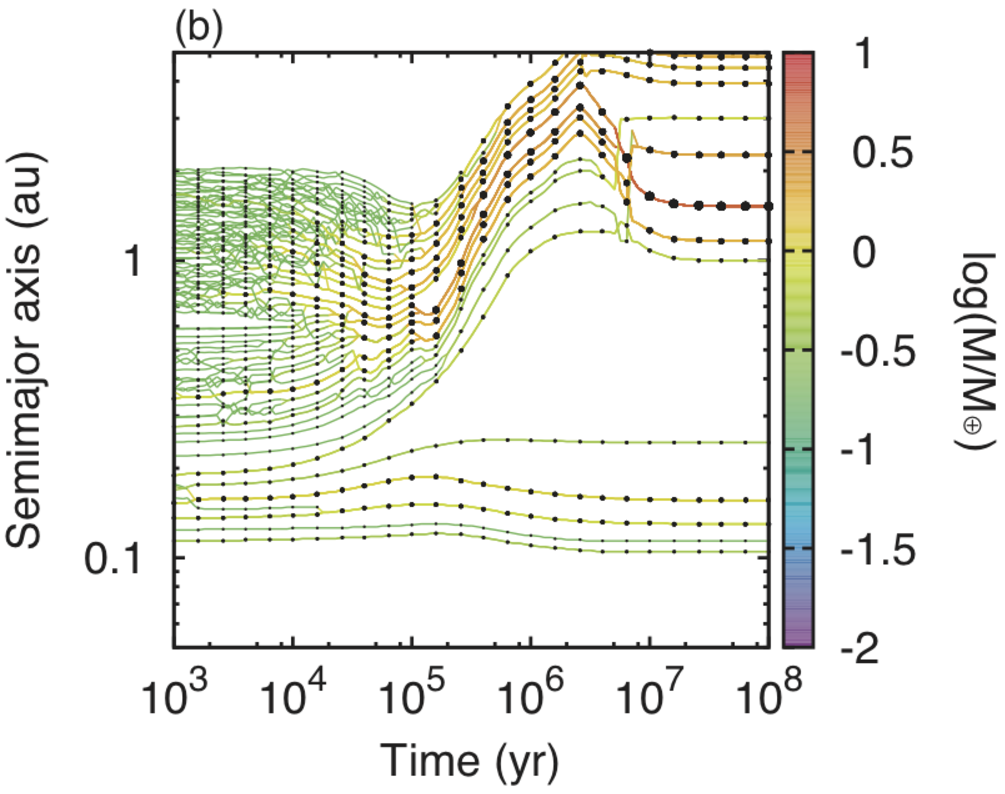}}
\caption{As in Fig.~\ref{fig:t_a11} but for model5 (a) and model6 (b) in MRI-inactive disks, in which the initial total mass is decreased.
}
\label{fig:t_a55}
\end{figure}

We also see results of simulations with smaller total mass in inactive disks. The typical time evolution of semimajor axis for model5($M_{\rm tot} = 40~M_\oplus$) and model6 ($M_{\rm tot} = 20~M_\oplus$) is shown in Fig.~\ref{fig:t_a55}. Results for model5 (Fig.~\ref{fig:t_a55}(a)) show a similar evolution for the semimajor axes as has been described for model2 in Fig.~\ref{fig:t_a33}. Planets grow to $M \simeq 2-5 ~M_\oplus$ inside $r = 1 {\rm ~au}$ within 0.1~Myr, which undergo inward migration. On the other hand, in results for model6 (Fig.~\ref{fig:t_a55}(b)), the mass of the largest planet is $\simeq 2 ~M_\oplus$ at the maximum, meaning that planets do not migrate inward but some planets move outward. This trend is also seen in Fig.~\ref{fig:t_a2} of Section~\ref{sec:various}.

%%FIg.10

Cumulative distributions (Fig.~\ref{fig:sum1}(d)-(f)) show similar trends to those for active disks (Sect.~\ref{sec:active_smaller}). Period ratio and mass distributions depend on the total mass; for the case with a smaller total mass (model6), the period ratio and the mass tend to have smaller values.

\subsection{Summary of supplementary simulations (model7, model8, model9, model10, model11)}
\label{sec:summary_app}
We have performed supplementary simulations that consider initially narrower solid distribution (model7 - model9) and initially steeper/shallower solid distribution (model10 and model11) in MRI-active disks. Simulation results are shown in Appendix~\ref{sec:app3}, but a brief summary is given here.

We find that results for model7 - model9 (initially narrow solid ring) are basically the same with results for our standard run (i.e., model1). See Fig.~\ref{fig:t_a66} for time evolution. The type I migration is significantly suppressed due to the effect of desaturated positive corotation torque in MRI-active disks. Planets are captured in a chain of resonant planets before the gas disk depletion. Whether the resonant chain undergoes late orbital instability after gas depletion depends on the width of initial solid distribution. For model1 and model7 with $r_{\rm out} = 2~{\rm au}$ and 1~au, the late orbital instability occurs for all ten runs of simulation. 
However, for model8 and model9 with $r_{\rm out} = 0.5~{\rm au}$ and 0.3~au, the resonant chain remains after gas depletion in some cases.

In model10 and model11, the solid surface density slope for initial solid distribution is changed. We confirm that results do not change from our standard run (model1). See Fig.~\ref{fig:sum3} for cumulative distribution of final orbital configurations.

\section{Discussion}
\label{sec:discussion}

\subsection{Avoiding mass ranking}
\label{sec:mass_ranking}
\citet{ogihara_etal15a} investigated the in-situ formation of close-in super-Earths using power-law disk profiles based on the MMSN model, and found that observed orbital configurations are not reproduced by their \textit{N}-body simulations. In particular, a clear mass ranking is observed, in which the planetary mass monotonically decreases when the semimajor axis increases (the black dashed line in Fig.~\ref{fig:sum1}(b); see also Fig.~3(a) of \citealt{ogihara_etal15a}). This is not consistent with observed distributions (see also Fig.~3(b) of \citealt{ogihara_etal15a}). In this paper, on the other hand, we do not see any clear mass ranking in our simulations and mass ratio distributions can be reproduced (e.g., Fig.~\ref{fig:sum1}(b)). Here we briefly discuss how to avoid mass ranking.

Forming the mass ranking in \citet{ogihara_etal15a} can be attributed to an edge effect. Let us consider the case in which there exists a disk inner edge and planets undergo inward migration towards the disk inner edge. After a planet stops its migration at the disk inner edge, migrating planets can subsequently be captured in mutual mean-motion resonances with the inner planet, depending on its migration rate. Slowly migrating planets can be trapped in resonances, while rapidly migrating planets go through resonant locations and undergo close encounters with the inner planet (e.g., \citealt{ogihara_kobayashi13}), leading to a collision and/or orbital rearrangement. As the type I migration speed is more rapid for massive planets, larger planets tend to go through resonances and undergo close encounters. By contrast, less massive planets that undergo slower migration are more likely to be captured in mean motion resonances with inner planets. As a result of these processes, a clear mass ranking forms at the disk inner edge in simulations of \citet{ogihara_etal15a}. Thus, if planets undergo rapid inward migration towards the disk inner edge, a mass ranking forms just outside the disk inner edge, which we call an ``edge effect.'' In other words, the existence of disk inner edge and rapid migration are required for the edge effect to operate. Observations indicate that close-in super-Earths formed in environments without the edge effect (and therefore without disk inner edge and/or rapid inward migration).
If we do not take into account the existence of the inner edge, the mass ranking should not form with the setting of \citet{ogihara_etal15a}; however, almost all planets are lost to the star due to rapid inward migration in the power-law disk, and no planets remain.

In the results of simulations presented here, we do not observe a clear mass ranking, simply because there is no disk inner edge in our disk model. However, even if we assume the existence of disk inner edge, migration speed is significantly reduced from that in previous studies. In \citet{ogihara_etal15a} in which the power-law disk model is used, the inward migration timescale is $\sim 0.01-0.1 {\rm ~Myr}$, while planets migrate inward on a timescale of $\sim 1-10 {\rm ~Myr}$ for model1 in this work. This means that super-Earths form almost in situ in active disks, and the edge effect would not occur even if there is a disk inner edge.

\subsection{Dependence of the total mass on period ratio distribution}
\label{sec:total_mass}
Comparing results for model1, model3, and model4, we find that the period ratio distribution depends on the initial total mass (Fig.~\ref{fig:sum1}(a)). As the initial total mass decreases, the period ratio of final planets also decreases. We briefly discuss this dependence.

%%Fig.11
\begin{figure}
\resizebox{1.0 \hsize}{!}{\includegraphics{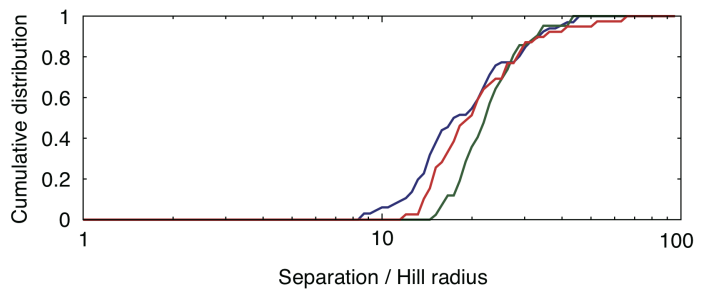}}
\caption{Cumulative distributions of orbital separations between planets in the final state. The red, green, and blue lines indicate results for model1 ($M_{\rm tot} = 80~M_\oplus$), model3 ($M_{\rm tot} = 40~M_\oplus$), and model4 ($M_{\rm tot} = 20~M_\oplus$), respectively.}
\label{fig:sum4}
\end{figure}

As we see in Sect.~\ref{sec:total_mass0}, the planet formation process is basically the same among simulations with different initial total mass. Planets undergo slow migration on a timescale of $\gtrsim 1 {\rm ~Myr}$ and they are tentatively trapped in mean motion resonances. During the depletion of the gas disk, their orbits become unstable and the planets undergo orbit crossings and/or collisions, which determines the final orbital configuration. Figure~\ref{fig:sum4} shows the cumulative distribution of the orbital separation normalized by the mutual Hill radius for model1, model3, and model4. We find that $\Delta / r_{\rm H} \simeq 10-30$ irrespective of the initial total mass. This fact has also been observed in previous \textit{N}-body simulations (e.g., \citealt{ogihara_ida09}). The mutual Hill radius has a dependence of $M^{1/3}$; thus the orbital separation (and hence the period ratio) depends on the planetary mass.

%%Fig.12
\begin{figure}
\resizebox{1.0 \hsize}{!}{\includegraphics{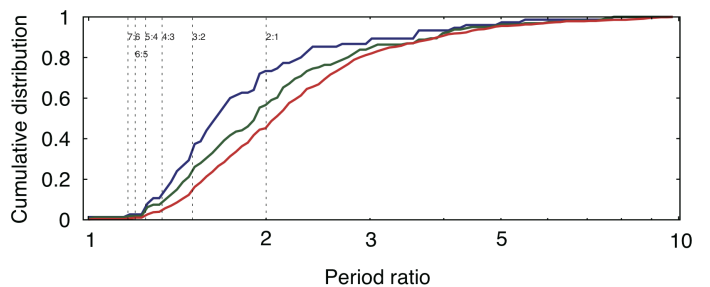}}
\caption{Cumulative period-ratio distributions of observed close-in super-Earths. The red line indicates the distribution of all close-in super-Earths, while the green and blue lines are distributions for smaller planets (green: $M \leq 5~M_\oplus$ or $R \leq 2~R_\oplus$) and (blue; $M \leq 2~M_\oplus$ or $R \leq 1.5~R_\oplus$).}
\label{fig:p_n_exo}
\end{figure}

Figure~\ref{fig:p_n_exo} shows cumulative distributions of period ratio of close-in super-Earths. The red line shows the distribution of all close-in super-Earths ($M \leq 100~M_\oplus$ or $R \leq 10~R_\oplus$), which is the same one shown by black lines in Fig.~\ref{fig:sum1}(a) and (d). Green and blue lines indicate distributions for close-in super-Earths but this is limited to small planets. The green distribution is limited to systems, in which all planets are relatively small ($M \leq 5~M_\oplus$ or $R \leq 2~R_\oplus$). In the blue line, only further smaller planets are considered ($M \leq 2~M_\oplus$ or $R \leq 1.5~R_\oplus$). Interestingly, we also see the dependence of the period-ratio distribution of observed close-in super-Earths on the planetary size. This would suggest that observed close-in super-Earths undergo giant impacts, which leads to orbital separations of $\Delta \simeq 10-30 ~r_{\rm H}$.

\subsection{Neighboring planets in the same system have similar masses}
\label{sec:similar_mass}
The California Kepler Survey probed precise properties of planets and host stars from high-resolution optical spectra obtained at the Keck Observatory. \citet{weiss_etal18} found that 
planets within a single system have similar sizes to what would be expected from drawing randomly from a collection of systems. Here we see whether or not this property is also obtained in our results of \textit{N}-body simulations.

\begin{figure}
\center{
\resizebox{1.0 \hsize}{!}{\includegraphics{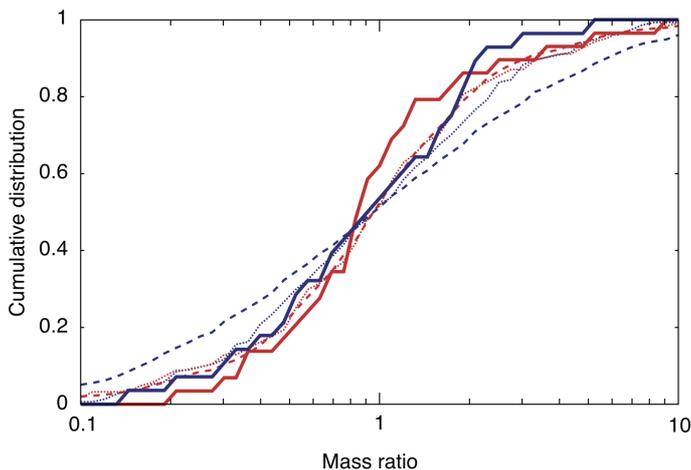}}
}
\caption{Cumulative distribution of mass ratio. Red and blue lines indicate results for model1 and model2, respectively. Solid lines are as in Figure~\ref{fig:sum1}, while dashed lines show cumulative distributions for all pairs of planets that formed ten runs of simulations for each model. Dotted lines represent cumulative distributions for all pairs from the same simulation (one simulation).
}
\label{fig:sum6}
\end{figure}

Solid lines  in Figure~\ref{fig:sum6} show the same mass ratio distributions for model1 and model2 shown in Figure~\ref{fig:sum1}, in which the mass ratio of two neighboring planets is accumulated. Dashed lines in Figure~\ref{fig:sum6}, on the other hand, indicate the mass ratio distribution of all planetary pairs, which are randomly drawn from all planets that formed in all ten runs of simulations for each model. We find from this figure that the mass ratio distribution for neighboring pairs has a sharper peak at a mass ratio of 1 than that for all pairs. This is consistent with results of \citet{weiss_etal18}.
In Fig.~\ref{fig:sum6}, we also plot the cumulative distribution of all pairs from the same simulation (i.e., one simulation) to determine whether or not any pairs of planets from the same system have similar masses. 
The dotted line shows a good agreement with the dashed line for model1, which means that neighboring planets are more likely to have similar masses. Comparing  the dashed line for model1 with the dashed line for model2, we see that the dashed line for model2 has a shallower profile, which indicates that all planets for model1 have more similar masses than all planets for model2; this trend is also seen in Fig.~\ref{fig:sum1}(c). Comparing the dotted line with dashed line for model2, we find a difference between these two distributions, which means that the variation in the mass is the largest for randomly chosen pairs of model2.

%%Fig.13

\subsection{Blending of several models}

In Section~\ref{sec:SE}, we compare our results of \textit{N}-body simulations with observed distributions (e.g., Fig.~\ref{fig:sum1}), and find that observed distributions can be matched by simulations without significant migration (e.g., model3). In each model, several parameters are set to fixed values and ten runs of simulations are performed. In reality, however, each parameter has a plausible range, which means that blending of models  would better reflect reality. In this section, we show several sets of blended models, which reproduce the observed distributions.

%%Fig.14
\begin{figure}
\resizebox{1.0 \hsize}{!}{\includegraphics{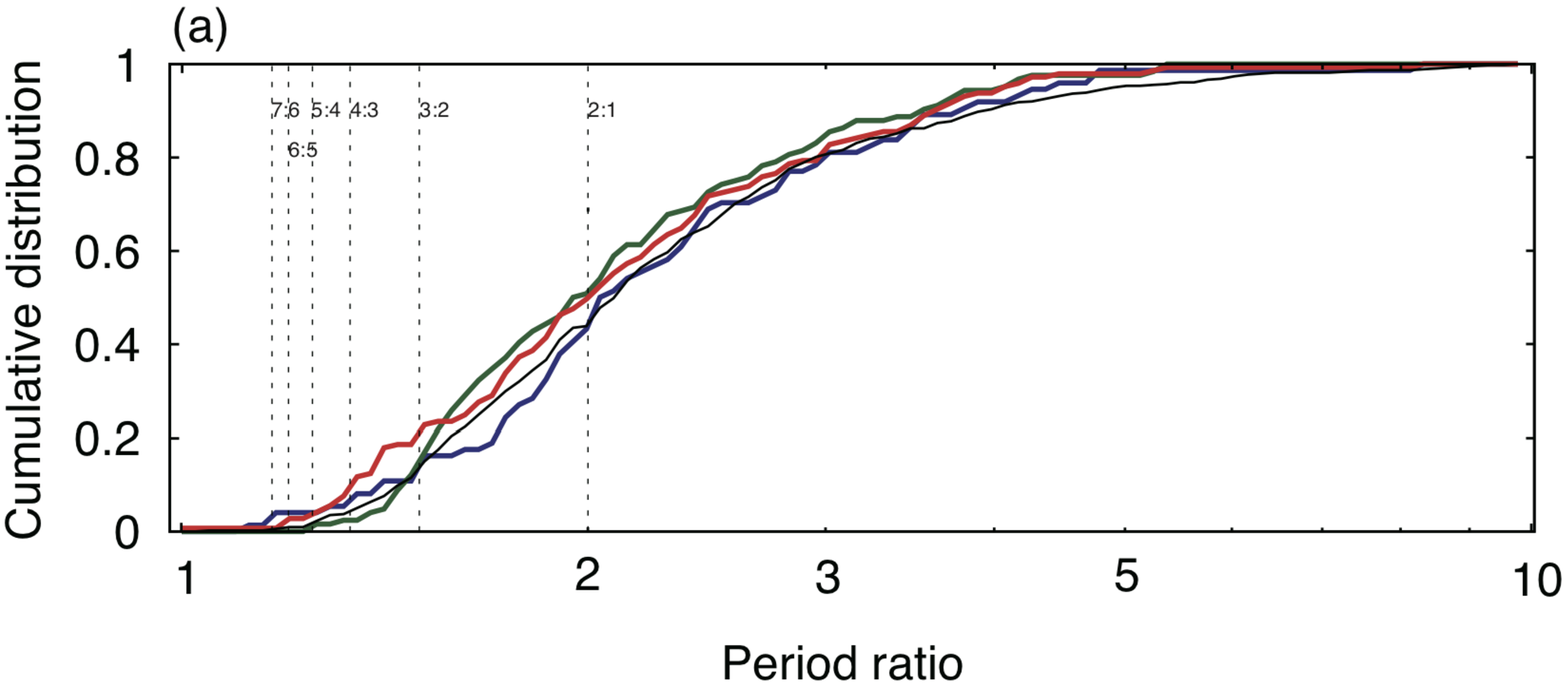}}
\resizebox{1.0 \hsize}{!}{\includegraphics{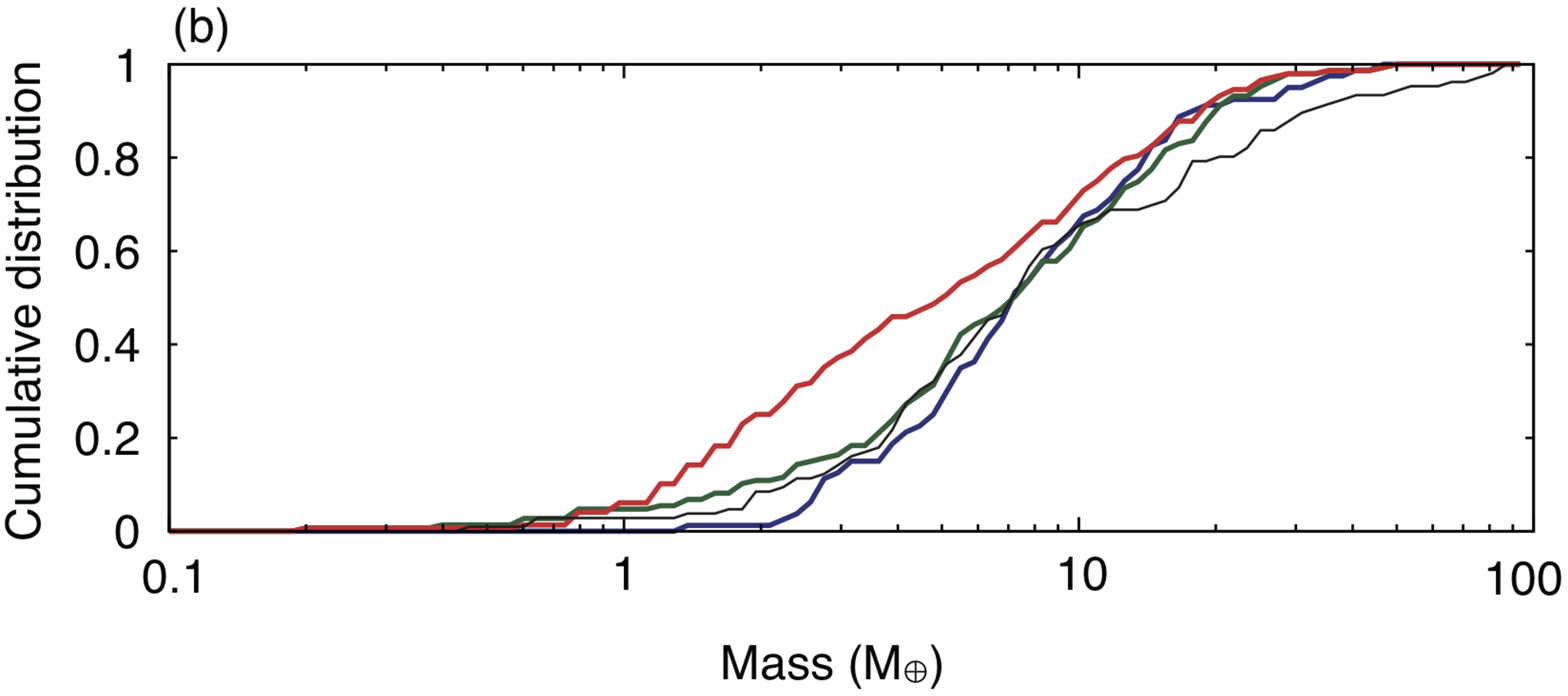}}
\caption{Cumulative distributions of period ratio (a) and mass (b). The black solid line represents observed distributions. The red line indicates distributions in which results of model1, model3, and model4 are superposed. The green line indicates blended distribution of model1, model7, and model8, while blue lines are for model2 and model3.
}
\label{fig:sum5}
\end{figure}

The red lines in Fig.~\ref{fig:sum5} represent the case in which results of model1, model3, and model4 are superposed. The period-ratio distribution is comparable to the observed distribution. For the mass distribution, the blended distribution is almost comparable to the observed distribution. We note that smaller planets are less likely to be observed; thus the difference in small-mass range would not be a serious problem.

In a series of \textit{N}-body simulations, we see that final orbital configurations of our standard models (e.g., model1 and model3) show smaller radial mass concentration than currently observed super-Earth systems. See Appendix~\ref{sec:app3} for a detailed discussion. Additional simulations that start from initially narrow solid ring (model7-model9) show that the relatively large radial mass concentration of observed super-Earths can be reproduced by these simulations. Here, the green lines in Fig.~\ref{fig:sum5} indicate the case in which simulations from a narrower ring are included (model1, model7, and model8). In this case, the observed distributions  of the period ratio and the mass are well reproduced. 

We also find that the blue lines in Fig.~\ref{fig:sum4} reproduce the observed distributions, in which results for MRI-active and inactive disks (model2 and model5) are blended. These blended models are simply examples. To constrain the formation model of exoplanet systems, a more detailed discussion would be required.

\subsection{A chain of resonant planets}
\label{sec:chain}

Most observed super-Earths are not in MMRs; however, there are some systems that have a chain of resonant planets. Here we look at several observed and simulated systems with a resonant chain.

%%Fig.15
\begin{figure}
\resizebox{0.8 \hsize}{!}{\includegraphics{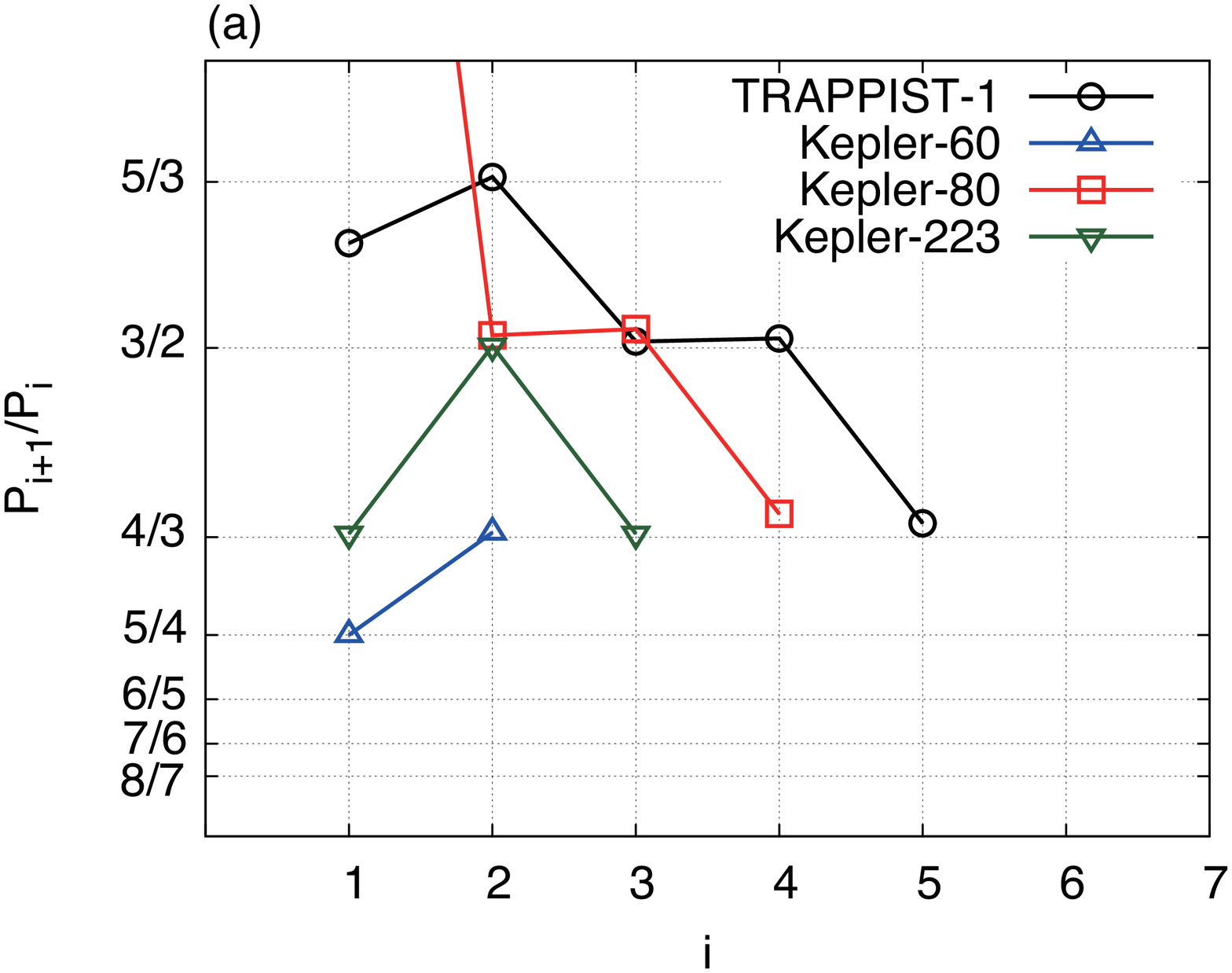}}
\resizebox{0.8 \hsize}{!}{\includegraphics{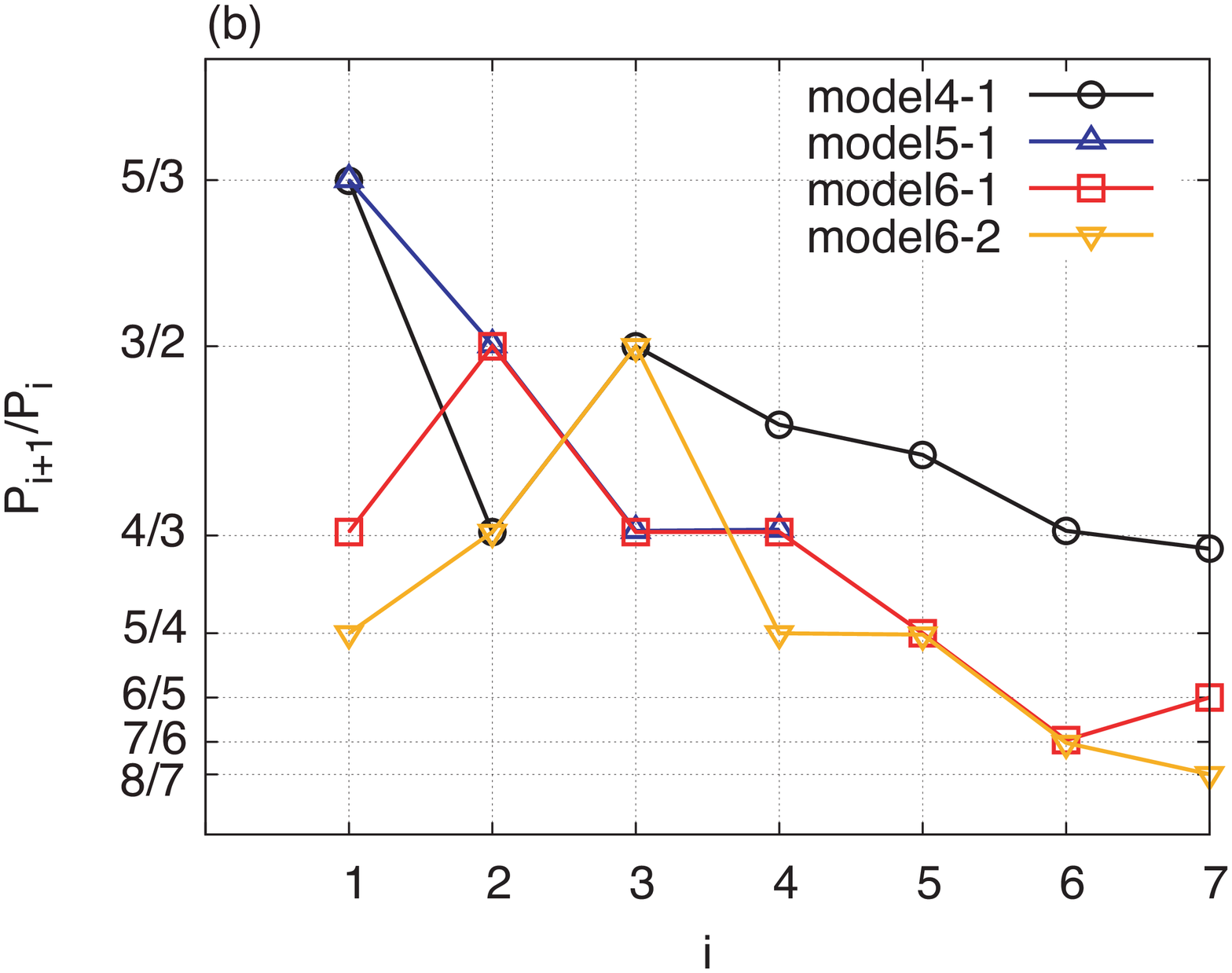}}
\caption{Period ratio of the $i$-th innermost pair ($i=1, 2, ...$). Panel~(a) shows observed close-in super-Earth systems, while results of \textit{N}-body simulations are shown in panel~(b). Horizontal lines indicate locations of mean-motion resonances.}
\label{fig:p_m_exo}
\end{figure}

Figure~\ref{fig:p_m_exo}(a) shows examples of super-Earth systems with a chain of resonant planets. Open circles represent planets of the TRAPPIST-1 system, in which the inner two pairs are close to a 5:3 MMR and the outer three pairs are near 3:2 and 4:3 MMRs. Several planetary systems from the Kepler survey have planets in a resonant chain, including Kepler-60, Kepler-80, and Kepler-223. The planets are in/near 3:2, 4:3, and 5:4 MMRs.

In our results of \textit{N}-body simulations shown in Section~\ref{sec:SE}, most planets lose their commensurabilities via late orbital instability during the gas depletion, which helps in reproducing the observed period-ratio distribution. On the other hand,  we observe that in some simulations, resonant planets remain at the end. Figure~\ref{fig:p_m_exo}(b) shows examples of such systems, in which both first-order MMRs (between 3:2 and 8:7) and a second-order MMR (5:3) are seen. We note that in this paper, we determine whether planets are in MMRs not by resonant angles but by period commensurabilities in the same way as for identifying observed resonant systems. 

In the result of model4-1, the innermost four planets are in mutual mean-motion resonances, which have 5:3, 4:3, and 3:2 commensurabilities.
In the result of model5-1 (triangles in Fig.~\ref{fig:p_m_exo}(b)), the innermost pair has a 5:3 commensurability and the outer three pairs are near 3:2 and 4:3 MMRs. The period-ratio structure is similar to the TRAPPIST-1 system because both systems have 5:3, 3:2, and 4:3 commensurabilities. The system is also similar to the Kepler-80 system, in which planets are new 3:2 and 4:3 MMRs. In the result of model6-1 (squares in Fig.~\ref{fig:p_m_exo}(b)), the innermost four planets are in a 8:6:4:3 Laplace resonance, which is the same relation as the four planets of the Kepler-223 system. 
In the result of model6-2 (inverted triangles in Fig.~\ref{fig:p_m_exo}(b)), the innermost three planets have a 5:4:3 relation, which is comparable to the Kepler-60 system. Therefore, we show in this paper that the observed cumulative distribution of period ratio, which shows that most planets are not in MMRs, can be reproduced by simulations, and at the same time, several systems with a chain of resonant planets can form.

In our simulation results, we identify several characteristics of the chain formation of resonant planets: (i) The resonant chain tends to form in MRI-inactive disks compared to MRI-active disks. (ii) The resonant chain tends to remain for smaller total mass cases. For example, no resonant planets form for model1 ($M_{\rm tot} = 80~M_\oplus$), but several resonant planets remain for model4 ($M_{\rm tot} = 20~M_\oplus$). (iii) The resonant chain tends to remain for models that consider an initially narrow solid ring. We saw this trend in Section~\ref{sec:summary_app}.

Here we briefly discuss the condition for forming the resonant chain. Planets can be captured in MMRs when they undergo convergent migration (e.g., \citealt{murray_dermott_99}). If the convergent migration rate is too high, planets go through resonant locations as stated in Section~\ref{sec:mass_ranking}. In all simulations of this work, migrating planets are basically captured in MMRs between neighboring planets because the type I migration is suppressed due to the effects of disk winds (see Figs.~\ref{fig:t_a11}, \ref{fig:t_a33}, and \ref{fig:t_a44} for example). This is true even in MRI-inactive disks, in which migration is more rapid than in MRI-active disks. Therefore, the condition for retaining the resonant chain depends on whether or not the system undergoes the late orbital instability after gas depletion; in other words, the orbital stability determines the condition.

The orbital stability of multiplanet systems is investigated by \textit{N}-body simulations \citep{chambers_etal96}. They showed that the orbital stability time decreases with increasing numbers of planets in the system and a decrease in the orbital separation scaled by their mutual Hill radii. The orbital stability time for systems with resonant planets has also been investigated \citep{matsumoto_etal12}. These latter authors confirmed that even though the system can be significantly stabilized depending on the resonant configuration, the orbital stability time still depends on the number of planets and the orbital separation divided by the mutual Hill radius. These dependencies can explain the three characteristics stated above. Regarding (i), the number of planets in a system for MRI-inactive disks is smaller than that for MRI-active disks, meaning that the system is more stable in inactive disks (e.g., Fig.~\ref{fig:t_a33}). Regarding (ii), the mutual Hill radius $r_{\rm H} \propto M^{1/3}$ increases with increasing the planet mass, meaning the orbital stability time is longer for cases with  a smaller total mass.
Regarding (iii), similar to trend (i), the number of planets in a system is smaller for narrower initial conditions.

\subsection{The origin of the different configurations of the solar system and close-in super-Earth systems}
\label{sec:origin}
In this study, we perform \textit{N}-body simulations of formation of close-in super-Earths in both MRI-active and inactive disks. As shown above, active disks are more suitable for reproducing the observed statistical properties of close-in super-Earths. In Paper~I, on the other hand, we aim to reproduce the localized orbital configuration of the terrestrial planets of  the solar system. We considered radial drift of planetesimals and type I migration of planetary embryos. We demonstrated that planetesimals undergo convergent radial drift towards 1~au from the Sun and embryos that formed from the narrow planetesimal ring do not undergo migration, which can reproduce the localized configuration of the terrestrial planets of the current solar system. This is different from results of this paper, in which planets remain in close-in orbits. Here we briefly discuss what makes such a difference in close-in orbits.

One possible explanation is that MRI-inactive disks produce systems concentrated at 1~au like the solar system, whereas MRI-active disks produce close-in super-Earth systems. The MRI-inactive disk has a pressure maximum at 1~au and planetesimals undergo convergent radial drift toward the region in Paper~I. In this paper, although we did not consider the stage of radial drift of planetesimals, we can assume that planetesimals do not undergo convergent radial drift in MRI-active disks. In fact, we performed \textit{N}-body simulations from embryos that are widely distributed in the initial state and found that active disks are more capable of reproducing the observed distribution of close-in super-Earths. We note that in this paper, we also started simulations from embryos that are widely distributed even in MRI-inactive disks, which is not really consistent with the radial drift of planetesimals.

Another possible explanation is the disk evolution. As we stated in Paper~I, it is likely that MRI activity increases with time. It would also be important to consider that it takes some time to form the pressure maximum even in MRI-inactive disks. As seen in Fig.~5 of \citet{suzuki_etal16}, it takes about 0.1 Myr to form the positive slope of gas surface density in MRI-inactive disks. Therefore the disk evolution may be divided into the initial evolution phase (i), the MRI-inactive phase (ii), and the MRI-active phase (iii). In phases (i) and (iii), no pressure maximum region is created.

For the solar system, we can consider that planetesimals formed in phase (ii) with pressure maximum at 1~au. On the other hand, for close-in super-Earth systems, we can assume that planetesimals rapidly form (e.g., streaming instability) and quickly grow to embryos in phase (i). When the disk evolves to phase (ii), planetary embryos already exist and have a wide range of radial distances. In addition, if the phase (ii) did not last long and orbital configurations of close-in super-Earths are mostly determined in phase (iii), observed properties of close-in super-Earths can be reproduced. Alternatively, if a lot of dust grains are still small in phase (ii) and did not undergo radial drift, then embryos that grow to planets form in phase (iii). Therefore, the difference would be in the mass of the objects at the time the disk started to develop a pressure bump in phase (ii).

\section{Conclusions}
\label{sec:conclusions}

Here we investigate the orbital evolution of planetary embryos under various disk conditions. There is presently an issue regarding the type I migration of protoplanets. If protoplanets larger than Mars form before the gas disk dispersal, they undergo rapid inward migration on a timescale of 0.1 Myr in the power-law disk model based on the MMSN, which causes several problems in reproducing orbital distributions of known planetary systems including the solar system and exoplanet systems. We demonstrate that the type I migration problem is mitigated in all cases of our simulations, and the problem can be resolved in many cases. That is, planets do not undergo significant migration in many cases, although super-Earth-mass planets $(M \gtrsim 2 ~M_\oplus)$ exhibit inward migration in some cases (e.g., disk2). Nevertheless, in these latter cases, the inward migration timescale is about 1~Myr, which is much slower than the migration speed in the power-law disk with a slope of -3/2. Our conclusion therefore is that type I migration can be suppressed in disks evolving via disk winds, which can be attributed to the decreased gas surface density and flat slope with desaturation of positive corotation torque in the close-in region.

We use the type I migration formula based on \citet{paardekooper_etal11} (see Section~\ref{sec:typeI}); significant modification of this formula would likely lead to different results. Recent three-dimensional hydro simulations show good agreement with the type I migration formula \citep{lega_etal15,fung_etal17}, which justifies the use of the formula of \citet{paardekooper_etal11}.
However, the dynamical torque may change the results. It is suggested that differences in radial velocities between planets and disk gas can change the corotation torque (e.g.,  \citealt{paardekooper14}). \citet{ogihara_etal17} considered the effects of dynamical torque and showed that this phenomenon plays a role in preventing inward migration of massive super-Earths in MRI-inactive disks, while the desaturation of corotation torque due to the dynamical torque is not important in MRI active disks. We note that the formula of dynamical torque used in \citet{ogihara_etal17} is derived assuming a locally isothermal disk. Therefore the actual dynamical torque in radiative disks can be altered, which should be further investigated.

We apply our disk evolution model to the formation of close-in super-Earth systems. We show that type I migration is significantly slowed down in disks with flat surface density profiles. In this case, the general orbital evolution is that planets are captured in mean motion resonances in a gas disk, and they undergo late orbital instability during the gas depletion, which leads to a non-resonant configuration.

The observed period-ratio distribution of close-in super-Earths can be reproduced when planets do not undergo significant migration and experience orbital instability during the gas depletion phase, which is also suggested by previous studies (e.g., \citealt{ogihara_ida09}; \citealt{izidoro_etal17}). In some simulations for MRI-inactive cases, planets are more prone to migration than in MRI-active disks with flat disk profiles; however, we also observe a case for an MRI-inactive disk in which planets do not undergo significant migration and the results of simulations are consistent with the observed period ratio distribution. We further find that the period ratio distribution depends on the planetary mass. The dependence of period ratio on planetary size is also seen in observed distribution, which suggests that observed close-in super-Earths underwent late orbital instability (and giant impacts) during the gas depletion phase. As stated above, we can explain the origin of the general trend of  non-resonant observed close-in super-Earths; at the same time, in several runs of our simulations, planetary systems create a chain of resonant planets, which are similar to observed systems that include the TRAPPIST-1 system. In almost all simulations, planets are captured in MMRs before gas depletion; some of the systems are stable even after gas depletion, which results in a chain of resonant planets in the final state.

For the mass ratio distribution, we do not see a mass ranking in our simulations, something that has been observed in previous \textit{N}-body simulations \citep{ogihara_etal15a}. This is because the type I migration is significantly suppressed and the edge effect, in which planetary migration is stopped at the disk inner edge (and orbital rearrangements take place), is not seen.

We observe that larger terrestrial planets can undergo inward migration in some cases. For example, the positive corotation torque for larger planets is not effectively desaturated in low-viscosity cases (MRI-inactive disks). As a side note, observations show that there may be a dependence of the size of planets on the location of the inner edge of planet population (Howard et al. 2012). There may also be a deficit in the occurrence rate distribution at $R \simeq 1.5-2.0 ~R_\oplus$ $(M  \simeq 3-5 ~M_\oplus)$ for close-in planets (Fulton et al. 2017). Our findings that larger planets can undergo inward migration may explain the origin of the observed features. This should be further investigated in a future work.

Our conclusion is that when the disk profile is altered due to the effect of disk winds, observed distributions of close-in super-Earths and several specific resonant systems can be produced. We note, however, that  we do not eliminate the possibility of a migration model for the origin of close-in super-Earths. In fact, it is shown that the period ratio distribution can be reproduced if the late orbital instability occurs \citep{izidoro_etal17}. The pros and cons of the formation model will need to be discussed in a separate paper.
We also note that the type I migration is suppressed due to the altered disk profile. Therefore, the results presented here can be applied more generally to disks with flat or positive density slopes.

\begin{acknowledgements}
We thank John Chambers for comments that improved the manuscript. 
Numerical computations were conducted on the general-purpose PC farm at the Center for Computational Astrophysics, CfCA, of the National Astronomical Observatory of Japan. This work was supported by JSPS KAKENHI Grant Numbers 16H07415 and 17H01105.
\end{acknowledgements}

%instead use BIBTEX (see p.12 in aa_instructions.pdf)
{}

%\bibliographystyle{aa}
%\bibliography{reference.bib} 

\appendix
\section{Supplementary simulations of orbital evolution under various disk models}
\label{sec:app1}

We see results of supplementary simulations of Section~\ref{sec:various} for various disk models (disk3-disk10) and initial solid distribution. Each disk model is summarized in Table~\ref{tbl:list1}.

\subsection{Model}
\begin{figure*}
\begin{center}
\resizebox{0.4 \hsize}{!}{\includegraphics{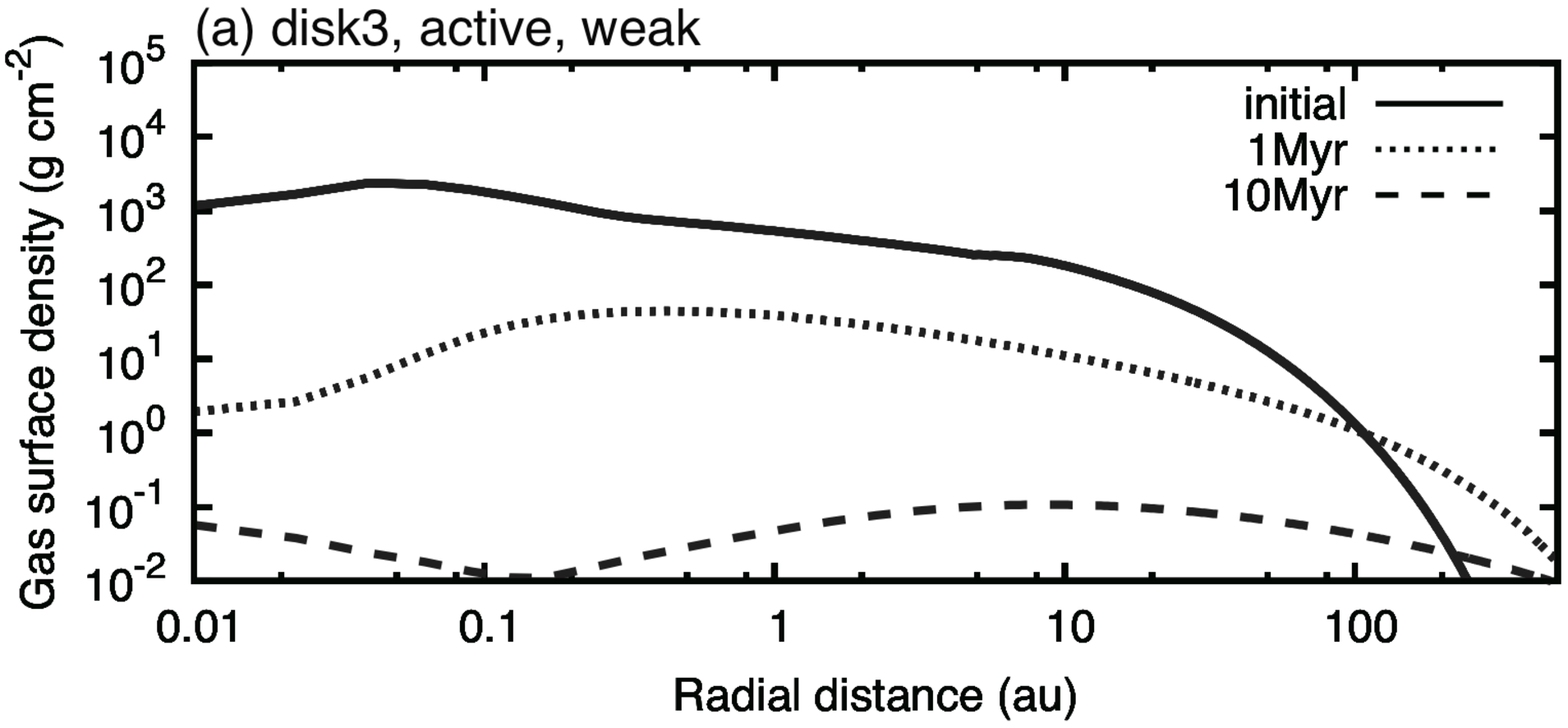}}
\resizebox{0.4 \hsize}{!}{\includegraphics{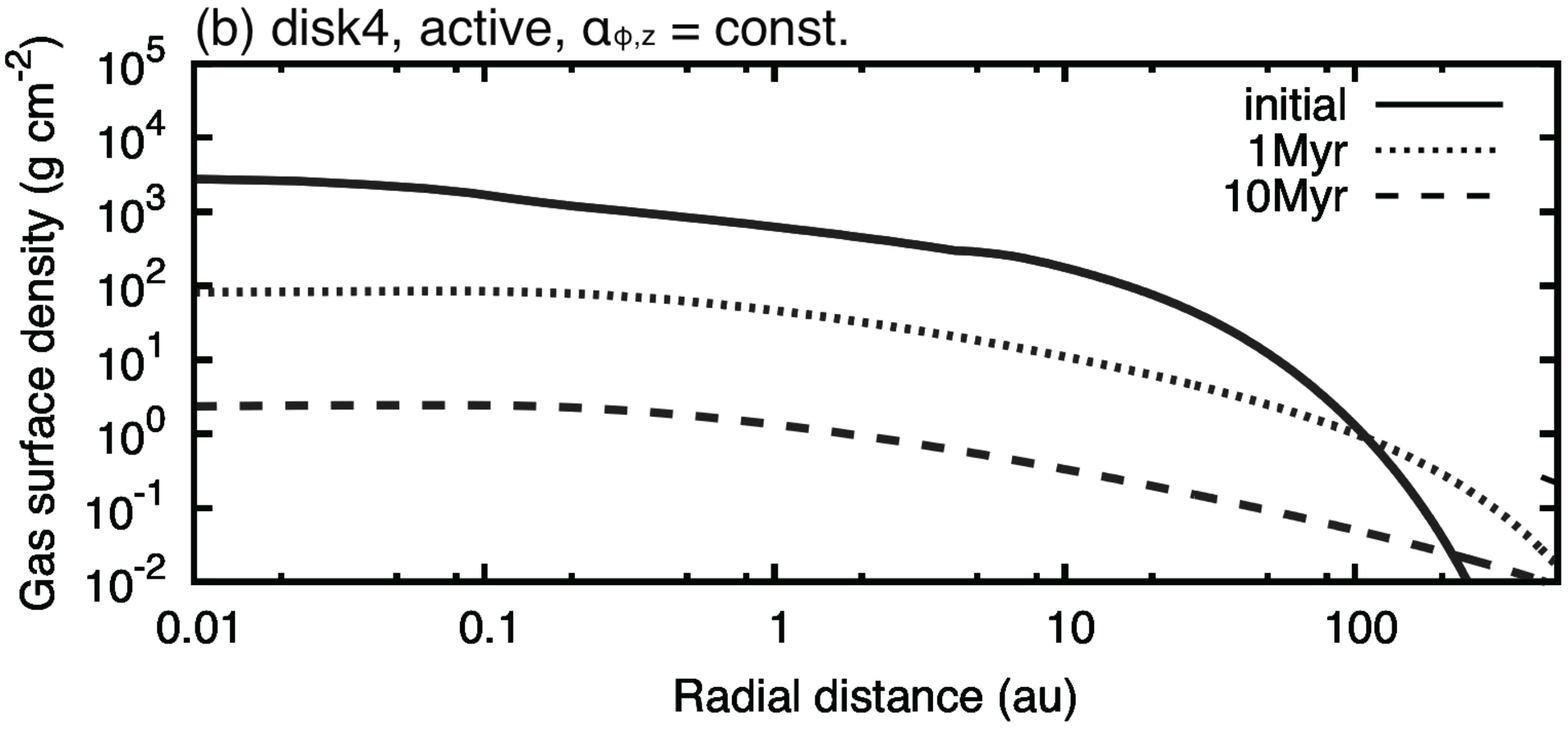}}
\resizebox{0.4 \hsize}{!}{\includegraphics{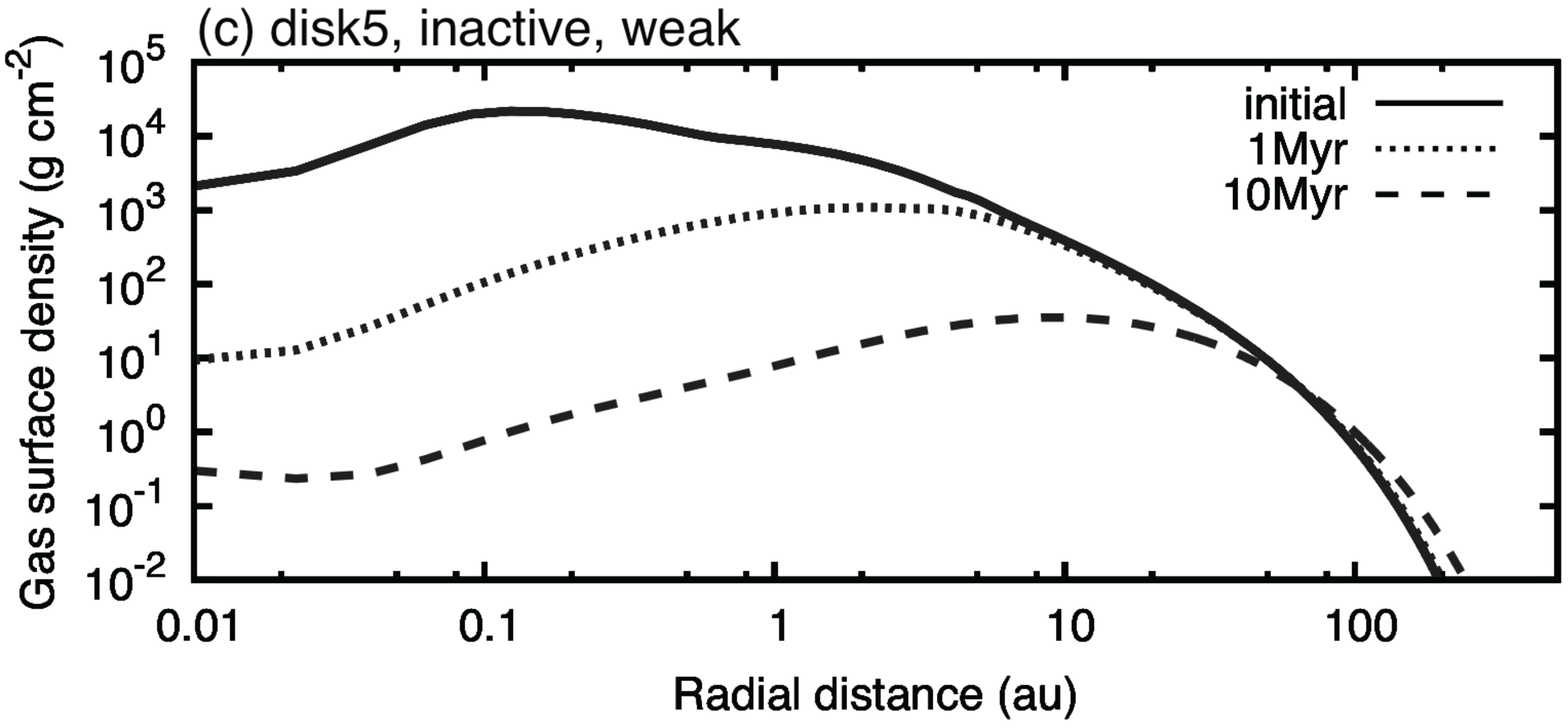}}
\resizebox{0.4 \hsize}{!}{\includegraphics{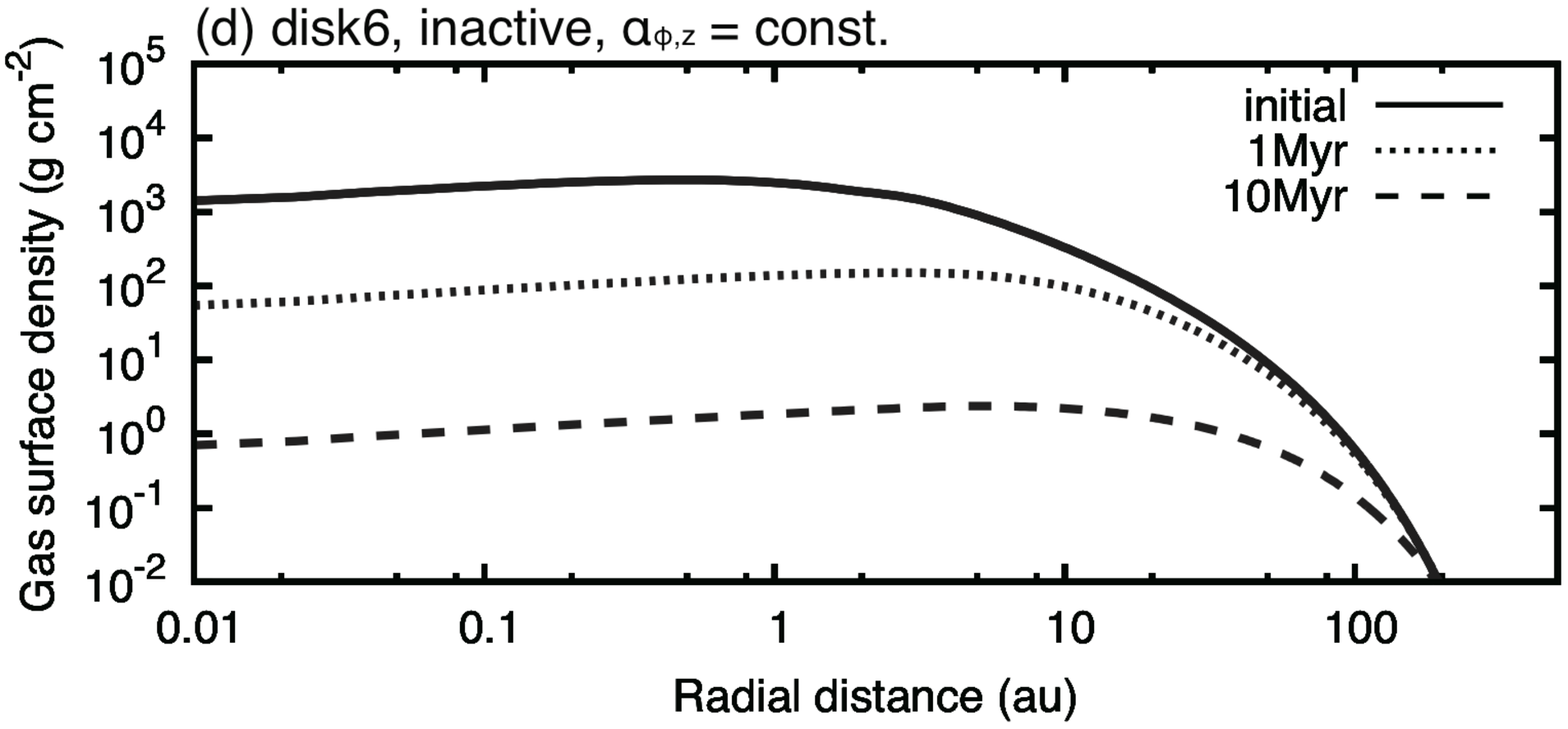}}
\resizebox{0.4 \hsize}{!}{\includegraphics{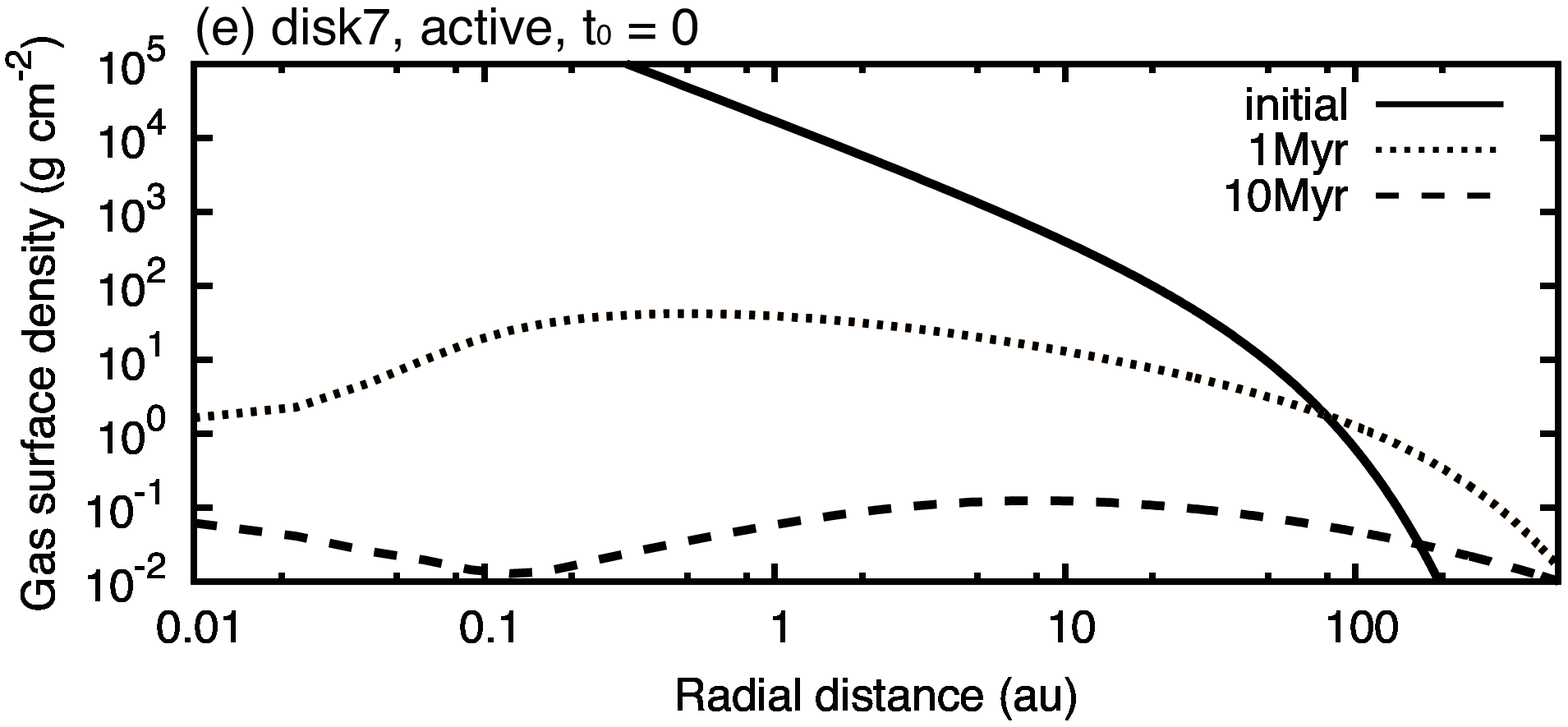}}
\resizebox{0.4 \hsize}{!}{\includegraphics{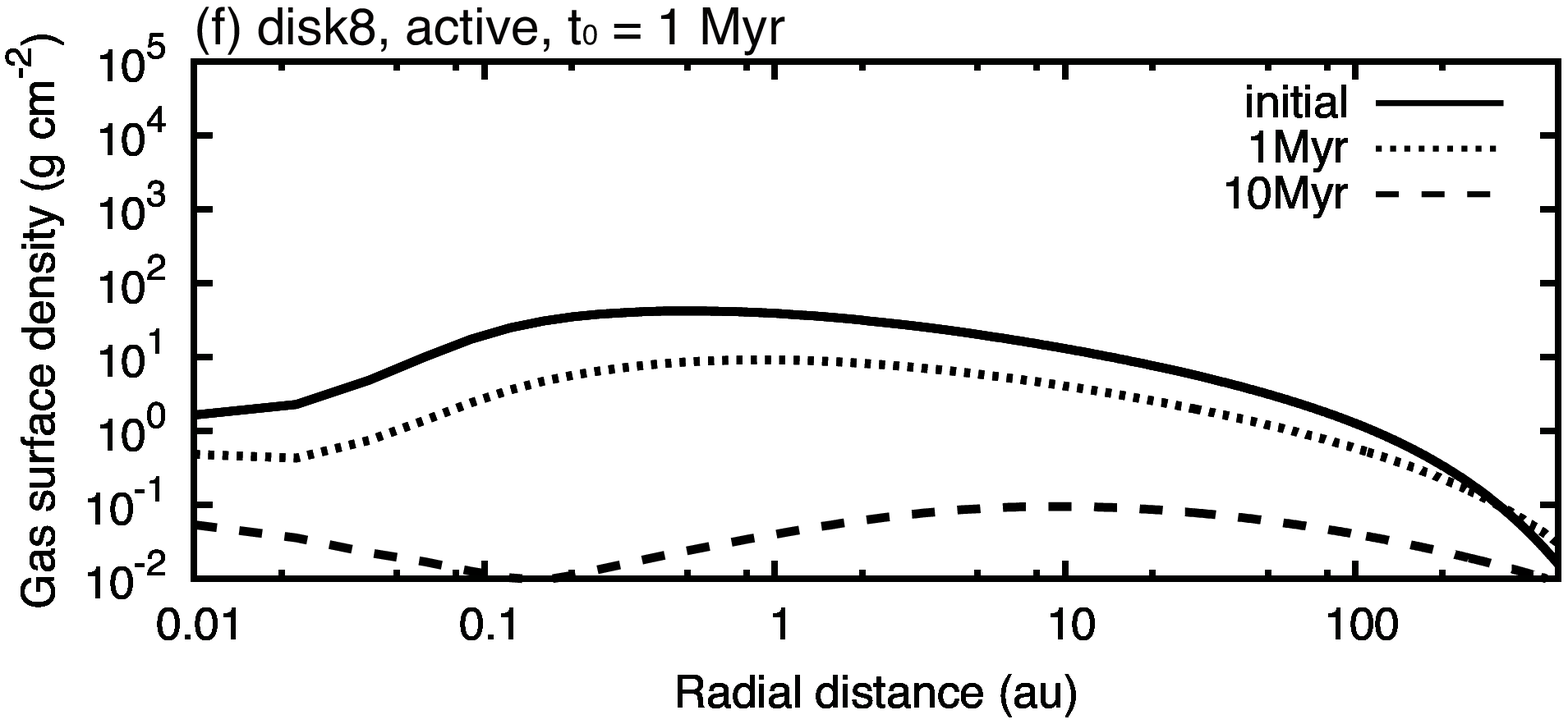}}\\
\resizebox{0.4 \hsize}{!}{\includegraphics{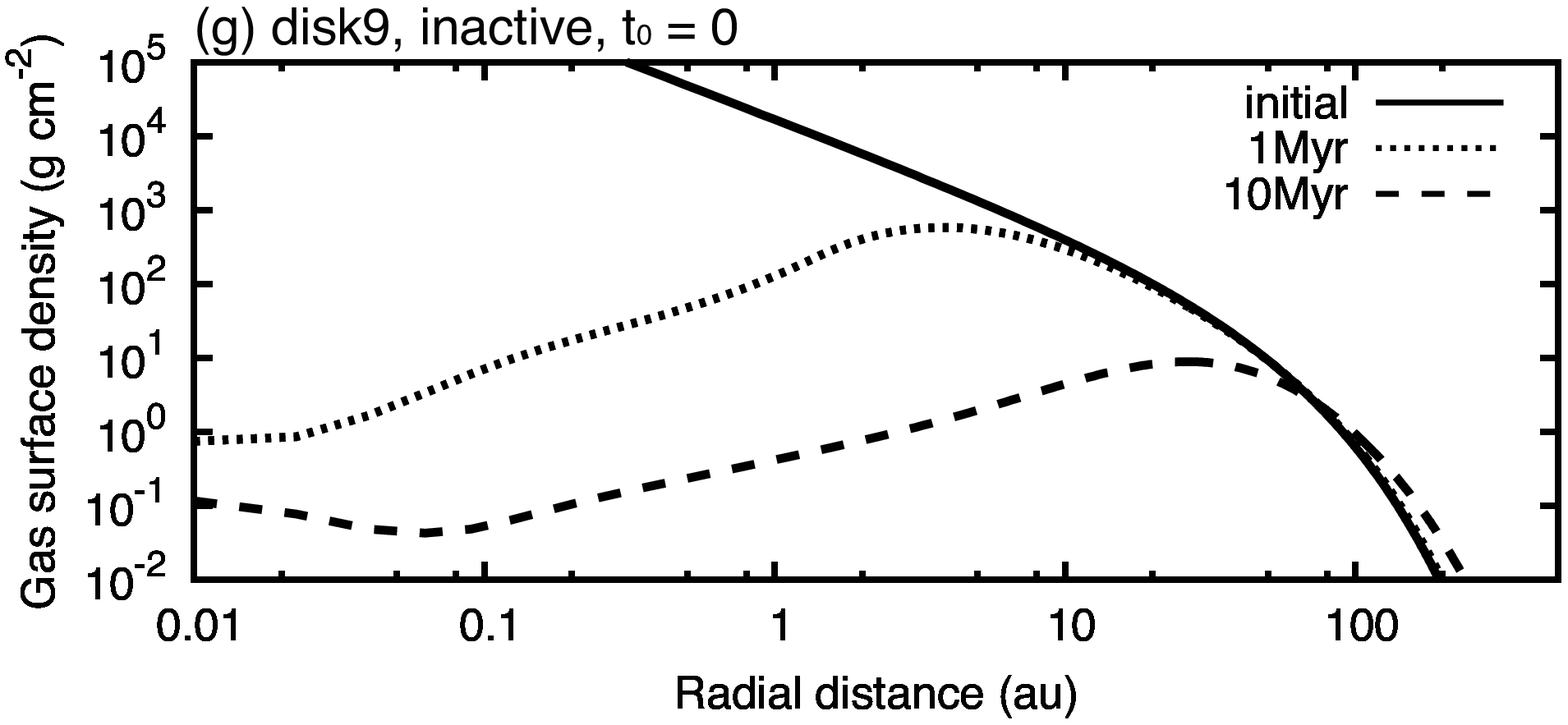}}
\resizebox{0.4 \hsize}{!}{\includegraphics{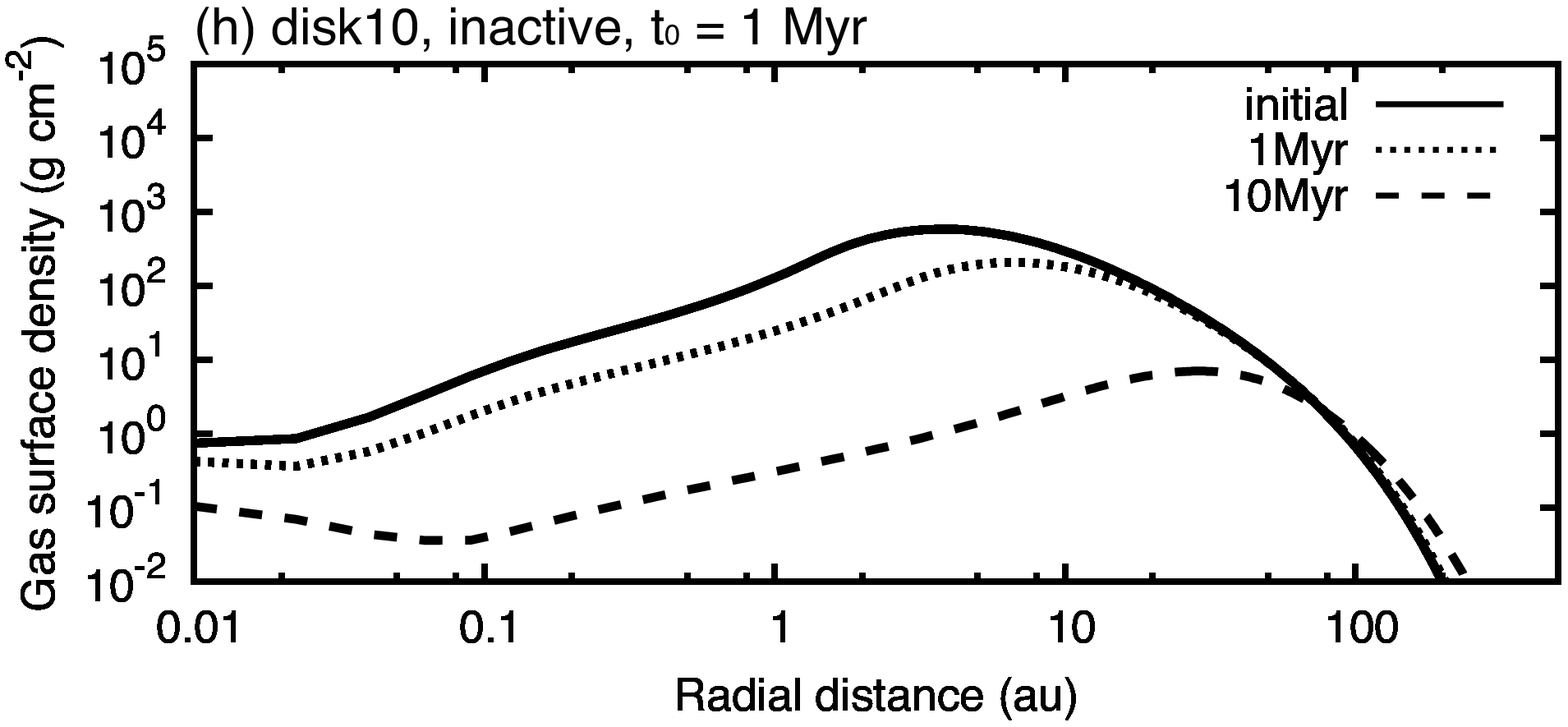}}
\end{center}
\caption{As in Fig.~\ref{fig:r_sigma} but for disk3 - disk10 models.
}
\label{fig:r_sigma_app}
\end{figure*}

The disk evolution for disk3-disk10 is shown in Fig.~\ref{fig:r_sigma_app}.
Figure~\ref{fig:r_sigma_app}(a) and (b) show the case for active disks (disk3 and disk4). In MRI-active disks, turbulence-driven accretion plays a certain role in disk evolution, and the region of positive slope ($p>0$) is limited to $r \lesssim 0.2 {\rm ~au}$. Compared to cases for MRI-inactive disks, disks give more flat distributions at $r \simeq 0.1-1 {\rm au}$. 

%%Fig.A1

For disk5 and disk6 models, the disk is considered to be inactive and $\overline{\alpha_{r,\phi}}= 8 \times 10^{-5}$ is assumed. For disk6 model in Fig.~\ref{fig:r_sigma_app}(d), we see that the density gradient is almost flat inside a few astronomical units. The disk evolution for disk5, in which weak DW regime is considered for the energetics, resembles something between that of disk2 and disk6, in which the density distribution exhibits the positive gradient at the late stage of disk evolution.

Figure~\ref{fig:r_sigma_app}(e)-(h) represent disk evolution assuming different ``initial'' states of gas disk for \textit{N}-body simulations. As stated in Section~\ref{sec:disk_model}, we consider $t_0 = 0.1 {\rm ~Myr}$ as a standard model. In addition to that, we start \textit{N}-body simulations from an earlier stage (disk7 and disk9) or a later stage (disk8 and disk10). The disk evolution at the early stage ($t \lesssim 0.1 {\rm ~Myr}$) is different from that for Fig.~\ref{fig:r_sigma}(a) and (b).

\subsection{Results}

\begin{figure*}
\begin{center}
\resizebox{0.24 \hsize}{!}{\includegraphics{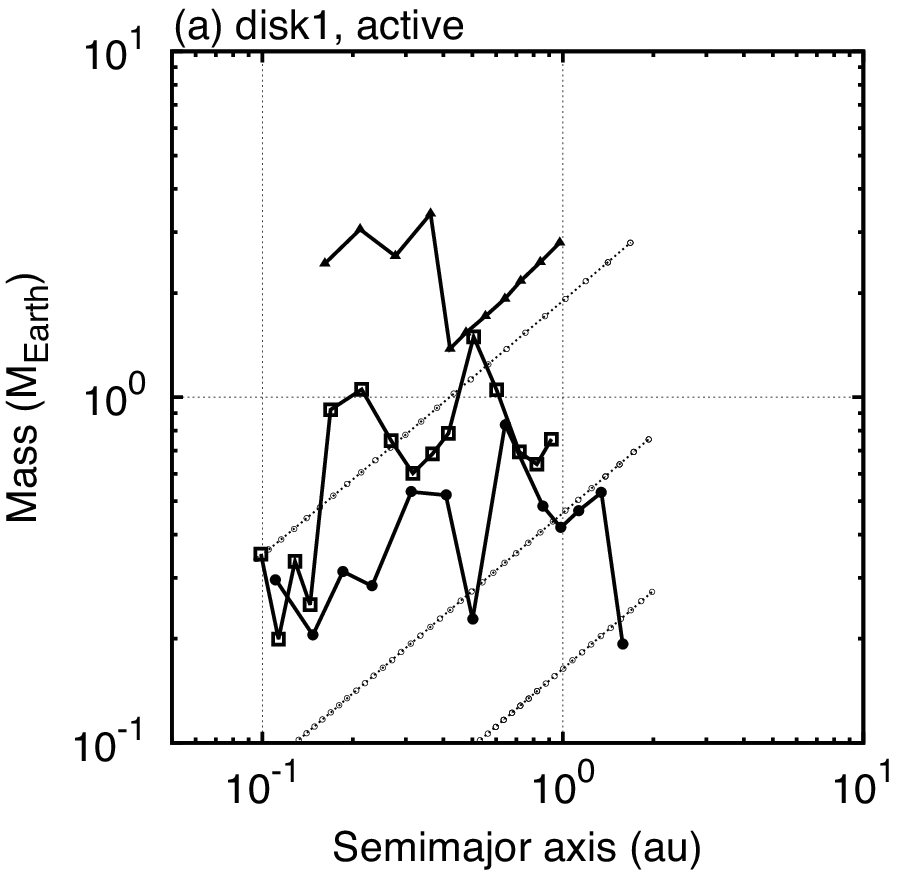}}
\resizebox{0.24 \hsize}{!}{\includegraphics{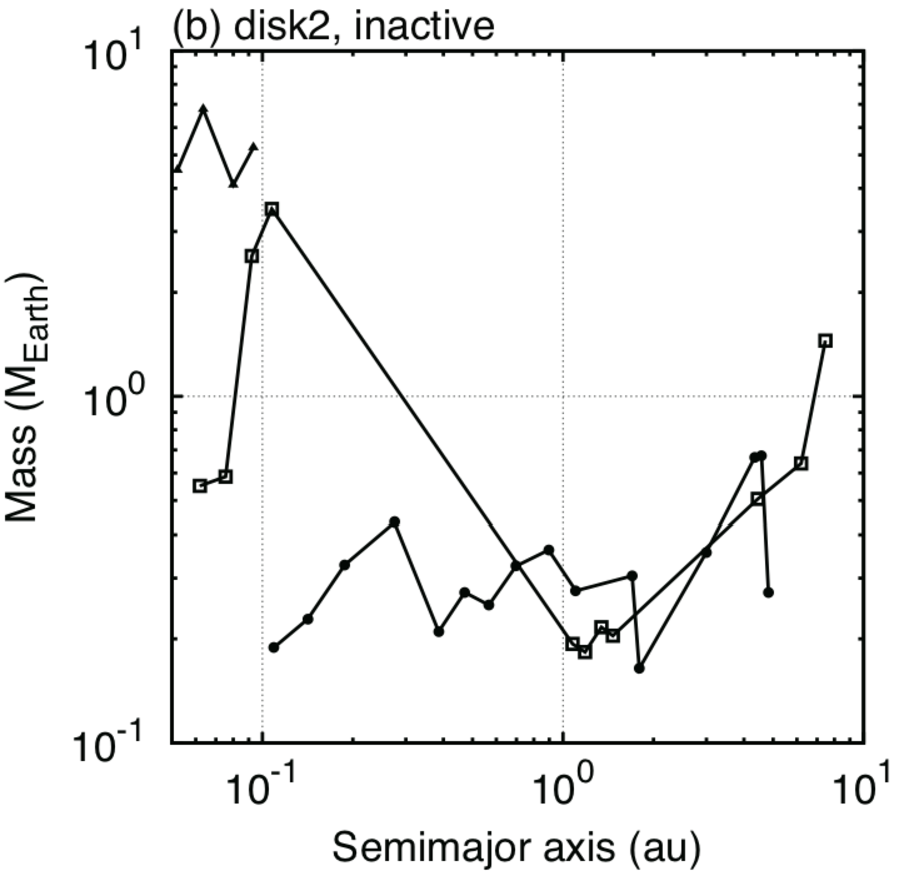}}
\\
\resizebox{0.24 \hsize}{!}{\includegraphics{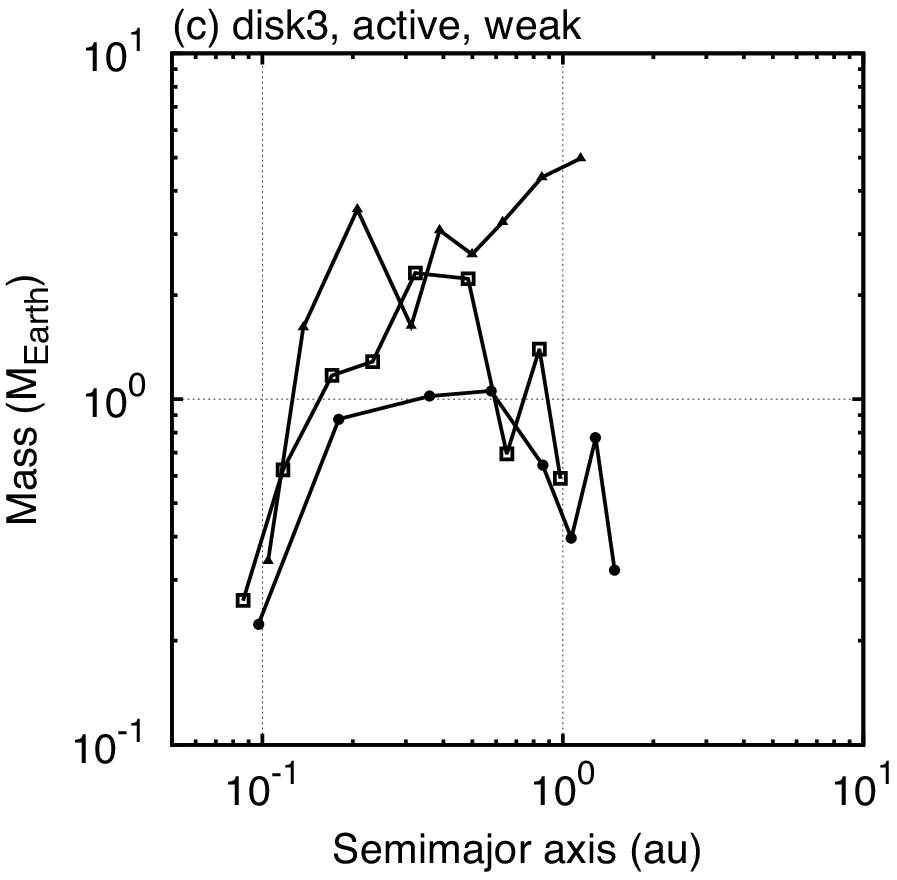}}
\resizebox{0.24 \hsize}{!}{\includegraphics{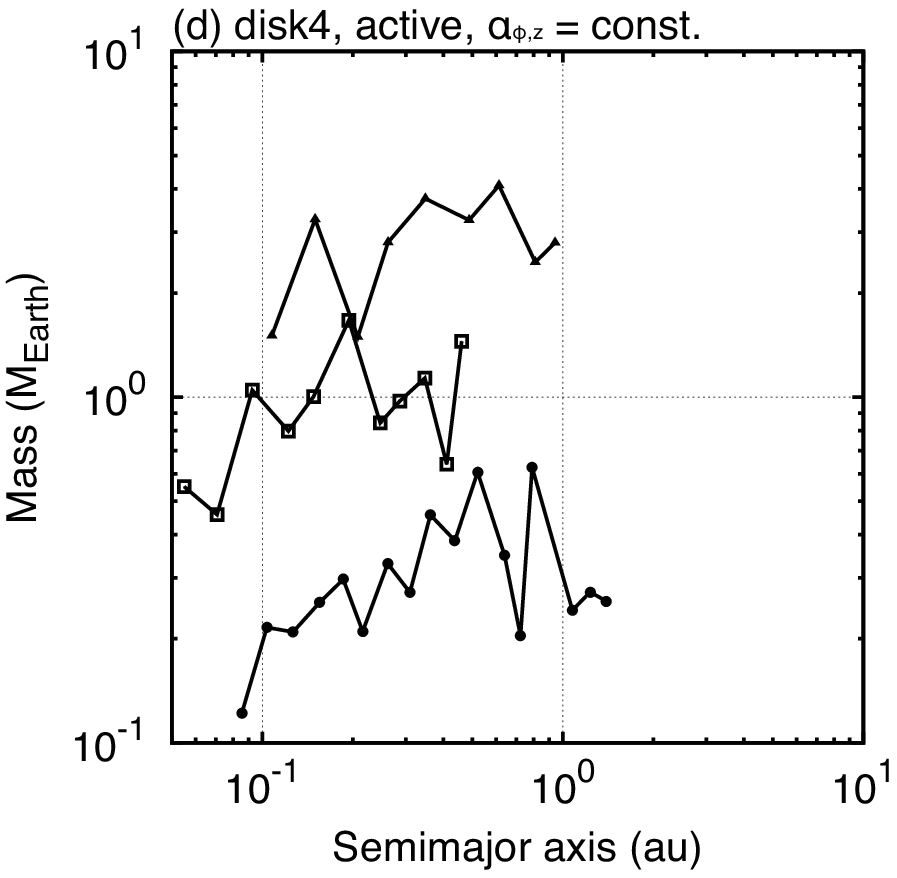}}
\resizebox{0.24 \hsize}{!}{\includegraphics{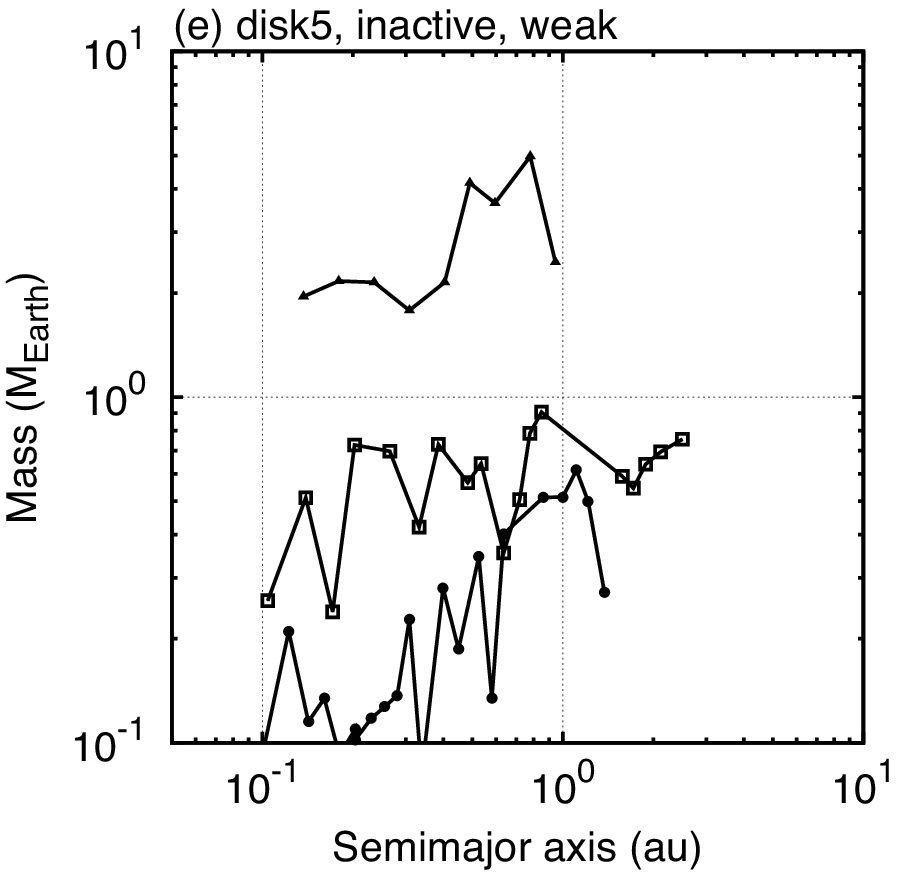}}
\resizebox{0.24 \hsize}{!}{\includegraphics{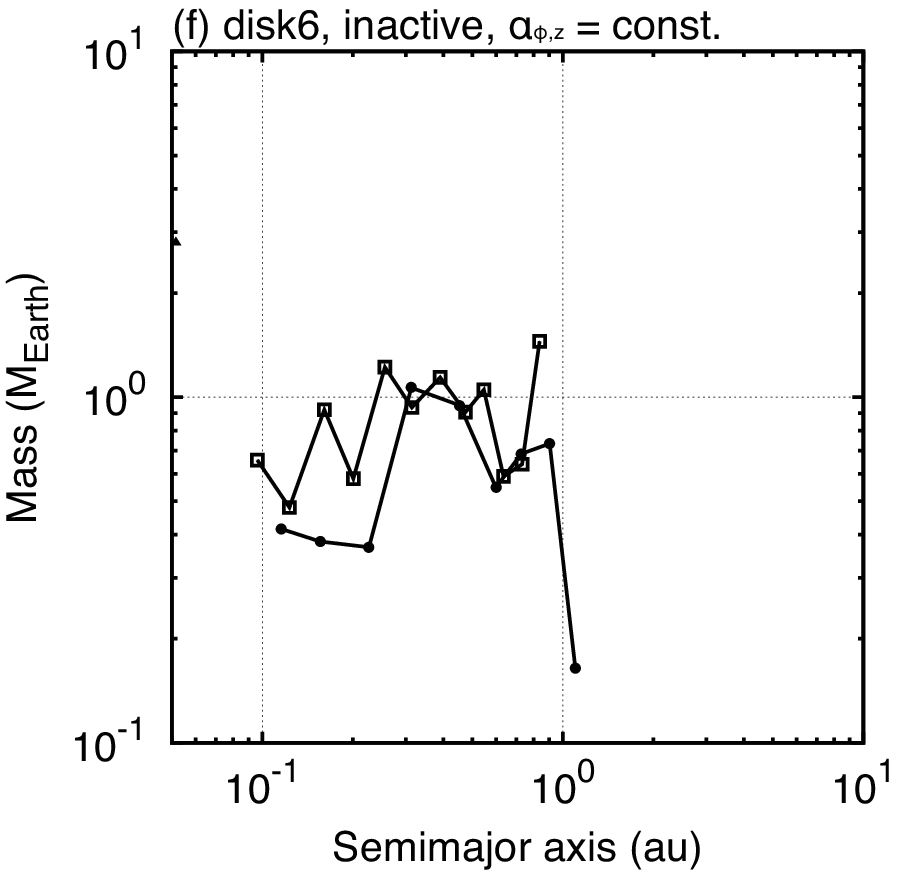}}
\\
\resizebox{0.24 \hsize}{!}{\includegraphics{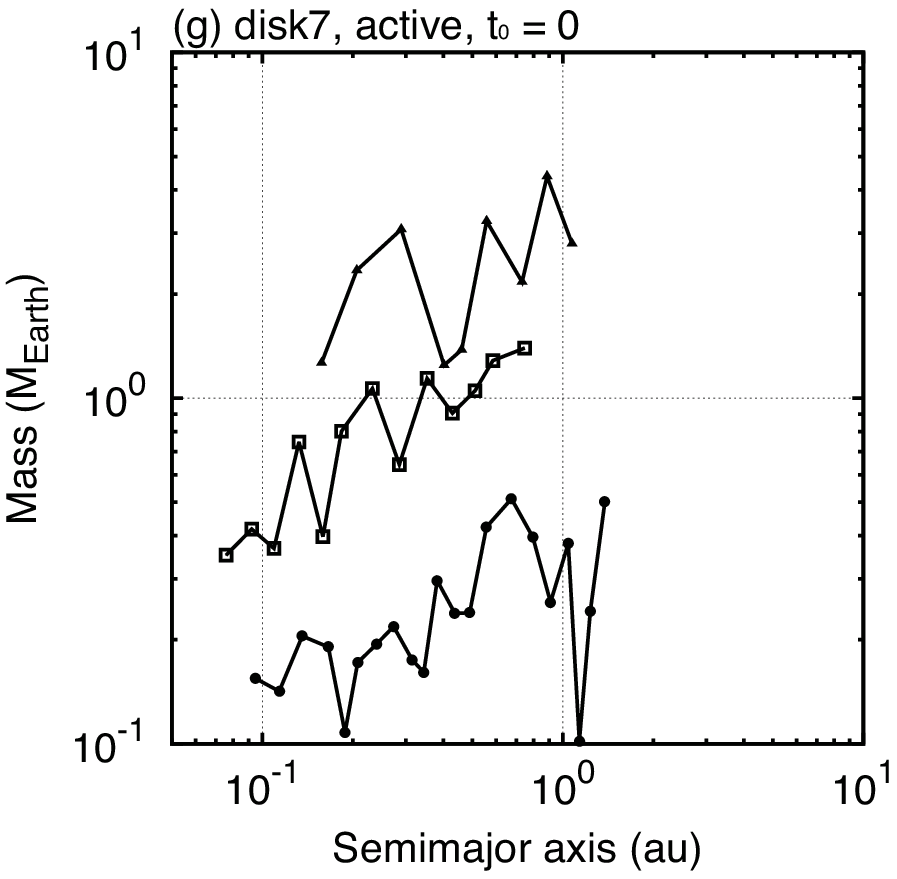}}
\resizebox{0.24 \hsize}{!}{\includegraphics{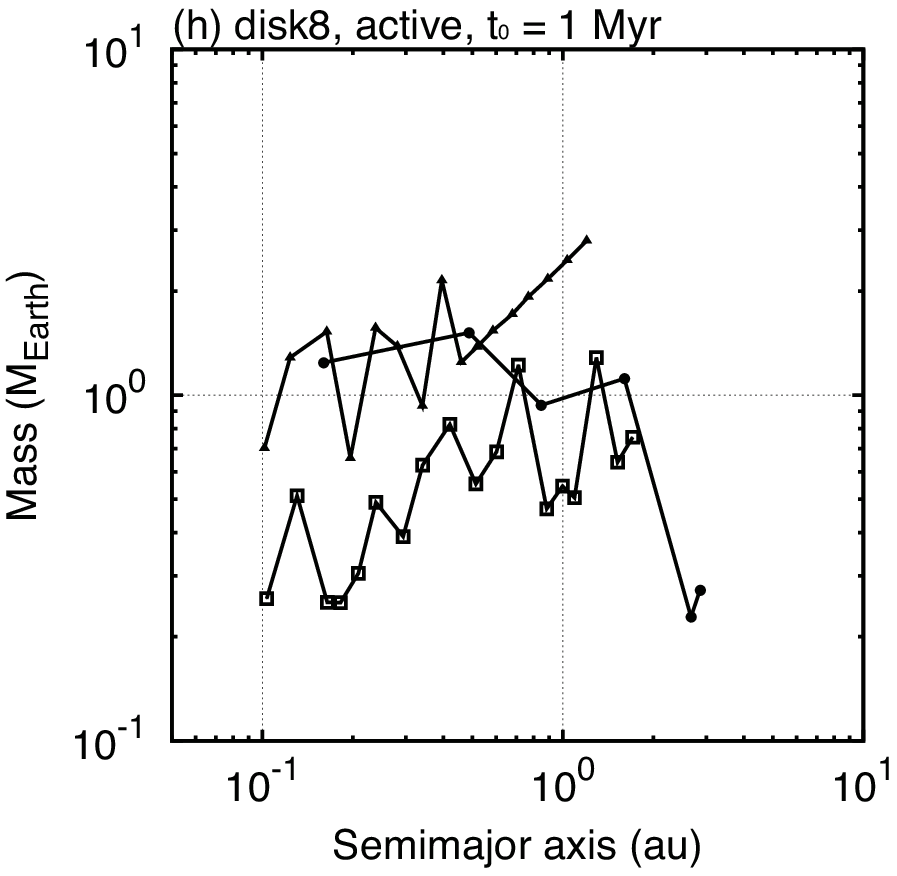}}
\resizebox{0.24 \hsize}{!}{\includegraphics{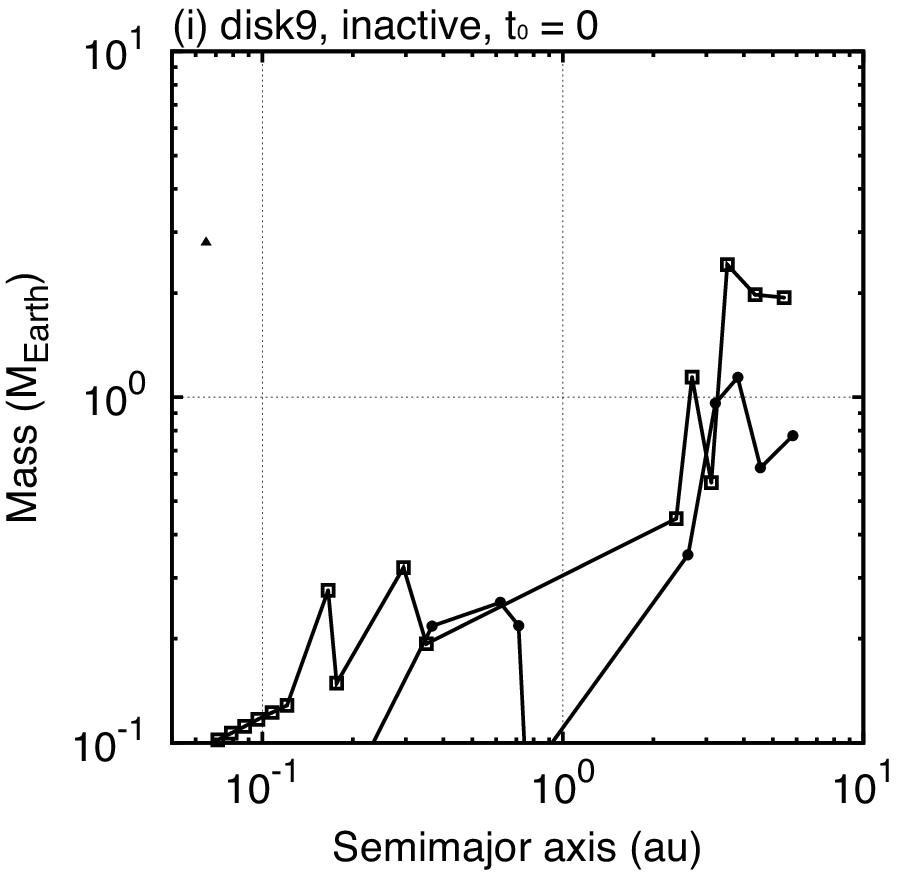}}
\resizebox{0.24 \hsize}{!}{\includegraphics{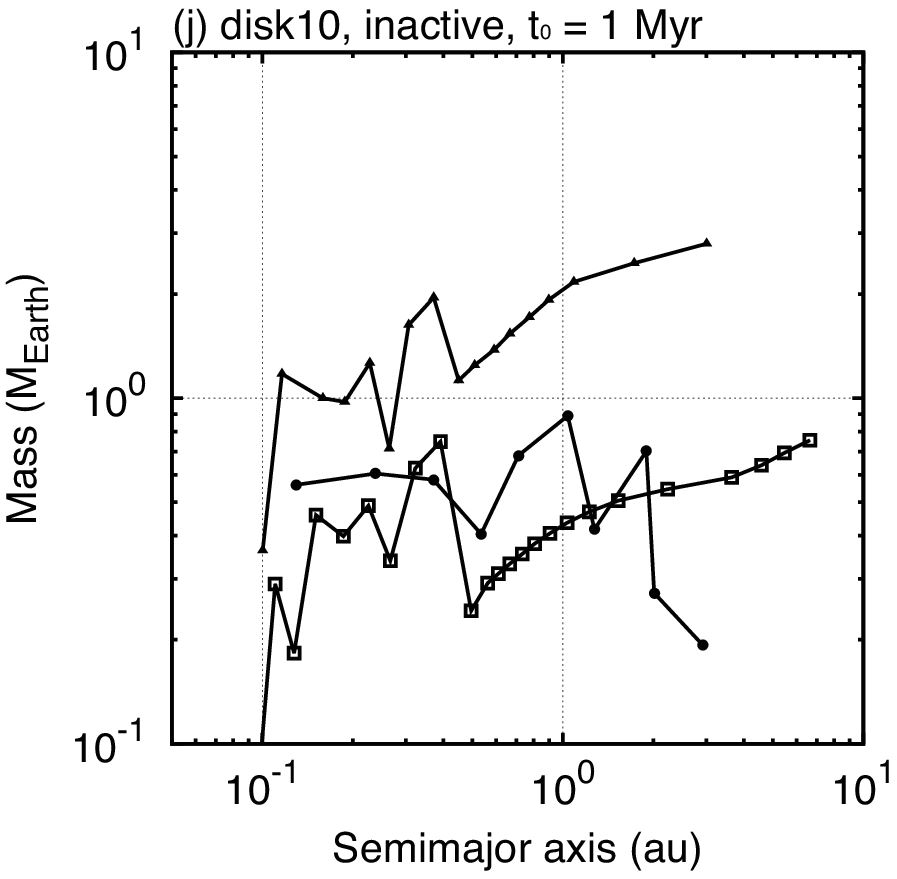}}
\end{center}
\caption{Final orbital configurations of simulations for disk1 - disk10 models. Filled circles, open squares, and filled triangles represent results for $\Sigma_0 = 15$, $\Sigma_0 = 30$, and $\Sigma_0 = 75$, respectively. Initial embryo distributions are also shown in panel~(a) by gray symbols.}
\label{fig:ain1_app}
\end{figure*}

Figure~\ref{fig:ain1_app} shows the final orbital configuration of all simulations both in Section~\ref{sec:various} and in this section. Planets that formed in the same run are connected with the line. Initial distributions of embryos are also shown in panel~(a) by gray symbols. In panel~(a) for disk1, we do not see significant migration as seen in Fig.~\ref{fig:t_a3}. In panel~(b) for disk2, some large planets ($M \gtrsim 2~M_\oplus$) migrate to close-in region, while small planets stay near their initial locations. We can also see that some planets with $M \simeq 1~M_\oplus$ undergo outward migration.

Figure~\ref{fig:ain1_app}(c) and (d) shows the summary for MRI-active disks (disk3 and disk4). There is no significant difference between the three disk models (disk1, disk3 and disk4); the disk evolution is similar in Fig.~\ref{fig:r_sigma}(a), Fig.~\ref{fig:r_sigma_app} (a) and (b).

%%Fig.A2

We look at results of simulations for disk5 and disk6, in which the effects of disk winds are weaker than in disk2. We find that in most cases migration speed is slower than in disk2. In Fig.~\ref{fig:ain1_app}(e) and (f), it is shown that planets do not undergo significant migration in all simulations except high-mass case ($\Sigma_{{\rm d},0}=75$) for disk6. In disk6 model, in which the surface density slope is almost flat in close-in region, migration is suppressed in most cases; however all planets are lost to the star before $t=1 {\rm ~Myr}$ for a high-mass case ($\Sigma_{{\rm d},0}=75$).

Figure~\ref{fig:ain1_app}(g)-(j) show the final state of each simulation for disk7-disk10, in which 
the initial state of the disk ($t_0$) is changed. As mentioned before, our standard model ($t_0 = 0.1 {\rm ~Myr}$) seems reasonable; however we see here the dependence of results on the initial disk condition. 
Figure~\ref{fig:ain1_app}(g) and (h) show results of simulations in MRI-active disks. As expected, in panel~(h), we do not see any significant migration. In panel~(g), some planets undergo migration, in particular high-mass case ($\Sigma_{{\rm d},0}=75$); however only some of the planets fall onto the star. There is no significant difference in final orbital locations between Fig.~\ref{fig:ain1_app}(a) and Fig.~\ref{fig:ain1_app} (g) and (h).
In the disk9 model ($t_0 = 0$) for MRI-inactive disks, we start \textit{N}-body simulations from the early stage of disk evolution when the disk is massive. In this case, planets are more prone to migration. On the other hand, in the disk10 model ($t_0 = 1 {\rm ~Myr}$), planets do not undergo significant migration because the gas surface density in close-in region is already small at the initial time (see Fig.~\ref{fig:r_sigma}). For the final orbital configuration, see Fig.~\ref{fig:ain1_app}(i) and (j). 

We also explore a range of initial conditions to test whether our results in Section~\ref{sec:various} can be modified. We change the slope of initial solid distribution. In some simulations, the mass of initial embryos are considered as the isolation mass but set at $0.1~M_\oplus$. As a conclusion, results have no clear dependence on the initial distribution. For MRI-inactive disks,  as described in the previous subsection, large planets with $M \gtrsim 2 ~M_\oplus$ migrate towards the central star by passing over the outward migration region in the migration map (Fig.~\ref{fig:map1}(b)). Smaller embryos with $M \lesssim 0.3 ~M_\oplus$ tend to stay at their initial locations. Some planets between these two planets can migrate outward as shown in Fig.~\ref{fig:t_a2}(b). For MRI-active disks, we confirm that planets do not undergo significant migration. The range of orbital locations of final planets is between about 0.1 and 2~au.

In summary, results of supplementary simulations shown in this section support our findings shown in Section~\ref{sec:various}. That is, type I migration can be significantly suppressed due to effects of disk winds. In some simulations, inward migration of high-mass embryos are observed in MRI-inactive disks, which can be explained by the fact that the large positive corotation torque is not desaturated enough owing to low viscosity.

\section{Supplementary simulations of close-in super-Earth formation}
\label{sec:app3}

Here we show results of additional simulations of Section~\ref{sec:SE} for various conditions.

\subsection{Initially narrower solid distribution (model7, model8, model9)}

%%Fig.B1
\begin{figure}
\resizebox{0.8 \hsize}{!}{\includegraphics{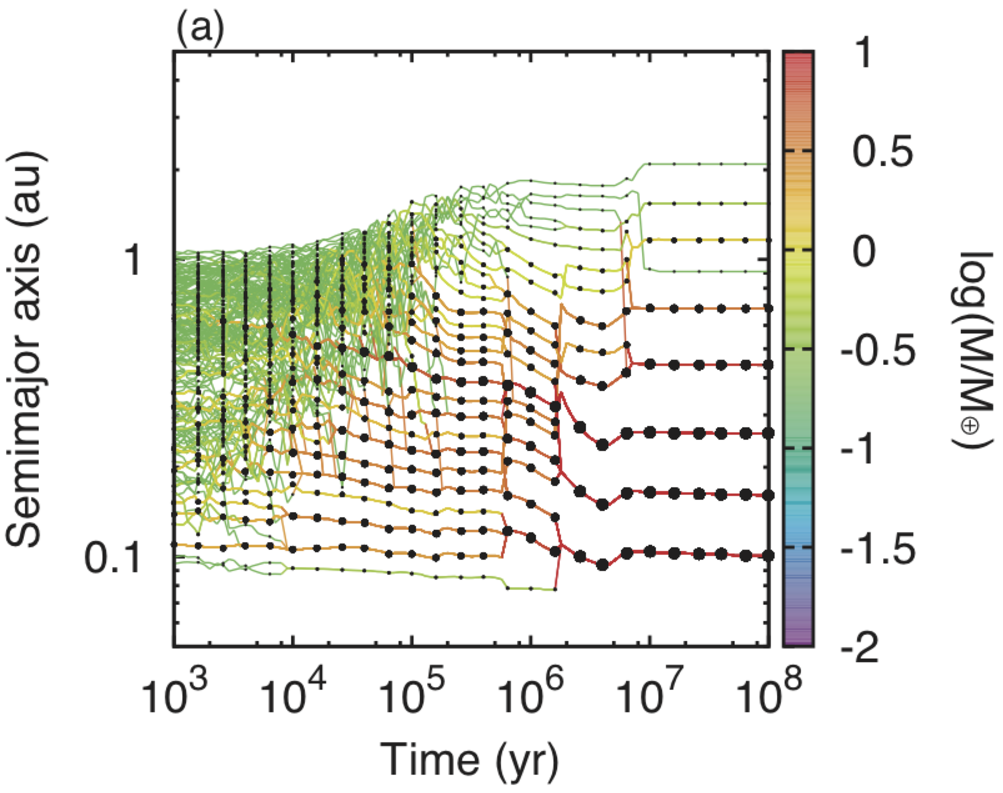}}
\resizebox{0.8 \hsize}{!}{\includegraphics{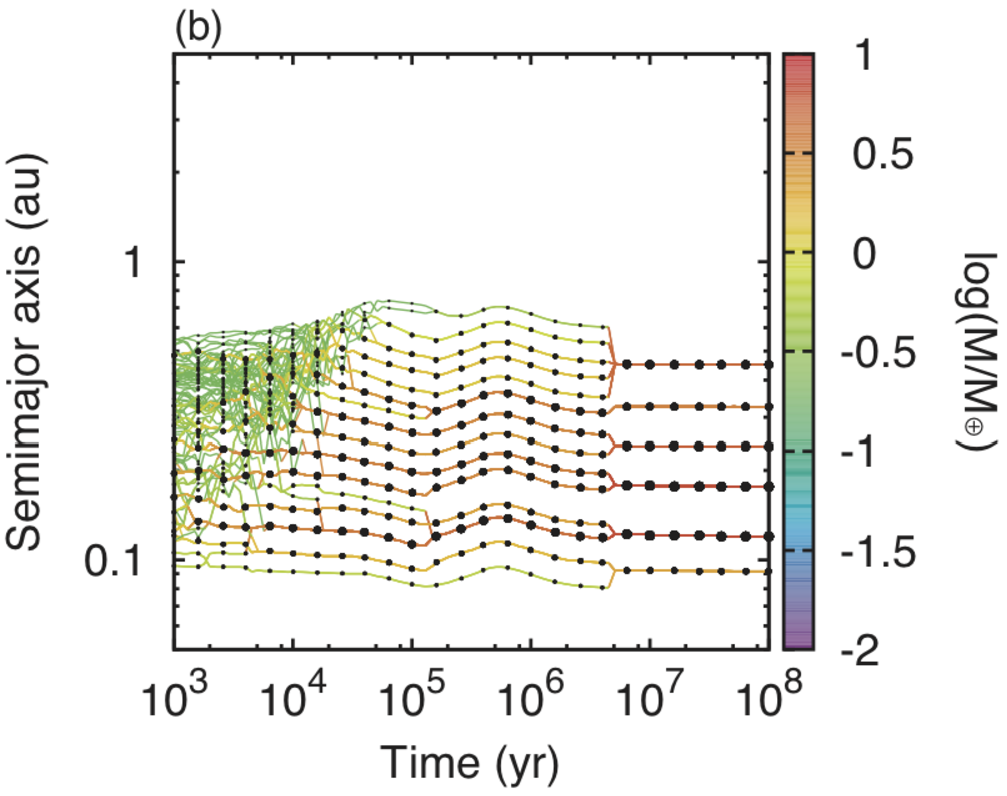}}
\resizebox{0.8 \hsize}{!}{\includegraphics{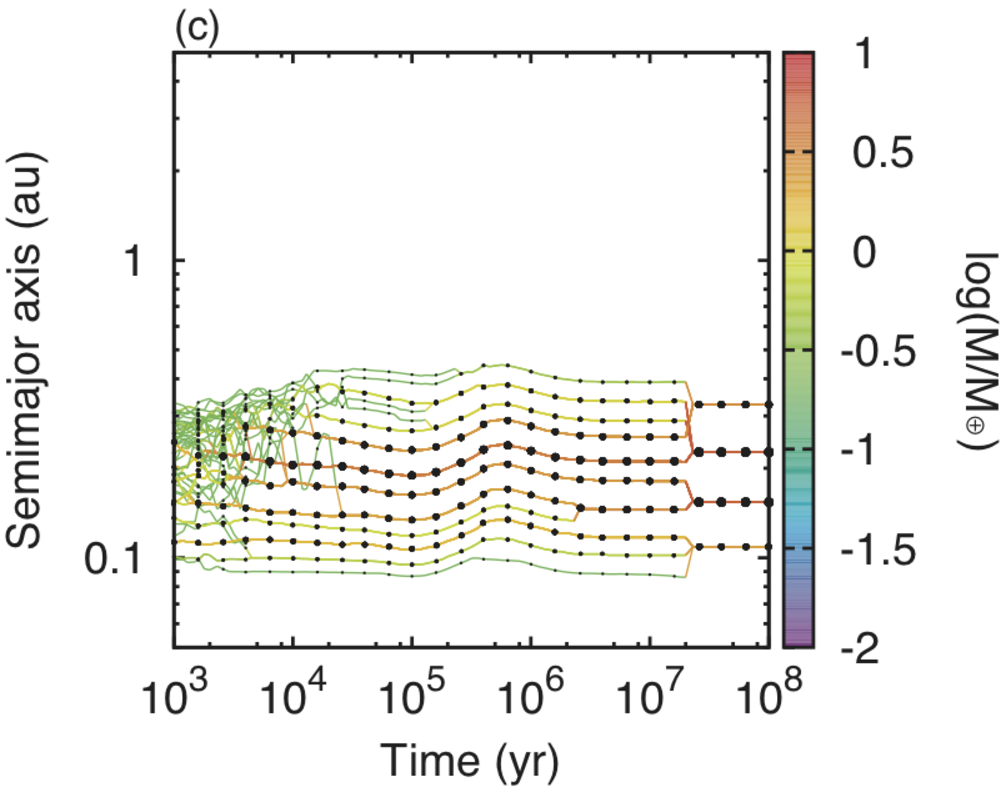}}
\caption{As in Fig.~\ref{fig:t_a11} but for model7 (a), model8 (b), and model9 (c).
}
\label{fig:t_a66}
\end{figure}

Fixing the solid surface density ($\Sigma_{{\rm d},0} = 159$ in Eq.~\ref{eq:solid}), the outer edge of the ring $r_{\rm out}$ is varied in model7 - model9. The outer edge is set at $r_{\rm out}=$ 1, 0.5, 0.3 au for model7, model8, and model9, respectively.
This initial condition can be justified by considering the evolution of smaller particles (i.e., planetesimals, dust aggregates). Disk winds can create a pressure maximum at $r \simeq 1 {\rm ~au}$ (Fig.~\ref{fig:r_sigma}(b)). Planetesimals and dust particles migrate outward due to gas drag inside the location of the pressure maximum, while they undergo inward migration outside the pressure maximum. As a result, planetesimals can form in a narrow ring-like region. 
We discussed this process in Paper~I. Although the pressure maximum can form in MRI-inactive disks, we also adopt initially narrower solid ring for MRI-active disks (model7 - model9).
As we stated in Section~\ref{sec:origin}, the MRI activity would increase with time. Thus it is reasonable to assume that the disk can be MRI-inactive in the phase of radial drift of planetesimals/dust and MRI-active in the phase of embryo migration.

Figure~\ref{fig:t_a66} shows the time evolution of semimajor axis. As stated in Section~\ref{sec:chain}, the orbital evolution is basically the same as that for model1 in Fig.~\ref{fig:t_a11}, except that the timing of giant impacts after gas depletion is delayed in models with smaller $r_{\rm out}$. The orbit crossing time is inversely proportional to the number of planets in systems (\citealt{chambers_etal96}; \citealt{matsumoto_etal12}). In Fig.~\ref{fig:t_a66}, the number of planets at $t = 1~{\rm Myr}$ is 20 (model7; panel~(a)), 14 (model8; panel~(b)) and 11 (model9; panel~(c)); thus the orbit crossing time is the longest for model9. In four of the ten runs for model9, the orbit crossing time is quite long and the global orbit crossing does not occur between $t = 0.1-10~{\rm Myr}$. 

\begin{figure}
\resizebox{1.0 \hsize}{!}{\includegraphics{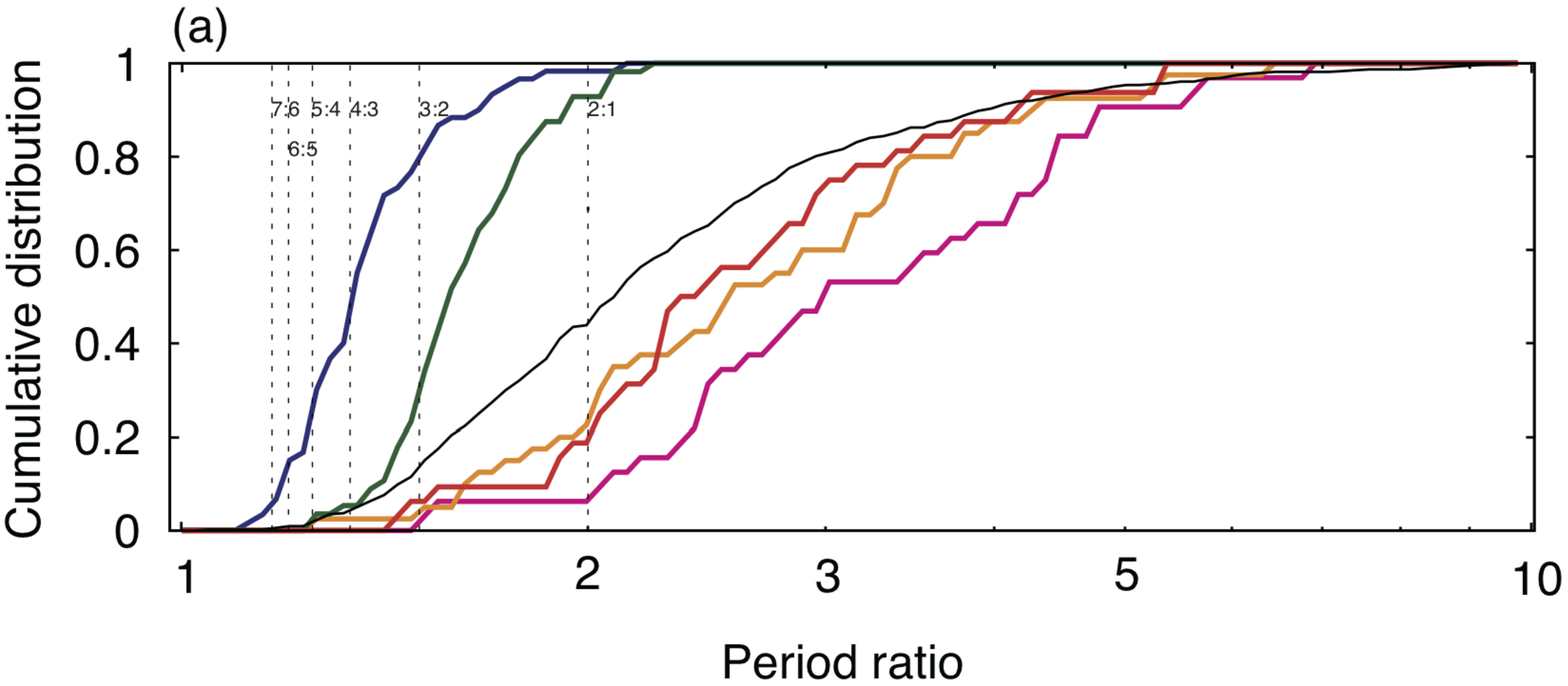}}
\resizebox{1.0 \hsize}{!}{\includegraphics{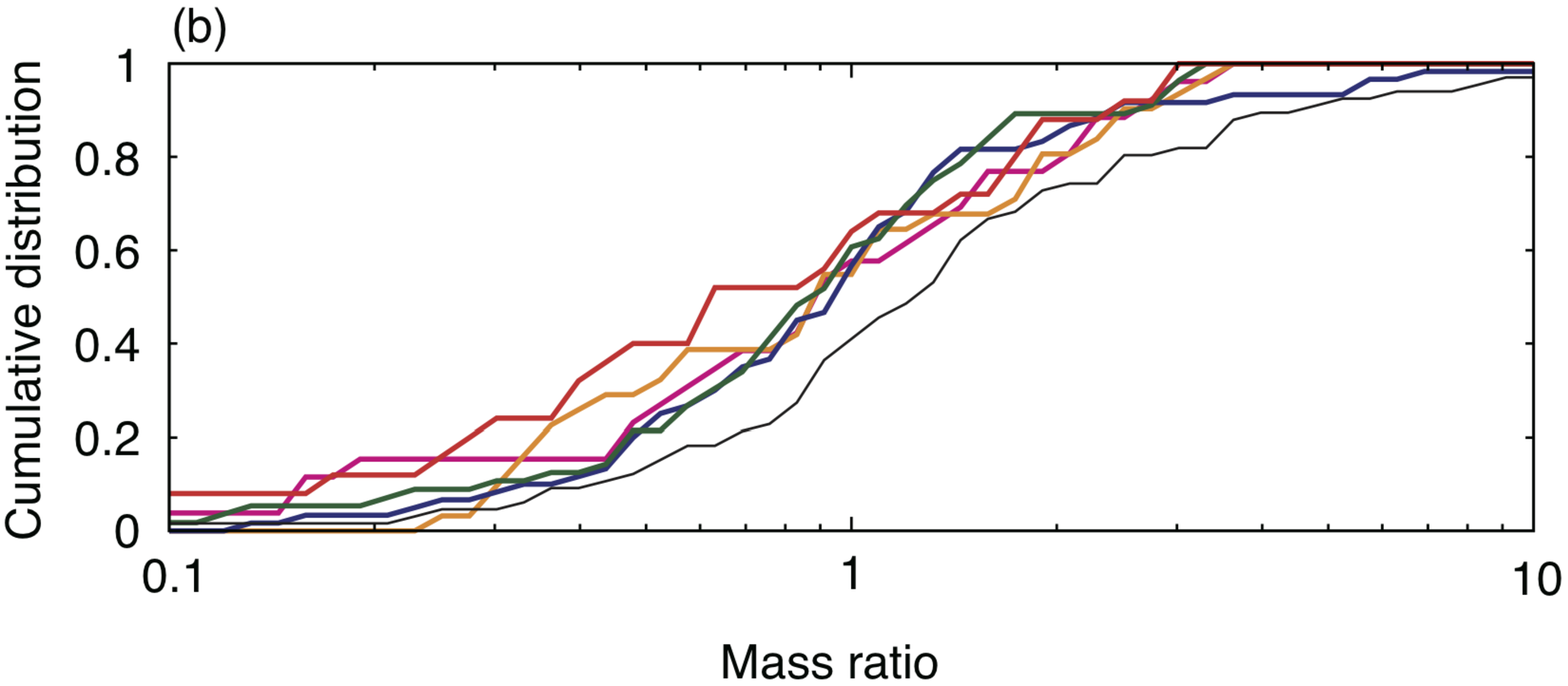}}
\resizebox{1.0 \hsize}{!}{\includegraphics{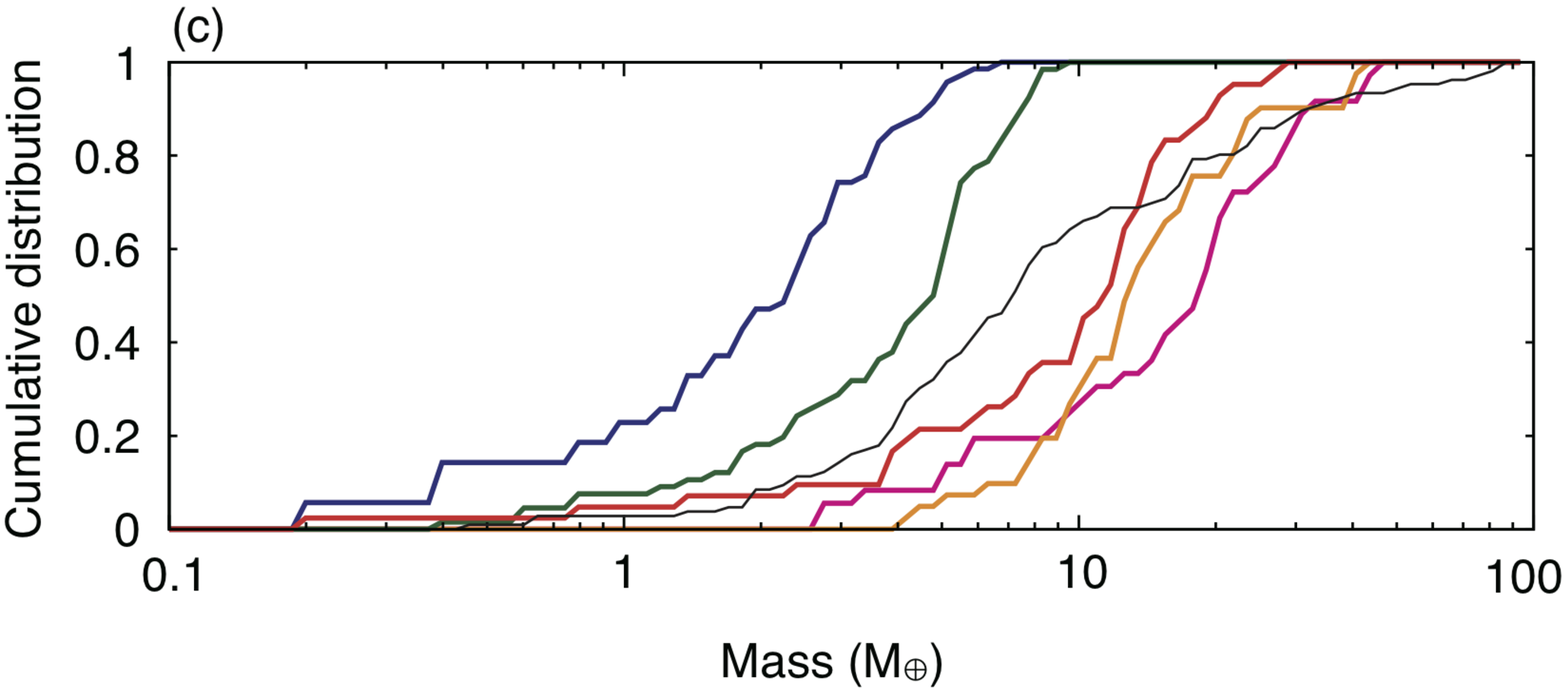}}
\caption{As in Fig.~\ref{fig:sum1} but for model7 - model11. Red, green, blue, orange, and magenta lines represent results for model7, model8, model9, model10, and model11.
}
\label{fig:sum3}
\end{figure}

Figure~\ref{fig:sum3} shows cumulative distributions. 
In panel~(a), we find that the period ratio decreasing with decreasing $r_{\rm out}$. This is because the late orbital instability during gas dissipation is less likely to occur for cases of smaller $r_{\rm out}$. In fact, many planets are in or near mean-motion resonances for model9 at the end (the blue line in Fig.~\ref{fig:sum3}(a)). For the mass ratio distribution in panel~(b), we do not see any significant differences between different models. The mass distribution depends on $r_{\rm out}$ in panel~(c). This is again because the number of planets in the system is larger and the system is likely to undergo late orbital instability for models with larger $r_{\rm out}$.

%%Fig.B2

\subsubsection{Radial mass concentration}

%%Fig.B3
\begin{figure}
\resizebox{1.0 \hsize}{!}{\includegraphics{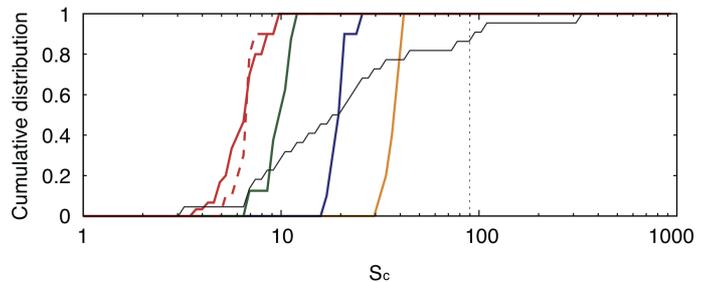}}
\caption{Cumulative distributions of a mass concentration statistic $S_{\rm c}$. The black line indicates the distribution for 23 close-in super-Earth systems. Red, red dashed, green, blue, and orange lines represent results for model1, model3, model7, model8, model9, respectively. The vertical dashed line at $S_{\rm c}=89.9$ corresponds to the terrestrial planets of the solar system.
}
\label{fig:sc}
\end{figure}

In Section~\ref{sec:SE}, we compare cumulative distributions of period ratio, mass ratio and planetary mass. Here, we look at another quantity. \citet{chambers01} introduced a mass-concentration statistic
\begin{eqnarray}
\label{eq:sc}
S_{\rm c} =  \max\left( \frac{\sum M_j}{\sum M_j [\log_{10}(a/a_j)]^2} \right),
\end{eqnarray}
where $j=1,2,...,N$. This quantity is determined by the maximum value of the function, which is a function of $a$, and represents the radial mass concentration of a planetary system. Larger values of $S_{\rm c}$ means that the system is more radially concentrated. Figure~\ref{fig:sc} shows cumulative distributions of $S_{\rm c}$. The black line represents the cumulative distribution of observed close-in super-Earth systems. The quantity $S_{\rm c}$ is calculated for 23 systems, in which the planetary mass is estimated for all planets in each system. The vertical dashed line at $S_{\rm c}=89.9$ indicates the radial mass concentration of the solar system's terrestrial planets, which shows high radial mass concentration because the masses of Mercury and Mars are small and the orbital separation between Venus and the Earth is not large. We see that $S_{\rm c}$ for close-in super-Earths is $\simeq 10-100$, which indicates that most close-in super-Earth systems are relatively compact. We note that only 23 systems are analyzed here and the distribution can be biased.

The red solid line and red dashed line represent the radial mass concentration in the final state for model1 and model3, respectively. Although results for model3 provide a good match to the observed distribution in Fig.~\ref{fig:sum1}, the radial mass concentration static $S_{\rm c}$ cannot be reproduced. 
We find two characteristics. First, the dispersion of $S_{\rm c}$ for each model is low. Second, values of $S_{\rm c}$ depend on $r_{\rm out}$ (i.e., initial $S_{\rm c}$). Both can be explained by the fact that planets do not undergo significant migration in active disks.

Although current observations may miss planets and the observed distribution of $S_{\rm c}$ shown in Fig.~\ref{fig:sc} can be biased, here we demonstrate that planetary systems in compact configuration can form. In model7 - model9, simulations start from embryos that are initially placed in a narrow ring-like region. In this case, the measure $S_{\rm c}$ is already large in the initial state. Then planetary embryos do not undergo migration in MRI-active disks and the compact orbital configuration is maintained to the end.

\subsection{Steeper/shallower solid density profile (model10, model11)}

\begin{figure}
\resizebox{0.8 \hsize}{!}{\includegraphics{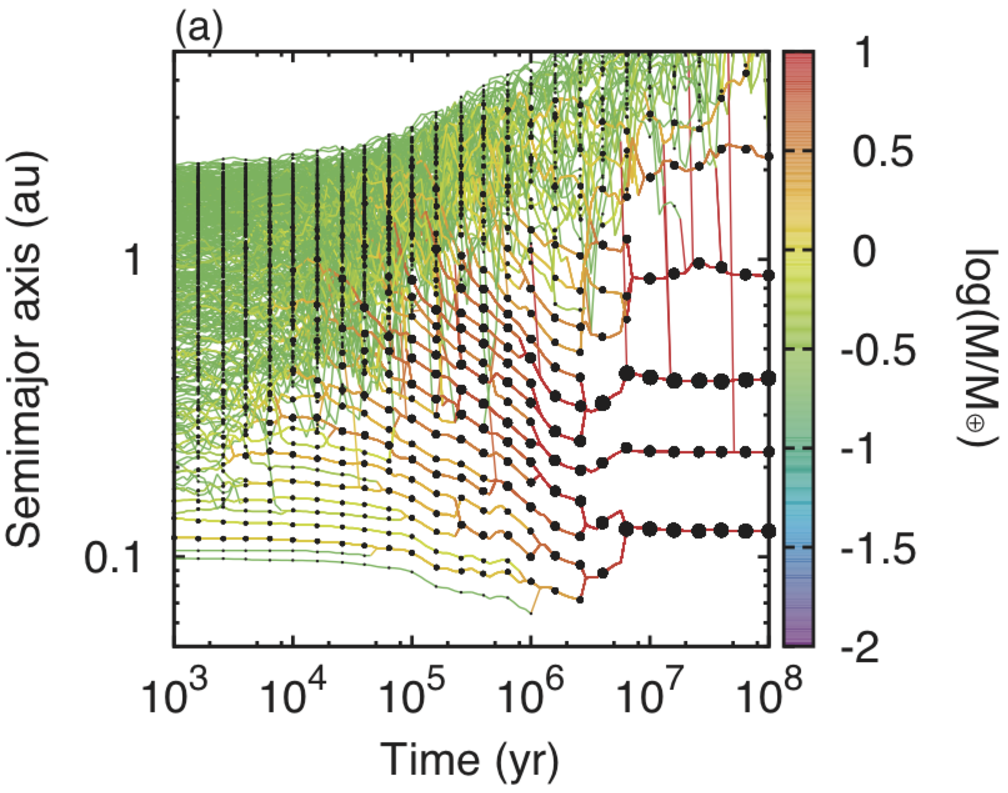}}
\resizebox{0.8 \hsize}{!}{\includegraphics{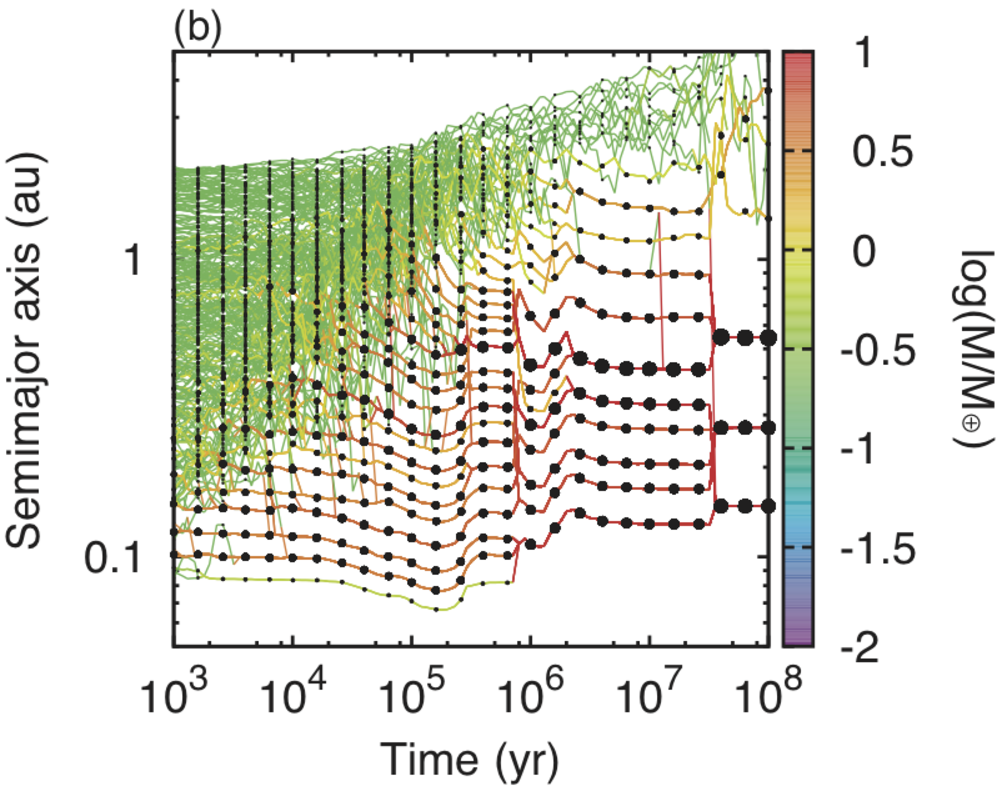}}
\caption{As in Fig.~\ref{fig:t_a11} but for 
model10 and model11.
}
\label{fig:t_a88}
\end{figure}

Now we present results of simulations in active disks that start simulations with different initial solid distributions, in order to see whether the initial slope of solid surface density affects the final configuration (especially the mass ratio distribution). The time evolution of the system for model10 and model11 is shown in Fig.~\ref{fig:t_a88}, in which the initial slope of solid surface density is -1 (model10) and -2 (model11). In panel~(a), the mass of grown planets increases with increasing $a$ for $a < 1 {\rm ~au}$. At $t = 0.1 {\rm ~Myr}$, the mass of the largest planet at $r = 0.1 {\rm ~au}$ is $M \simeq 1 ~M_\oplus$, while the largest planet at $r = 0.5 {\rm ~au}$ has a mass of $\simeq 5 ~M_\oplus$. The outer (larger) planets undergo faster migration than less massive planets; as a result, larger planets move to closer region. On the other hand, in panel~(b), planets at $r \simeq 0.1 {\rm ~au}$ grow to $M \simeq 5 ~M_\oplus$, and the system does not undergo significant inward migration. As a result, no significant difference is observed between models that start from different initial solid distributions.

Cumulative distributions of final configurations are shown in Fig.~\ref{fig:sum3}. For the mass-ratio distribution, we find, in Fig.~\ref{fig:sum1}(b) and Fig.~\ref{fig:sum3}(b), that there is no significant difference between model1, model10 and model11. So initial solid distributions do not affect the final configuration. In Fig.~\ref{fig:sum3}(a) and (c), although simulations for model11 obtain slightly larger period-ratio distribution, we do not see clear trends.

\end{document}